\documentclass[12pt,english,floatfix,superscriptaddress,aps,prd,preprint,showkeys]{revtex4}
\usepackage{amsmath}
\usepackage{amssymb}
\usepackage{amsbsy}
\usepackage{amsfonts}
\usepackage{amsopn}
\usepackage{amstext}
\usepackage{graphicx}
\usepackage{amssymb}
\usepackage{amsfonts}
\usepackage{amsmath}
\usepackage{graphicx}
\usepackage[english]{babel}
\usepackage{color}
\usepackage{slashed}
\usepackage{esint}
\usepackage[dvips]{epsfig}
\usepackage[dvips]{graphicx}
\usepackage{float}
\usepackage{units}
\usepackage{textcomp}

\usepackage{slashed}
\usepackage{graphicx}
\usepackage{amssymb}
\usepackage{amsmath}
\usepackage{color}

\begin{document}



\title{First-order formalism for thick branes in $f(T,\mathcal{T})$ gravity}


\author{A.R.P. Moreira}
\email{allan.moreira@fisica.ufc.br}
\affiliation{Universidade Federal do Cear\'a (UFC), Departamento de F\'isica,\\ Campus do Pici, Fortaleza - CE, C.P. 6030, 60455-760 - Brazil.}

\author{F.C.E. Lima} 
\email{cleiton.estevao@fisica.ufc.br}
\affiliation{Universidade Federal do Cear\'a (UFC), Departamento de F\'isica,\\ Campus do Pici, Fortaleza - CE, C.P. 6030, 60455-760 - Brazil.}

\author{J.E.G. Silva}
\email{euclides.silva@ufca.edu.br}
\affiliation{Universidade Federal do Cariri(UFCA), Av. Tenente Raimundo Rocha, \\ Cidade Universit\'{a}ria, Juazeiro do Norte, Cear\'{a}, CEP 63048-080, Brazil.}


\author{C.A.S. Almeida}
\email{carlos@fisica.ufc.br}
\affiliation{Universidade Federal do Cear\'a (UFC), Departamento de F\'isica,\\ Campus do Pici, Fortaleza - CE, C.P. 6030, 60455-760 - Brazil.}

\begin{abstract}


In this paper we study the thick brane  scenario constructed in the recently proposed $f(T,\mathcal{T})$ theories of gravity, where $T$ is called the torsion scalar and $\mathcal{T}$ is the trace of the energy-momentum tensor. We use the  first-order formalism to find analytical solutions for models that include a scalar field as a source.   In particular, we describe two interesting case in which in the first  we obtain a double-kink solution,  which generates a splitting in the brane. In the second case, proper management of a kink solution obtained generates a splitting in the brane intensified by the torsion parameter, evinced by the energy density components satisfying the weak and strong energy conditions. In addition, we investigate the behavior of the gravitational perturbations in this scenario. The parameters that control the  torsion and the trace of the energy-momentum tensor tend to shift the massive modes to the core of the brane, keeping a gapless non-localizable and stable tower of massive modes and producing more localized massless modes.

\end{abstract}


\maketitle

\section{Introduction}

This work deals with braneworld models \cite{Akama1982,Visser1985,ArkaniHamed1998,rs,rs2}, which has as its fundamental idea that the visible universe is localized on a 3-brane which is embedded in a higher-dimensional bulk. The  braneworld model proposed by Randall and Sundrum (RS) is a thin brane model \cite{rs,rs2}, where the five-dimensional geometry is a slice of $AdS_5$ spacetime due to the negative cosmological constant in the bulk.  The RS model allowed a new approaches to address of  outstanding issues such as the hierarchy problem \cite{rs2}, the cosmological problem \cite{cosmologicalconstant}, the nature of dark matter \cite{darkmatter} and dark energy, which made this theory gain a lot of attention since it was proposed. Adding a scalar field, in a modification of the RS model, makes the warp function behave smoothly, leading to a thick braneworld scenario \cite{Goldberger1999,Gremm1999, DeWolfe1999,Bazeia2008,Dzhunushaliev2009}. This possibility has opened a new area of study, where various thick brane models have been investigated, as seen in Refs. \cite{Charmousis2001, Arias2002, Barcelo2003, Bazeia2004, CastilloFelisola2004, Navarro2004, BarbosaCendejas2005, Bazeia2007, Koerber2008,deSouzaDutra2008,conifold,Almeida2009, Cruz2013, Liu2011, Dutra2014}.

Known as $f(R)$ gravity \cite{DeFelice2010, Nojiri2011}, a change in the usual gravity theory could be considered as an extension of General Relativity (GR),  where $R$ is the scalar of curvature. Another possibility would be $f(R,\mathcal{T})$ gravity, where $\mathcal{T}$ is the trace of the energy-momentum tensor \cite{Harko2011k}. The gravitational modification $ f (R, \mathcal {T}) $ motivated the investigation of several models of thick branes, as can be seen in Refs. \cite{Bazeia2015,Moraes2015d,Correa2015q,Gu2016o}. As a matter of fact, all of these works do not consider the contribution of spacetime torsion. However, torsion can leads to the so-called teleparallel equivalent of general relativity (TEGR) \cite{Hayashi1979, deAndrade1997, deAndrade1999, Aldrovandi}. The TEGR  is constructed using the Weitzenb\"{o}ck connection instead of the  Levi-Civita connection of GR, which leads to a vanishing curvature but a non-vanishing torsion, where the fundamental dynamical quantity of the theory is a tetrad field. In such a formulation, the contribution of torsion in the gravitational Lagrangian results from contractions of the torsion tensor and it is called the torsion scalar $T$.

Similarly to the $f(R)$ extensions of GR, one can construct $f(T)$ extensions of TEGR \cite{Ferraro2007, Ferraro2011}. An interesting feature in this extension is that although TEGR is equivalent to GR, $f(T)$ is not equivalent to $f(R)$ gravity. This feature guarantees that the field equations in $f(T)$ theories are second-order ones, which represent an advantage over the $f(R)$ theories that have fourth-order equations. This gravity modification motivated the investigation of several models of thick brane, as can be seen in Refs. \cite{Yang2012, Capozziello, Menezes, tensorperturbations, ftnoncanonicalscalar, ftborninfeld, ftmimetic}. Thus, a new possibility of modified teleparallel gravity arose, namely, $f(T,\mathcal{T})$ gravity \cite{Harko2014}. This new modified gravity model has gained a lot of attention recently due significant results obtained in cosmological perturbations and thermodynamics \cite{Harko2014,Momeni2014,Nassur2015,Ganiou2016,Ganiou2015, Junior2015,Farrugia2016,Rezaei2017,Arouko2020g}. Also, interesting astrophysical applications  \cite{Saez2016,Pace2017,Pace2017d,Salako2020t}, new developments in dark energy \cite{Salako2015,Mirzaei2020}, and in a viable alternative to black hole (called Gravstar) \cite{Ghosh2020}, were obtained through ideas of $f(T,\mathcal{T})$ gravity.

The increasing interest in modified teleparallel gravity and in the significant results obtained in $f(T,\mathcal{T})$ gravity, that seems as an alternative to GR, inspired us to investigate the impact of torsion (T) and of  trace of the energy-momentum tensor ($\mathcal{T}$) on the structure of branes. Also, the role of T and $\mathcal{T}$ in the localization of gravity on the  branes is another interesting point to be studied. Both issues are investigated with the help of first-order formalism.


The paper is organized as follows. In Section (\ref{sec1}) we review the main definitions of the teleparallel gravity theory and we introduce the $f(T,\mathcal{T})$ theory. Then we give the field equations for the five-dimensional braneworld.  In Sec. (\ref{sec2}), we obtain analytical brane solutions by considering different forms of super-potentials and we examine the energy density behavior in the brane. In section (\ref{sec3}), we focus on tensor and scalar perturbations, and explore the gravitational Kaluza-Klein (KK) modes. Finally, additional comments are discussed in Section (\ref{sec4}).

\section{Brane in $f(T,\mathcal{T})$ gravity}
\label{sec1}

In this section we present the main concepts of the modifield teleparallel $f(T,\mathcal{T})$ gravity and obtain the modified gravitational equations for the braneworld scenario.

 As it is done in teleparallel gravity, the modified versions of this theory are also described by the vielbeins and its components are defined on the tangent space of each point of the manifold. The spacetime metric can be constructed from the vielbein 
\begin{eqnarray}\label{01}
g_{MN}=\eta_{ab}h^a\ _M h^b\ _N, 
\end{eqnarray}
where the capital latin index $M={0,...,D-1}$ are the bulk coordinate indexes and the latin index $a={0,...,D-1}$ is a vielbein index, so $\eta_{ab}=diag(-1,1,1,1,1)$ is the metric for the tangent space. From the relation (\ref{01}), we have
\begin{eqnarray}\label{02}
h_a\ ^M h^a\ _N=\delta^M_N,\ \    h_a\ ^M h^b\ _M=\delta^b_a.    
\end{eqnarray}

In the teleparallel theory and its extended versions, one keeps the scalar torsion by using a curvature free connection, known as the Weitzenb\"{o}ck connection, defined by
\begin{eqnarray}\label{a.685}
\widetilde{\Gamma}^P\ _{NM}=h_a\ ^P\partial_M h^a\ _N.
\end{eqnarray}
From the above connection, one obtains the geometric objects of the formalism. The torsion tensor is defined by
\begin{eqnarray}
T^{P}\  _{MN}= \widetilde{\Gamma}^P\ _{NM}-\widetilde{\Gamma}^P\ _{MN},
\end{eqnarray}
from which we define the contorsion tensor as
\begin{eqnarray}
K^P\ _{NM}\equiv\widetilde{\Gamma}^P\ _{NM}- \Gamma^P\ _{NM}=\frac{1}{2}\Big( T_N\ ^P\ _M +T_M\ ^P\ _N - T^P\ _{NM}\Big),
\end{eqnarray}
where  $\Gamma^P\ _{NM}$ stands for the Levi-Civita connections \cite{Aldrovandi}.

The torsion and contorsion tensors are used to define so-called superpotential torsion tensor as
\begin{eqnarray}
S_{P}\ ^{MN}=\frac{1}{2}\Big( K^{MN}\ _{P}-\delta^N_P T^{QM}\ _Q+\delta^M_P T^{QN}\ _Q\Big),
\end{eqnarray} 
such that the torsion scalar can be constructed from torsion and superpotential as
\begin{eqnarray}
T\equiv T_{PMN}S^{PMN}=\frac{1}{2} T^{P}\  _{MN} T_{P}\ ^{MN} +T^{P}\ _{MN}T^{NM}\ _{P}-2T^{P}\ _{MP}T^{NM}\ _{N}.
\end{eqnarray} 
Therefore, the Lagrangian of TEGR reads $\mathcal{L}= -hT/4\kappa_g$, where $h=\sqrt{-g}$, with $g$ the determinant of the metric and $\kappa_g=4\pi G/c^4$ is the gravitational constant \cite{Aldrovandi}.

A modified gravity theory can be accomplished by considering as the gravitational Lagrangian a function of $T$ and the term proportional to
the trace of the energy-momentum tensor $\mathcal{T}$, leading to $f(T,\mathcal{T})$ gravity \cite{Harko2014}. We assume a five-dimensional bulk $f(T,\mathcal{T})$ gravity in the form  
\begin{eqnarray}\label{55.5}
\mathcal{S}=\int h\Big( \frac{1}{4\kappa_g}f(T,\mathcal{T})+\mathcal{L}_m\Big)d^5x,
\end{eqnarray}
where $\mathcal{L}_m$ is the matter Lagrangian.
By varying the action with respect to the vierbein we
obtain the following field equations \cite{Harko2014}
\begin{eqnarray}\label{3.36}
\Big(f_{TT}\partial_QT+f_{T\mathcal{T}}\partial_Q\mathcal{T} \Big)S_N\ ^{MQ}+\frac{1}{h}f_T\Big[\partial_Q(h S_N\ ^{MQ})-h\widetilde{\Gamma}^R\ _{SN}S_R\ ^{MS}\Big]& &\nonumber\\
+\frac{1}{4}f\delta_N^M-\frac{1}{2}f_{\mathcal{T}}\Big(\mathcal{T}_N\ ^M+p\delta_N^M\Big)&=&\kappa_g\mathcal{T}_N\ ^M,
\end{eqnarray}
where $f\equiv f(T,\mathcal{T})$, $f_T\equiv\partial f(T,\mathcal{T})/\partial T$, $f_{\mathcal{T}}\equiv\partial f(T,\mathcal{T})/\partial \mathcal{T}$, $f_{T\mathcal{T}}\equiv\partial^2 f(T,\mathcal{T})/\partial T\partial \mathcal{T}$  and $\mathcal{T}_N\ ^M$ is the stress-energy tensor, which in terms of the matter Lagrangian is given by 
\begin{eqnarray}
\mathcal{T}_a\ ^M=-\frac{\delta\mathcal{L}_m}{\delta h^a\ _M}.
\end{eqnarray}

In our model, we take a Lagrangian given by
\begin{equation}
\mathcal{L}_m=-\frac{1}{2}\partial^M\phi\partial_M\phi-V(\phi),
\end{equation}
where $\phi\equiv \phi(y)$ is a background scalar
field that generates the brane. From the equation above, the energy-momentum tensor of the scalar field in this theory is given by
\begin{equation}\label{03}
\mathcal{T}_{MN}=\partial_M\phi\partial_N\phi-\frac{1}{2}g_{MN}g^{PQ}\partial_P\phi\partial_Q-g_{MN}V(\phi).
\end{equation}

In order to construct a thick braneworld model, we use the metric ansatz
\begin{equation}\label{45.a}
ds^2=e^{2A(y)}\eta_{\mu\nu}dx^\mu dx^\nu+dy^2,
\end{equation}
where $\eta_{\mu\nu}=(-1, 1, 1, 1)$ is the four-dimensional
Minkowski metric and $e^{A(y)}$ is the so-called warp factor. Since $f(T)$
gravity is not invariant under local Lorentz transformation, different choices of the vielbein will correspond to different solutions. So the choice of the vielbein is an important issue. Accordingly, adopting the \textit{f\"{u}nfbein} in the form 
\begin{eqnarray}\label{04}
h^a\ _M=diag(e^A, e^A, e^A, e^A, 1). 
\end{eqnarray}
The torsion scalar is
\begin{eqnarray}
 T=-12A'^2,
\end{eqnarray}
and with Eq.(\ref{03}) we get the trace of the energy-momentum tensor
\begin{eqnarray}
 \mathcal{T}=-\frac{3}{2} \phi'^2-5V,
\end{eqnarray}
where the prime $(\ '\ )$ denotes differentiation with respect to $y$. 

 The explicit field equations (\ref{3.36}) and the equation of motion for
the scalar field are given by

\begin{eqnarray}
\label{scalarfieldeom}
\Big(1+\frac{3}{4}f_{\mathcal{T}}\Big)\phi''+\Big[(4+3f_{\mathcal{T}})A'+\frac{3}{4}f_{\mathcal{T}}'\Big]\phi'&=&\Big(1+\frac{5}{4}f_{\mathcal{T}}\Big)\frac{\partial V}{\partial \phi}\\
\label{e.1}
\frac{3}{2}(A''+4A'^2)f_{T}+\frac{3}{2}A'f_{T}'+\frac{1}{4}f&=& \frac{\phi'^2}{2}+V,\\
\label{e.2}
\frac{1}{4}f+6A'^2f_T&=&-\frac{\phi'^2}{2}(1+f_{\mathcal{T}})+V,
\end{eqnarray}
where $\kappa_g=1$ for simplicity.
The equations (\ref{scalarfieldeom}), (\ref{e.1}) and (\ref{e.2})  form a quite intricate system of coupled equations. Note that we can rewrite equations (\ref{e.1}) and (\ref{e.2}) as
\begin{eqnarray}
3A''+12A'^2&=& \frac{2}{f_T}\Big(\rho-\rho_{T\mathcal{T}} \Big),\label{0.00005}\\
6A'^2&=&\frac{1}{f_T}\Big(p-p_{T\mathcal{T}} \Big),\label{0.00006}
\end{eqnarray}
where
\begin{eqnarray}
 \rho_{T\mathcal{T}}&=&\frac{3}{2}A'f_{T}'+ \frac{1}{4}f,\\
 p_{TB}&=&\frac{1}{4}f+\frac{\phi'^2}{2}f_{\mathcal{T}}.
\end{eqnarray}
Note that the left side of equations (\ref{0.00005}) and (\ref{0.00006}) are equivalent to that obtained in TEGR. So, we can state that modified gravity equations of motion of $f(T,\mathcal{T})$ gravities are similar to an inclusion of an additional source with $\rho_{T\mathcal{T}}$ and $p_{T\mathcal{T}}$. 

The diagonal tetrad (\ref{04}) represents a good choice among all the possible vielbeins giving metric (\ref{45.a}). In fact, the gravitational field equations do not involve any additional constraints on the function $f(T,\mathcal{T})$ or the scalar $T$ and the trace of the energy-momentum $\mathcal{T}$.
For a braneworld scenario, the diagonal tetrad (\ref{04}) turned out to be a good tetrad for both a $f(T)$ gravity \cite{Yang2012, Capozziello, Menezes, tensorperturbations, ftnoncanonicalscalar, ftborninfeld, ftmimetic}, and a $f(T,B)$ gravity \cite{Moreira2021, Moreira2021a}.
 Thus, the choice in Eq. (\ref{04}) can be regarded as a "good vielbein". Similarly, in the FRW cosmological models the $f(T,\mathcal{T})$ gravitational dynamics preserves the form of the usual Friedmann equations (two equations) \cite{Harko2014,Momeni2014,Saez2016,Mirzaei2020,Cai2015}. 

In fact, the problem of frame dependence and violation of location Lorentz invariance, which is related to good and bad choice of tetrad, is a consequence of neglecting the role of spin connection \cite{Krssak2015}. This issue is resolved introducing covariant $f(T)$ gravity, which uses both the tetrad and the spin connection as variables. That is, the covariant gravity $f(T)$ allow the use of an arbitrary tetrad in an arbitrary coordinate system, along with the corresponding spin connection, always resulting in the same physically relevant field equations. For more details see Ref.\cite{Krssak2015}.

In this work we consider two power-law modified gravity in the form $f(T,\mathcal{T})=k_0\mathcal{T}+kT^{n}$, where $k$ and $n$ are parameters controlling the influence of torsion and $k_0$ controls the influence of the trace of the energy-momentum tensor, and $f(T,\mathcal{T})=-T-k_1T^{2}+k_2\mathcal{T}$, where $k_1$ is parameter that controls the influence of torsion and $k_2$ controls the influence of the trace of the energy-momentum tensor.  We chose these particular $f(T,\mathcal{T})$ models because of their simplicity and inspired by the models proposed in gravity $f(R,\mathcal{T})$, which can be seen in Refs. \cite{Bazeia2015,Moraes2015d,Correa2015q,Gu2016o}.

\section{First-order formalism for thick brane models}
\label{sec2}

The first-order formalism is a very powerful tool to obtain
analytical brane solutions \cite{Gremm1999,Afonso2006,Janssen2007}. This formalism appears for the first time in the study of supergravity domain walls \cite{Cvetic1992} and was generalized in Refs. \cite{DeWolfe1999,Skenderis1999} to include non-supersymmetric domain walls in several space-time dimensions.  With this formalism, the second-order coupled field equations can be written as a set of first-order ones by introducing one or more auxiliary super-potentials. Very recently, first order formalism has been used to find brane solutions in $f(T)$ gravity models \cite{Menezes,ftborninfeld}. Here we investigate applications for generalized models within the context of modified teleparallel gravity $f(T,\mathcal{T})$.

\subsection{$f(T,\mathcal{T})=k_0\mathcal{T}+kT^{n}$}
We can write equations (\ref{e.1}) and (\ref{e.2}) for the case where $f(T,\mathcal{T})=k_0\mathcal{T}+kT^{n}$, which take the form
\begin{eqnarray}
\label{e.10}
\frac{1}{4(2+k_0)}\Big(B_n k n A'^{2n-2}\Big)A''&=& \phi'^2,\\
\label{e.20}
B_nkA'^{2n}&=&-\frac{\phi'^2}{2}(4+k_0)+V(4+5k_0),
\end{eqnarray}
where $B_n=(-1)^{n-1}12^n(2n-1)$. With the aim of introducing the first-order formalism, we choose the derivative of the warp factor with respect to the extra dimension to be a function of the scalar field, namely,
\begin{eqnarray}\label{09a}
A'=-\alpha W(\phi).
\end{eqnarray}
In this case, from the first-order equation (\ref{e.10}) we get
\begin{eqnarray}\label{07}
\phi'=\frac{1}{4(2+k_0)}\Big[B_n k n (-\alpha)^{2n-1}W^{2n-2}\Big] W_\phi.
\end{eqnarray}
The potential can be found by substituting this equation in Eq.(\ref{e.20}) such that
\begin{eqnarray}\label{08}
V(\phi)&=&\Bigg(\frac{1}{4+5k_0}\Bigg)\Bigg\{B_n k n (-\alpha W)^{2n}\nonumber\\&+&\frac{4+k_0}{2}\Bigg[\frac{1}{4(2+k_0)}B_n k n (-\alpha)^{2n-1}W^{2n-2}\Bigg]^2W_\phi^2\Bigg\}.
\end{eqnarray}
By making the parameters $k_{0}=0$, $k=-1$, $n=1$ and $\alpha=\frac{1}{3}$, the equations (\ref{07}) and (\ref{08}) take the form 
\begin{eqnarray}
\label{oo.3}\phi'&=&\frac{ W_\phi}{2},\\ 
\label{oo.4}V(\phi)&=&\frac{W_\phi ^2}{8}-\frac{W^2}{3},
\end{eqnarray}
and the standard scenario is restored \cite{Gremm1999}.

We can obtain the energy density through the expression
\begin{eqnarray}\label{05}
\rho(y)=-e^{2A(y)}\mathcal{L}_m.
\end{eqnarray}
In our case, the energy density (\ref{05}) is
\begin{eqnarray}\label{06}
\rho(y)&=&\Bigg(\frac{e^{2A}}{4+5k_0}\Bigg)\Bigg\{B_n k n (-\alpha W)^{2n}\nonumber\\ &+&(4+3k_0)\Bigg[\frac{1}{4(2+k_0)}B_n k n (-\alpha)^{2n-1}W^{2n-2}\Bigg]^2W_\phi^2\Bigg\}.
\end{eqnarray}
Now, the solutions of the thick brane system are completely determined by the so-called superpotential function $W(\phi)$ required to be specified. We concentrate on the simplest case, with $n = 1$ and $k_0=1$. The first example is the periodical superpotential
\begin{eqnarray}
W(\phi)=\beta^2\sin\Big(\frac{\phi}{\beta}\Big).
\end{eqnarray}
In this way the equations (\ref{07}) and (\ref{08}) give us the  thick brane solution
\begin{eqnarray}
\phi(y)&=&\beta\ \mathrm{arcsin}\Big[\tanh(-12k\alpha y)\Big],\\
V(\phi)&=&\frac{1}{9}\Big\{12 k\alpha^2\beta^4 \sin^2\Big(\frac{\phi}{\beta}\Big)+\frac{5}{3}\Big[ k \alpha\beta\cos\Big(\frac{\phi}{\beta}\Big)\Big]^2\Big\}.
\end{eqnarray}
Then we get a different form of the warp factor  through  Eq.(\ref{09a}), i. e.
\begin{eqnarray}\label{09}
A(y)=-\frac{\beta^2}{k}\ln[\mathrm{sech}(k\alpha y)].
\end{eqnarray}
Now, the energy density for the above brane system can be expressed as
\begin{eqnarray}
\rho(y)=\frac{k}{54}\alpha^2\beta^2\cosh(k\alpha y)^{2\left(\frac{\beta^2}{k}-1\right)}\Big\{37k-36\beta^2[1+\coth(2k\alpha y)]\Big\}.
\end{eqnarray}

\begin{figure}
\begin{center}
\begin{tabular}{ccc}
\includegraphics[height=5cm]{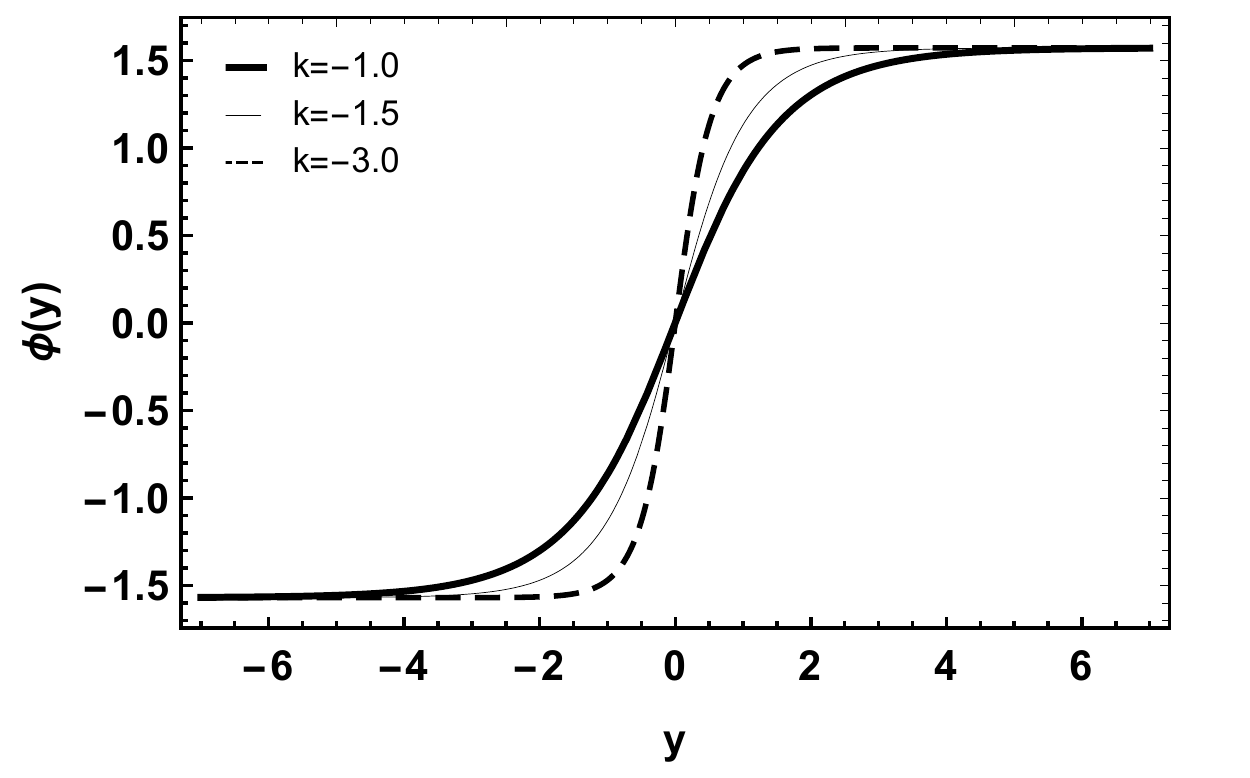}
\includegraphics[height=5cm]{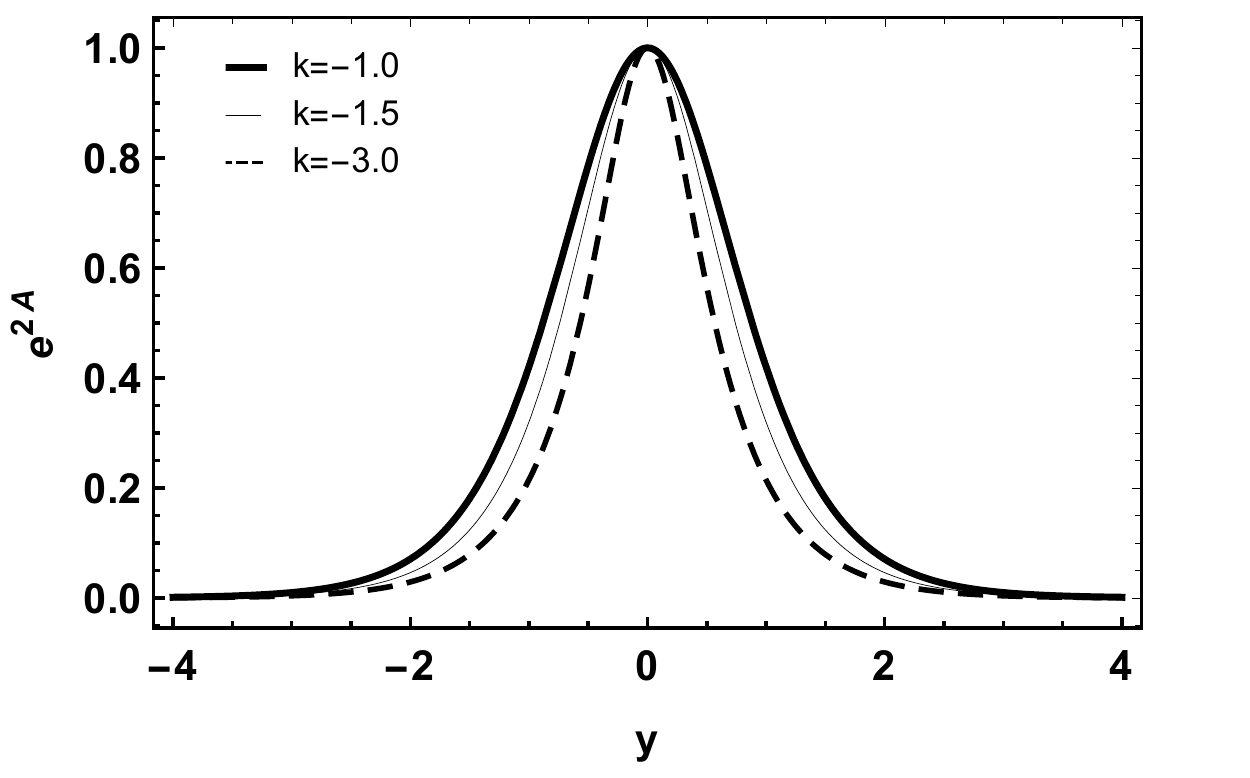}\\ 
(a) \hspace{8 cm}(b)\\
\includegraphics[height=5cm]{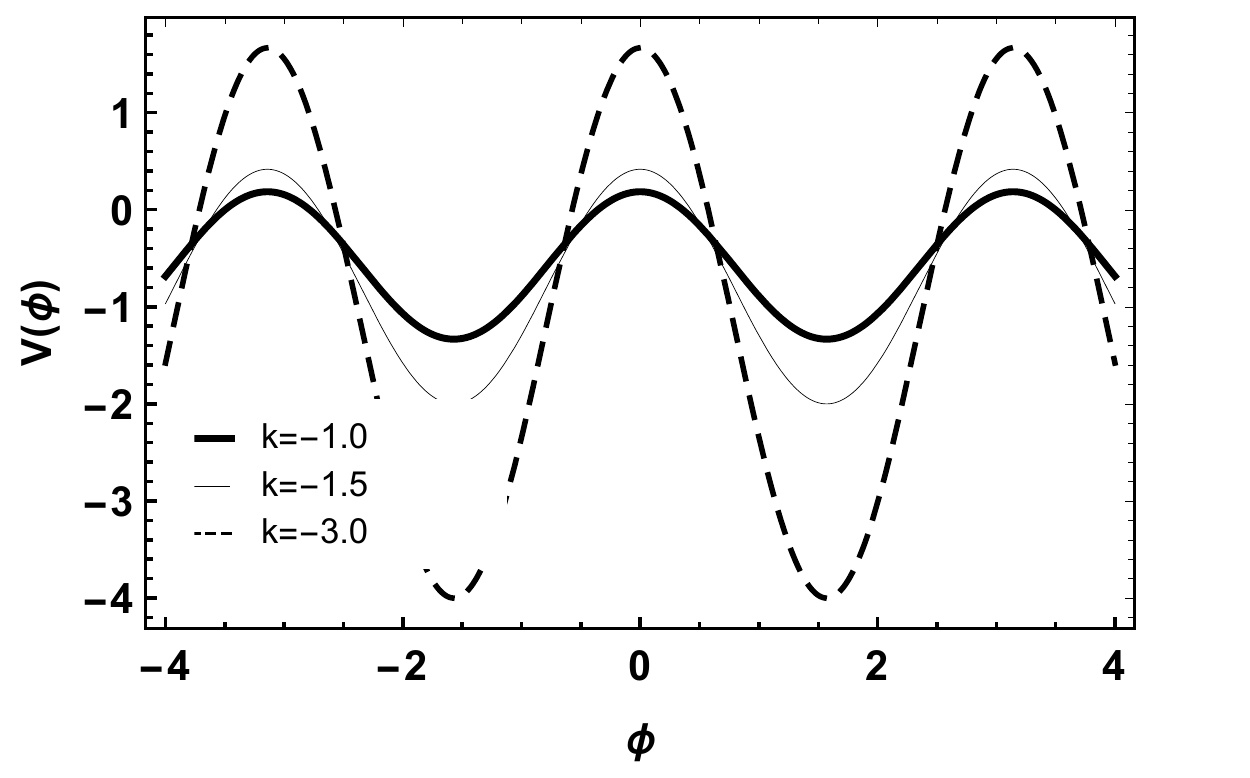}
\includegraphics[height=5cm]{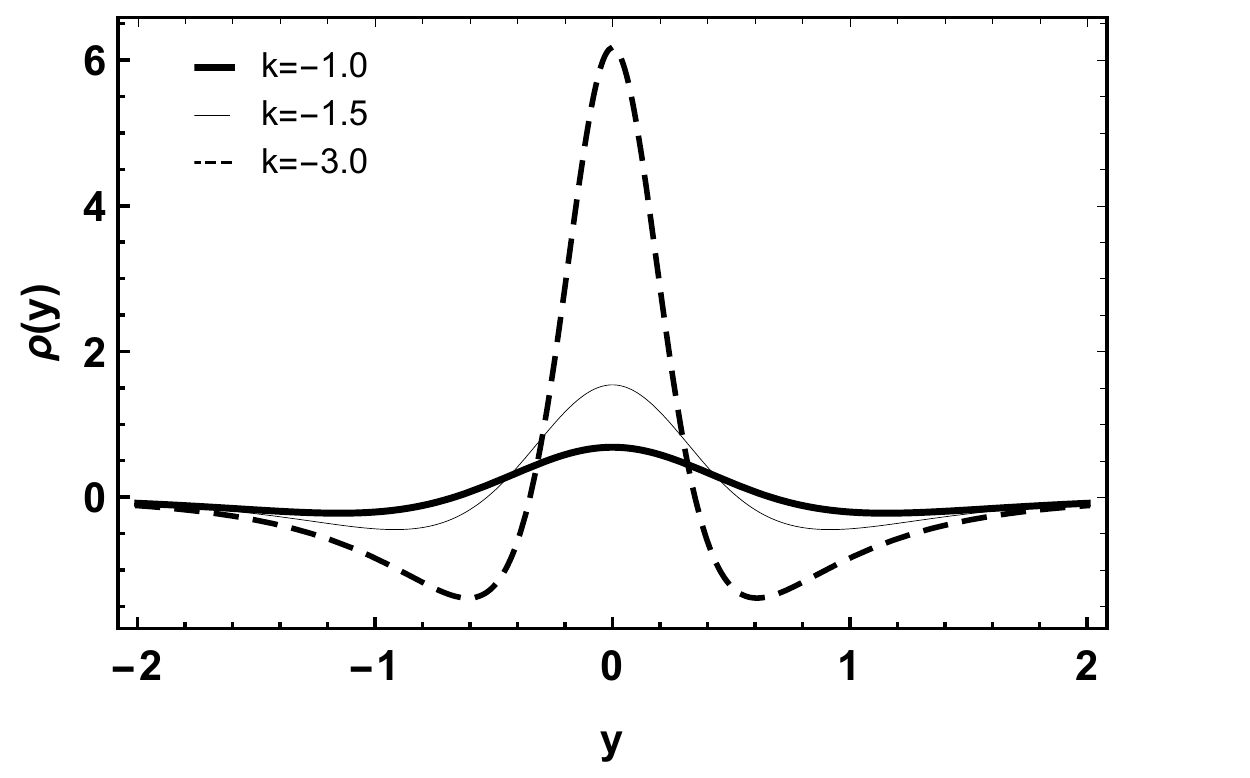}\\ 
(c) \hspace{8 cm}(d)
\end{tabular}
\end{center}
\caption{Plots of the kink solution (a), warp factor (b), potential (c), and energy density (d), for the periodical superpotential, where $\alpha=\beta=1$. 
\label{figene1}}
\end{figure}

In Fig.\ref{figene1}, we plotted the  profile of the kink solution $\phi(y)$, the warp factor $e^{2A}$, the potential $V(\phi)$, and the energy density $\rho(y)$ for the periodical superpotential varying the parameter $k$.  We note that the thickness
of the solution is controlled essentially by $k$ (Figure \ref{figene1} $a$), where the warp factor narrows
as $k$ decreases (Figure \ref{figene1} $b$), modifying also the potential (Figure \ref{figene1} $c$) and the energy density (Figure \ref{figene1} $d$). When $k$ is negative, the scalar field profile guarantees topological stability. For positive $k$, we cannot guarantee this topological stability, making the energy density of the brane physically unpleasant.

We can notice that when the extra dimension $y$ runs from one boundary $y \rightarrow -\infty$ to the other $y \rightarrow \infty$, the scalar field $\phi(y)$ runs smoothly from $\phi(-\infty) \rightarrow-\frac{\pi\beta}{2}$ to $\phi(\infty) \rightarrow\frac{\pi\beta}{2}$ with $k\alpha<0$, where the vacuum potentials $V\left(\pm\frac{\pi\beta}{2}\right)=\frac{4}{3}k\alpha^2\beta^4$ are just located. So the scalar field is indeed a kink solution. From the asymptotic behaviors of the warp factor
$A(y\rightarrow\pm\infty)\rightarrow\alpha\beta^2|y|$,
one can conclude that the spacetimes for the brane system are asymptotically anti-de Sitter along the fifth dimension.

The second example are the polynomial superpotentials
\begin{eqnarray}
W(\phi)=\beta\phi-\frac{\phi^3}{3\beta}.
\end{eqnarray}
The equations (\ref{07}) and (\ref{08}) give us the  thick brane solution
\begin{eqnarray}
\label{11}\phi(y)&=&-\beta\ \tanh(k\alpha y),\\
V(\phi)&=&\frac{1}{9}\Big\{12 k\alpha^2\Big(\beta\phi-\frac{\phi^3}{3\beta}\Big)+\frac{5}{3} \Big(\frac{k\alpha}{\beta}\Big)^2(\beta^2-\phi^2)^2\Big\}.
\end{eqnarray}
The warp factor is obtained from Eq.(\ref{09a}), such that
\begin{eqnarray}\label{10}
A(y)=-\frac{\beta^2}{6k}\Big\{\mathrm{sech}^2(k\alpha y)+4\ln[\mathrm{sech}(k\alpha y)]\Big\}.
\end{eqnarray}
The energy density of the brane system above can be expressed as
\begin{eqnarray}
\rho(y)&=&\frac{k}{54}\alpha^2\beta^2\exp\Big[-\frac{\beta^2}{3k}\mathrm{sech}^2(k\alpha y)\Big]\cosh(k\alpha y)^{\frac{4\beta^2}{3k}}\Big[32\beta^2\nonumber\\&+&(37k-24\beta^2\mathrm{sech}^4(k\alpha y))-8\beta^2\mathrm{sech}^6(k\alpha y)\Big].
\end{eqnarray}

\begin{figure}
\begin{center}
\begin{tabular}{ccc}
\includegraphics[height=5cm]{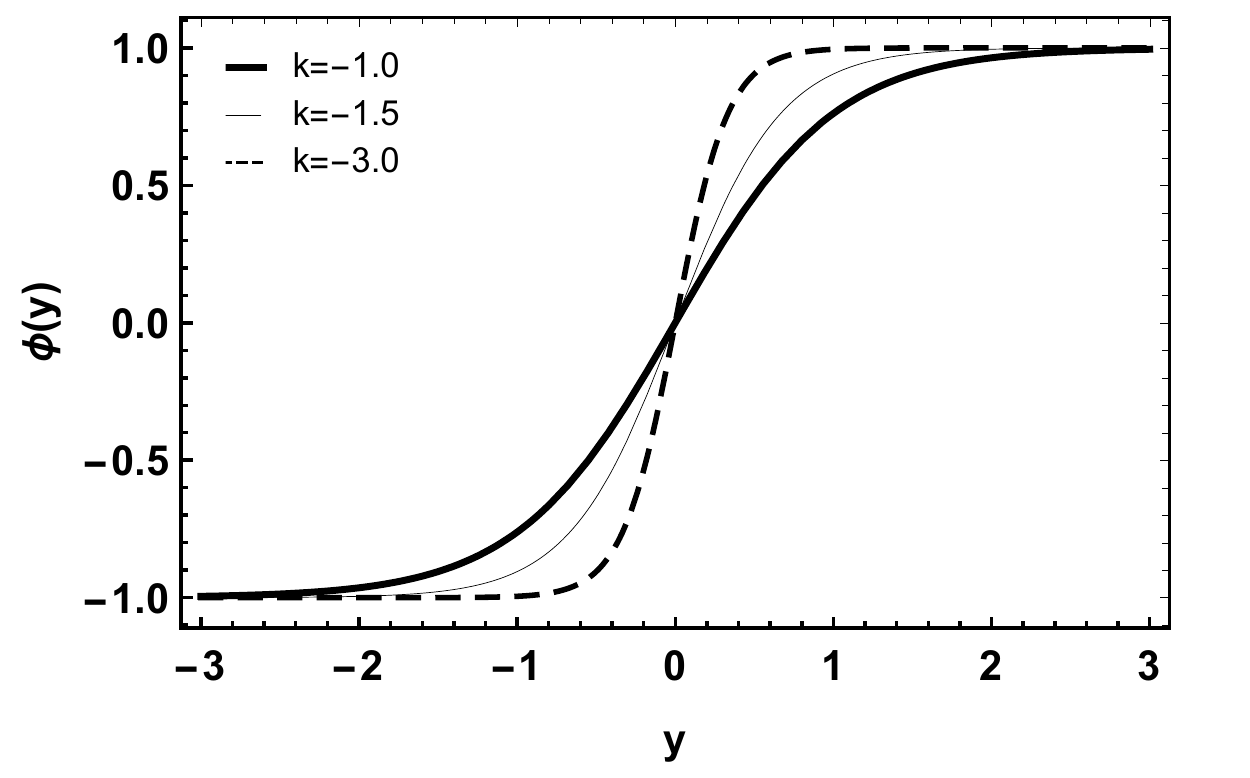}
\includegraphics[height=5cm]{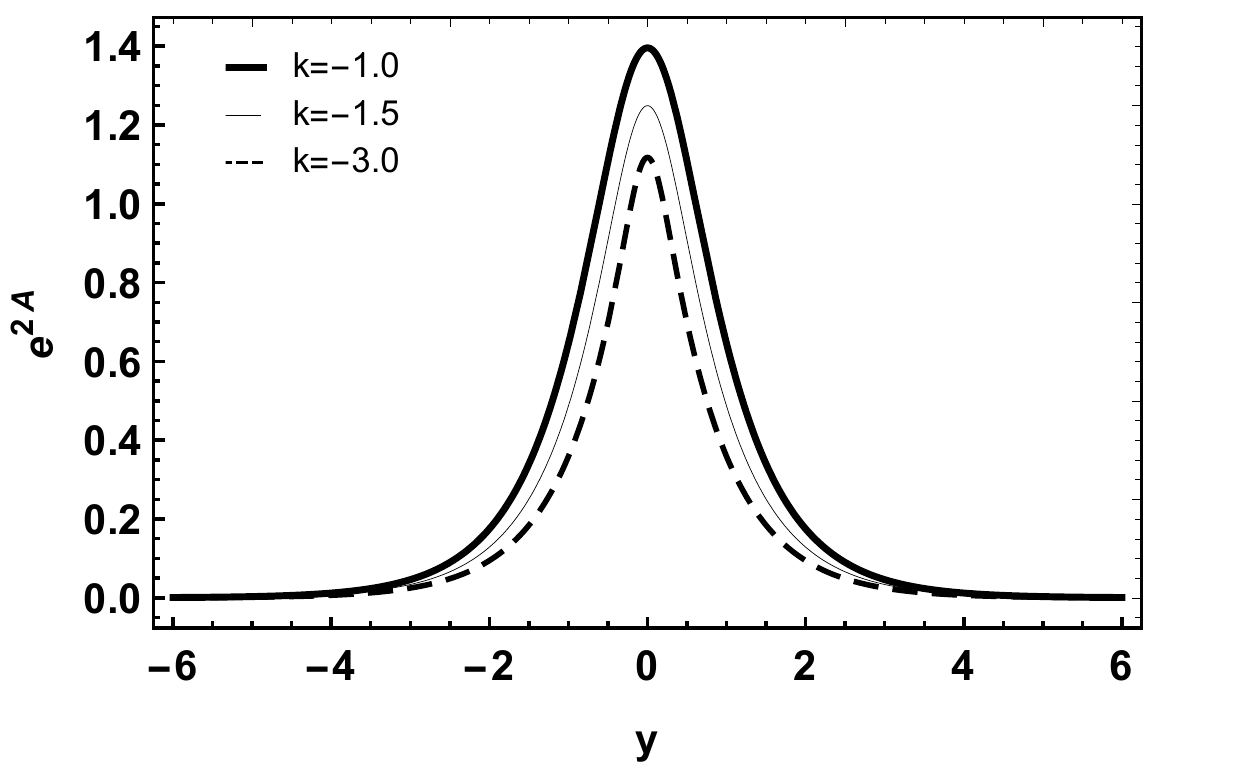}\\ 
(a) \hspace{8 cm}(b)\\
\includegraphics[height=5cm]{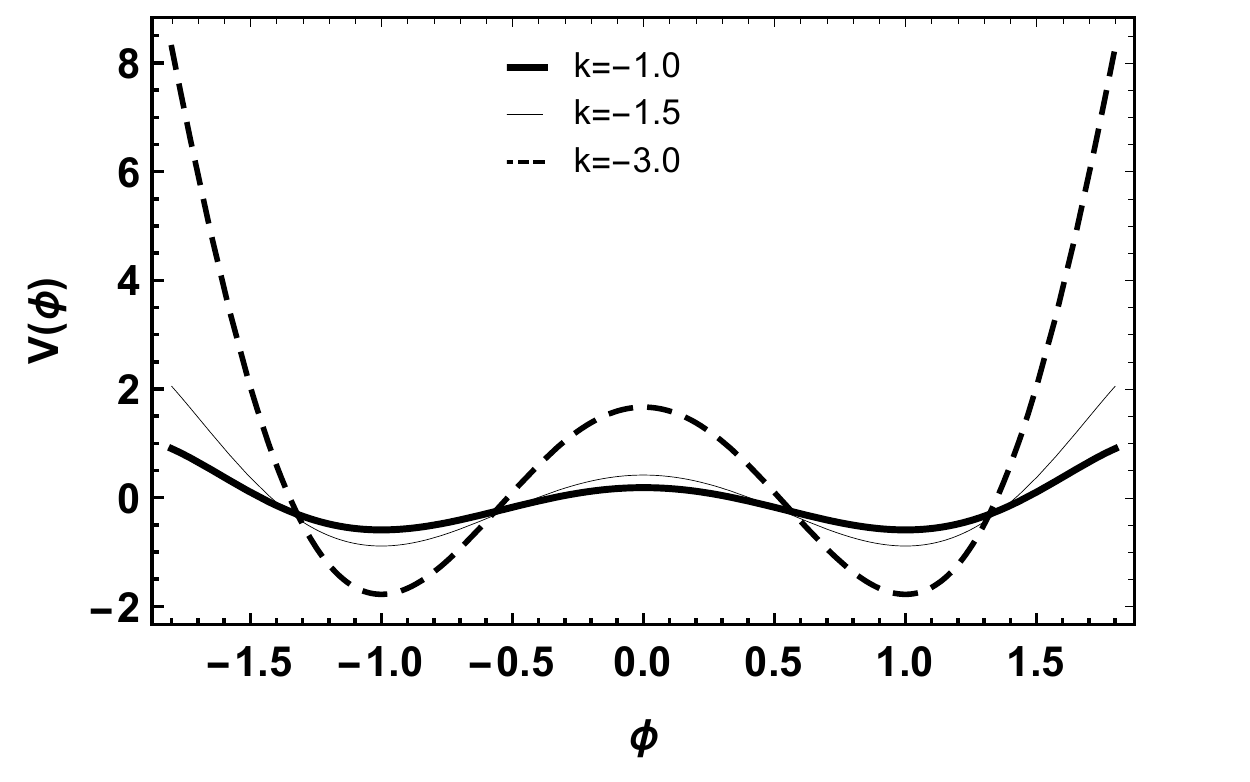}
\includegraphics[height=5cm]{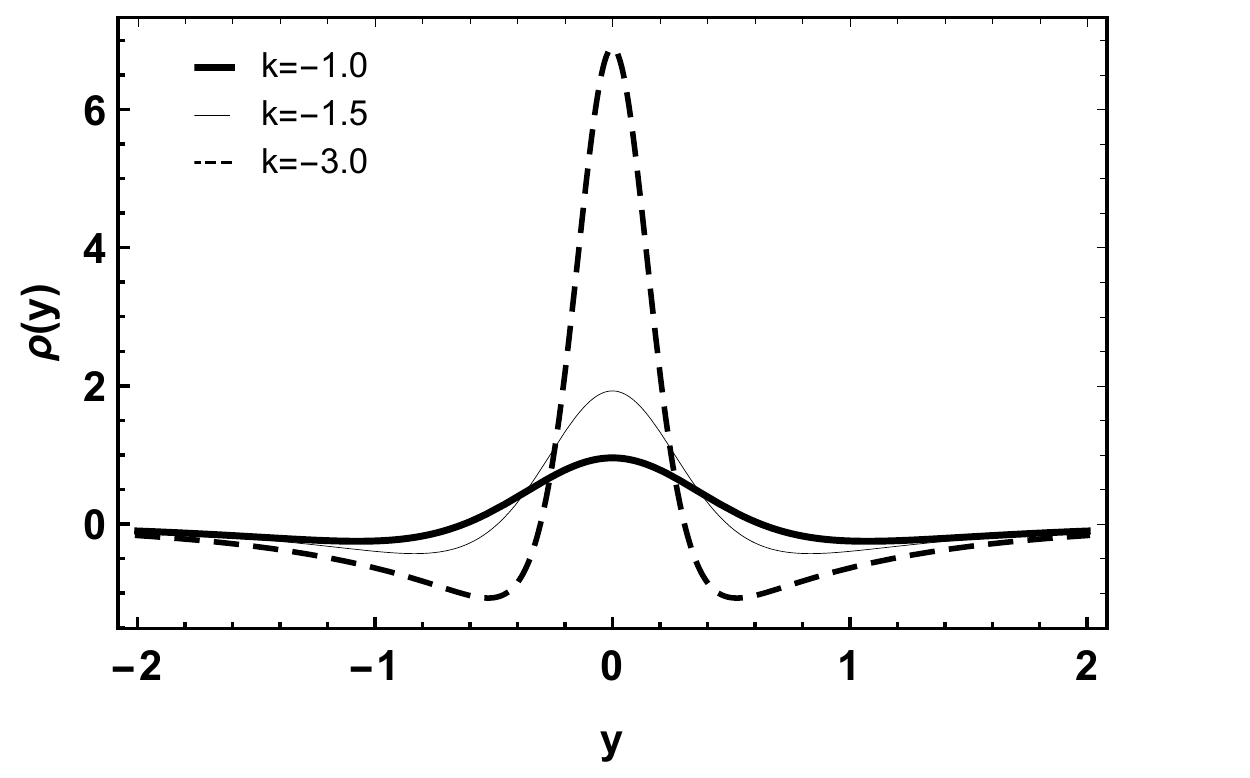}\\ 
(c) \hspace{8 cm}(d)
\end{tabular}
\end{center}
\caption{Plots of the kink solution (a), warp factor (b), potential (c), energy density (d), for the polynomial superpotential, where $\alpha=\beta=1$. 
\label{figene2}}
\end{figure}

In Fig.\ref{figene2}, we plotted the  profile of the kink solution $\phi(y)$, the warp factor $e^{2A}$, the potential $V(\phi)$, and the energy density $\rho(y)$ for the polynomial superpotential varying the parameter $k$. Similar to the periodical superpotential, we notice that the thickness of the solution is essentially controlled by $k$ (Figure \ref{figene2} $a$), where the warp factor narrows and decreases its amplitude as $k$ decreases (Figure \ref{figene2} $b$), modifying also the potential (Figure \ref{figene2} $c$) and the energy density (Figure \ref{figene2} $d$). The scalar field profile only guarantees topological stability for negative values of $k$.

From Eq.(\ref{11}) we can notice that when the extra dimension $y$ runs from one boundary $y \rightarrow -\infty$ to the other $y \rightarrow \infty$, the scalar field $\phi(y)$ runs smoothly from $\phi(-\infty) \rightarrow-\beta$ to $\phi(\infty) \rightarrow\beta$ with $k\alpha<0$, where the vacuum potentials $V\left(\pm\beta\right)=\frac{16}{27}k\alpha^2\beta^4$ are just located. So, the scalar field is indeed a kink solution. From the asymptotic behavior of the warp factor
$A(y\rightarrow\pm\infty)\rightarrow-\frac{\beta^2}{6k}\left(e^{2k|y|}-4k\alpha|y|\right)\rightarrow\frac{2}{3}\alpha\beta^2|y|$,
one can conclude that the spacetimes for the brane system are asymptotically anti-de Sitter along the fifth dimension.

As a third example, we propose the so called fractional superpotentials  in the form
\begin{eqnarray}
W(\phi)=3\beta\Big(\frac{3}{5}\phi^{\frac{5}{3}}-\frac{3}{7}\beta\phi^{\frac{7}{3}}\Big).
\end{eqnarray}
Again, from equations (\ref{07}) and (\ref{08}), we have the  thick brane solution
\begin{eqnarray}
\label{12}\phi(y)&=&\frac{1}{\beta^{\frac{3}{2}}}\tanh(-k\alpha y),\\
V(\phi)&=&\frac{1}{9}\Big\{\frac{972}{1225} k\alpha^2\beta^2\Big(7-5\beta\phi^{\frac{2}{3}}\Big)^2+\frac{5}{3} \Big[3k\alpha\beta\Big(\beta\phi^{\frac{2}{3}}-1\Big)\phi^{\frac{2}{3}}\Big]^2\Big\}.
\end{eqnarray}
The warp factor from Eq.(\ref{09a}) is
\begin{eqnarray}\label{13}
A(y)=-\frac{3}{70k\beta^{\frac{3}{2}}}\Big\{12\ln[\mathrm{sech}(k\alpha y)]+3\mathrm{sech}^2(k\alpha y)-12\mathrm{sech}^4(k\alpha y)+5\mathrm{sech}^6(k\alpha y)\Big\}.
\end{eqnarray}
The energy density of the brane system above can be expressed as
\begin{eqnarray}
\rho(y)&=&\frac{k\alpha^2}{7350\beta^{\frac{3}{2}}}\exp\Big\{-\frac{3\mathrm{sech}^2(k\alpha y)}{280k\beta^{\frac{3}{2}}}\Big[1-36\cosh(2k\alpha y)+3\cosh(4k\alpha y)\Big]\Big\}\nonumber\\
&\times&\Big\{45325 k\beta^3+648[6+\cosh(2k\alpha y)]^2\tanh^6(k\alpha y)\Big\}\nonumber\\
&\times&\cosh(k\alpha y)^{\frac{36}{35k\beta^{\frac{3}{2}}}}\tanh^4(k\alpha y)\ \mathrm{sech}^4(k\alpha y).
\end{eqnarray}

\begin{figure}
\begin{center}
\begin{tabular}{ccc}
\includegraphics[height=5cm]{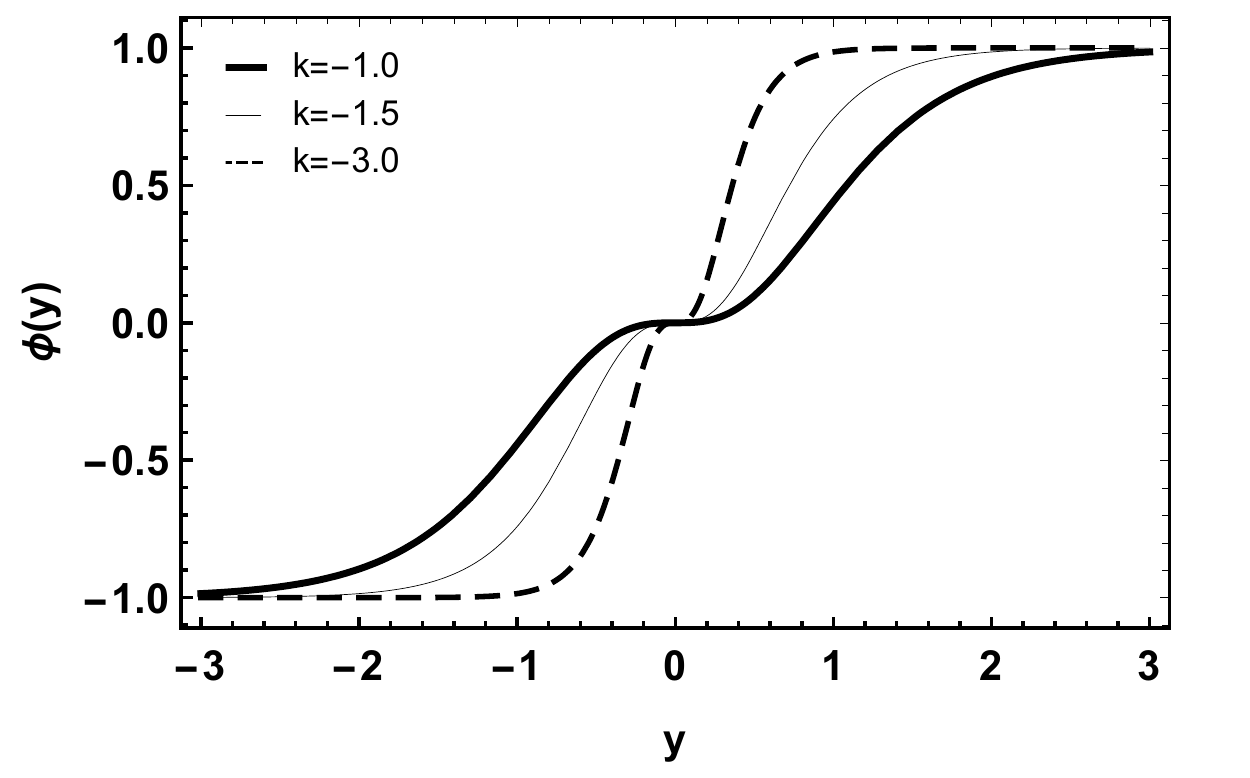}
\includegraphics[height=5cm]{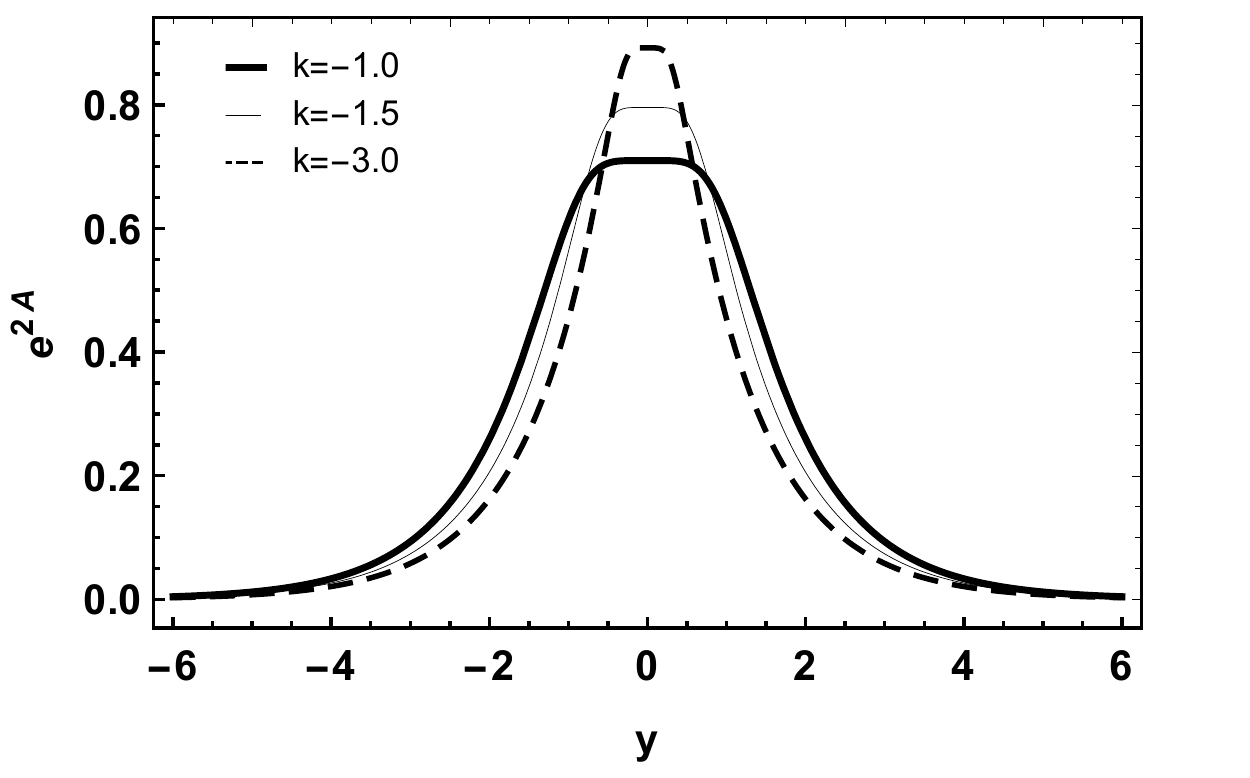}\\ 
(a) \hspace{8 cm}(b)\\
\includegraphics[height=5cm]{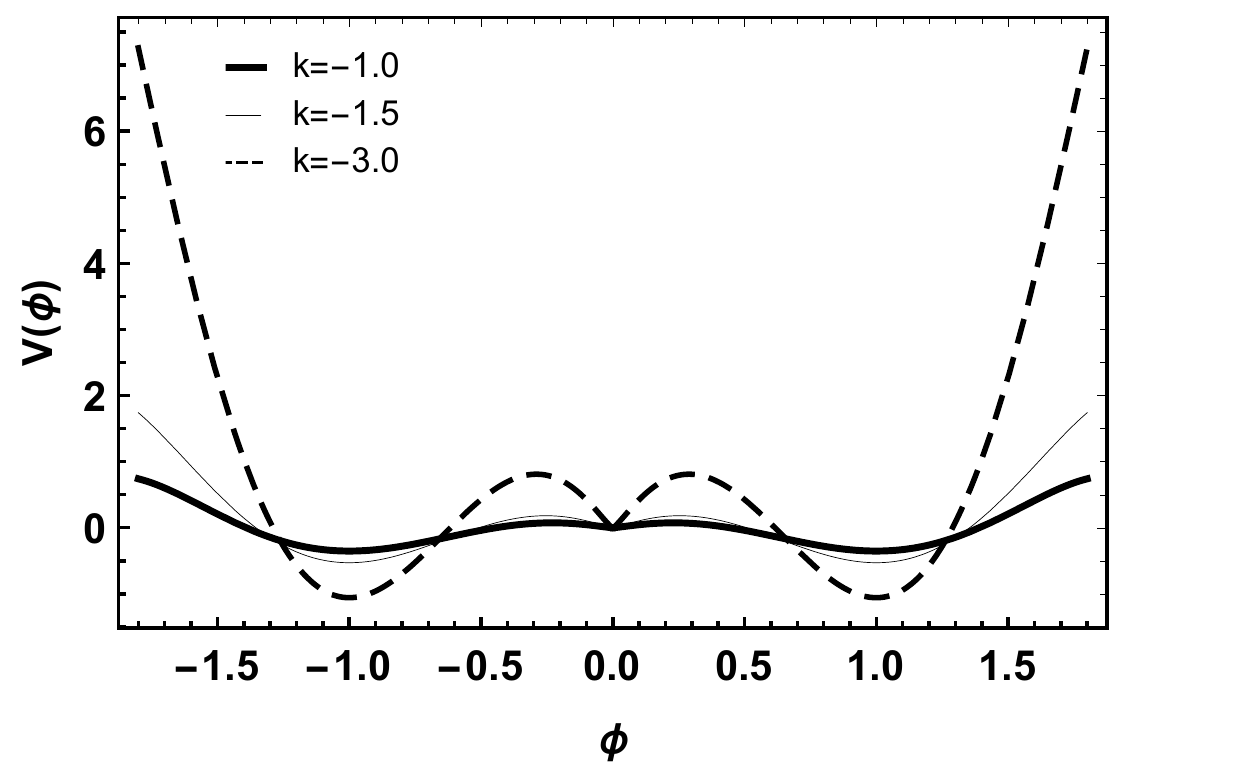}
\includegraphics[height=5cm]{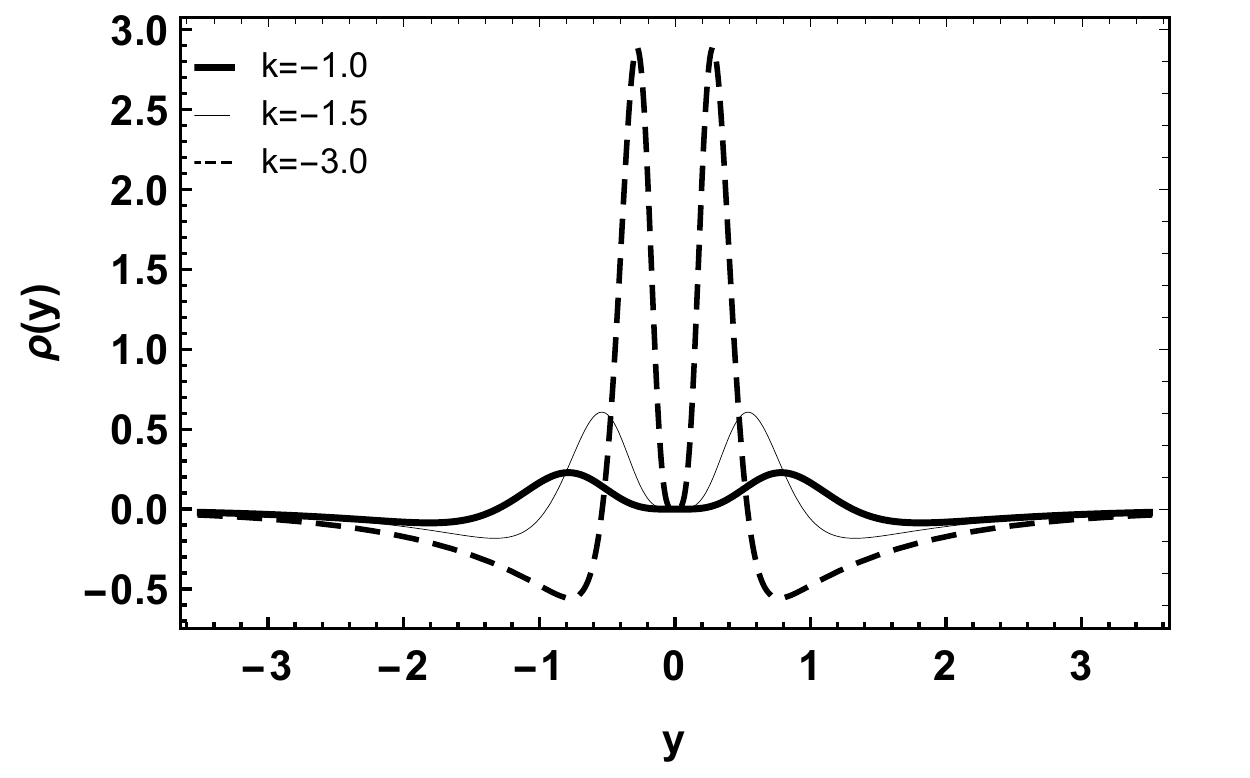}\\ 
(c) \hspace{8 cm}(d)
\end{tabular}
\end{center}
\caption{Plots of the kink solution (a), warp factor (b), potential (c), energy density (d), for the fractional superpotential, where $\alpha=\beta=1$. 
\label{figene3}}
\end{figure}

In Fig.\ref{figene3}, we plotted the  profile of the kink solution $\phi(y)$, the warp factor $e^{2A}$, the potential $V(\phi)$, and the energy density $\rho(y)$ for the fractional superpotential varying the parameter $k$. The scalar field profile only guarantees topological stability for negative values of $k$. For this case we have a double-kink solution, where the thickness of the solution is controlled  by $k$ (Figure \ref{figene3} $a$). The warp factor has a flattened peak, which decreases  as $k$ decreases (Figure \ref{figene3} $b$), modifying also the potential (Figure \ref{figene3} $c$). The energy density (Figure \ref{figene3} $d$) has two peaks, which is intensified as parameter $k$ decreases, representing a splitting of the brane.

With the Eq.(\ref{12}) we can notice that when the extra dimension $y$ runs from one boundary $y \rightarrow -\infty$ to the other $y \rightarrow \infty$, the scalar field $\phi(y)$ runs smoothly from $\phi(-\infty) \rightarrow-\frac{1}{\beta^{3/2}}$ to $\phi(\infty) \rightarrow\frac{1}{\beta^{3/2}}$ with $k\alpha<0$, where the vacuum potentials $V\left(\pm\frac{1}{\beta^{3/2}}\right)=\frac{432}{1225}\frac{k\alpha^2}{\beta^3}$ are just located. So the scalar field is indeed a double-kink solution. From the asymptotic behavior of the warp factor
$A(y\rightarrow\pm\infty)\rightarrow-\frac{3}{70k\beta^{3/2}}\left(3e^{2k|y|}-12e^{4k|y|}+5e^{6k|y|}-12k\alpha|y|\right)\rightarrow\frac{18}{35}\frac{\alpha}{\beta^{3/2}}|y|$,
one can conclude that the spacetimes for the brane system are asymptotically anti-de Sitter along the fifth dimension.

For a more complicated case, with $n=2$, considering $k_0=1$ for simplicity, we can propose an example of the superpotential such that
\begin{eqnarray}
W(\phi)=\frac{1}{2^{\frac{1}{3}}}\Big(\beta\phi-\frac{1}{3\beta}\phi^3\Big)^{\frac{1}{3}}.
\end{eqnarray}
In this way the equations (\ref{07}) and (\ref{08}) give us the  thick brane solution
\begin{eqnarray}
\label{14}\phi(y)&=&\beta\ \tanh(k\alpha^3 y),\\
V(\phi)&=&\frac{1}{9}\Big\{\frac{5}{3}\Big[\frac{k \alpha^3(\beta^2-\phi^2)}{3\beta}\Big]^2-(12)^{\frac{1}{3}}8 k\alpha^4\Big(3\beta\phi-\frac{\phi^3}{\beta}\Big)^{\frac{4}{3}} \Big\}.
\end{eqnarray}

\begin{figure}
\begin{center}
\begin{tabular}{ccc}
\includegraphics[height=5cm]{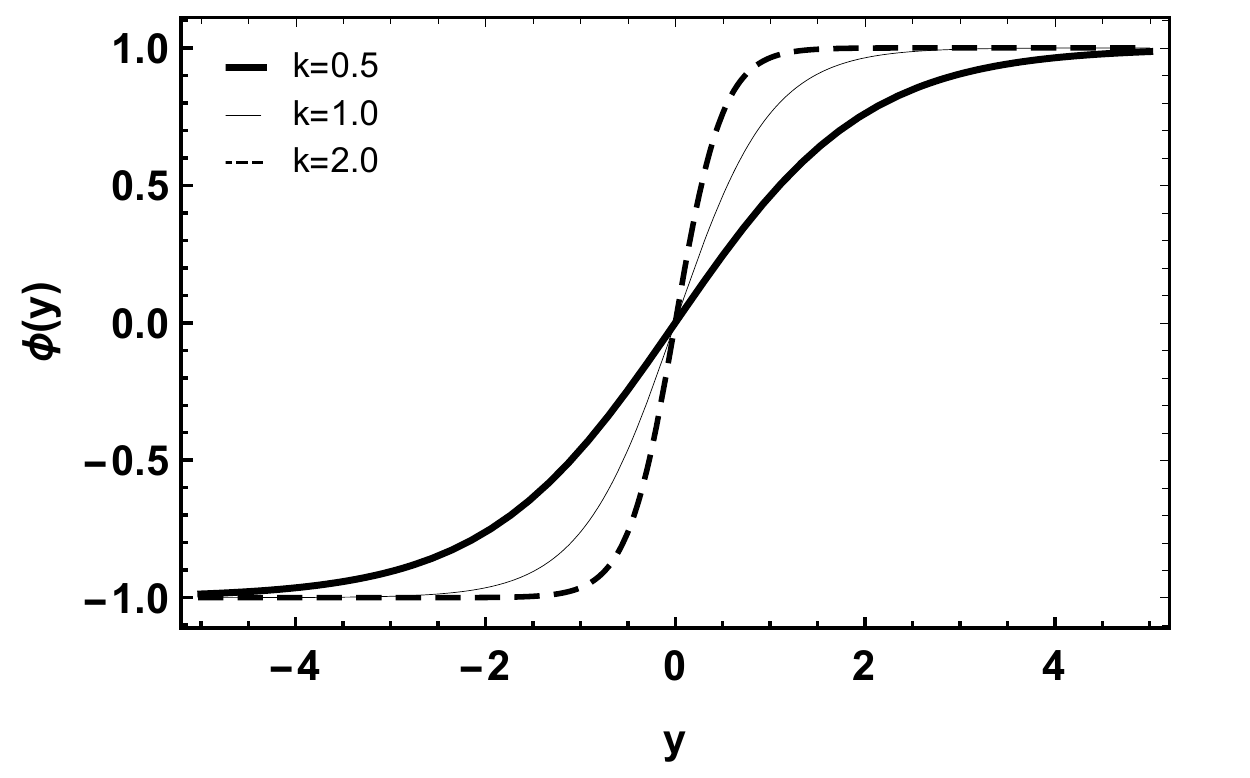}
\includegraphics[height=5cm]{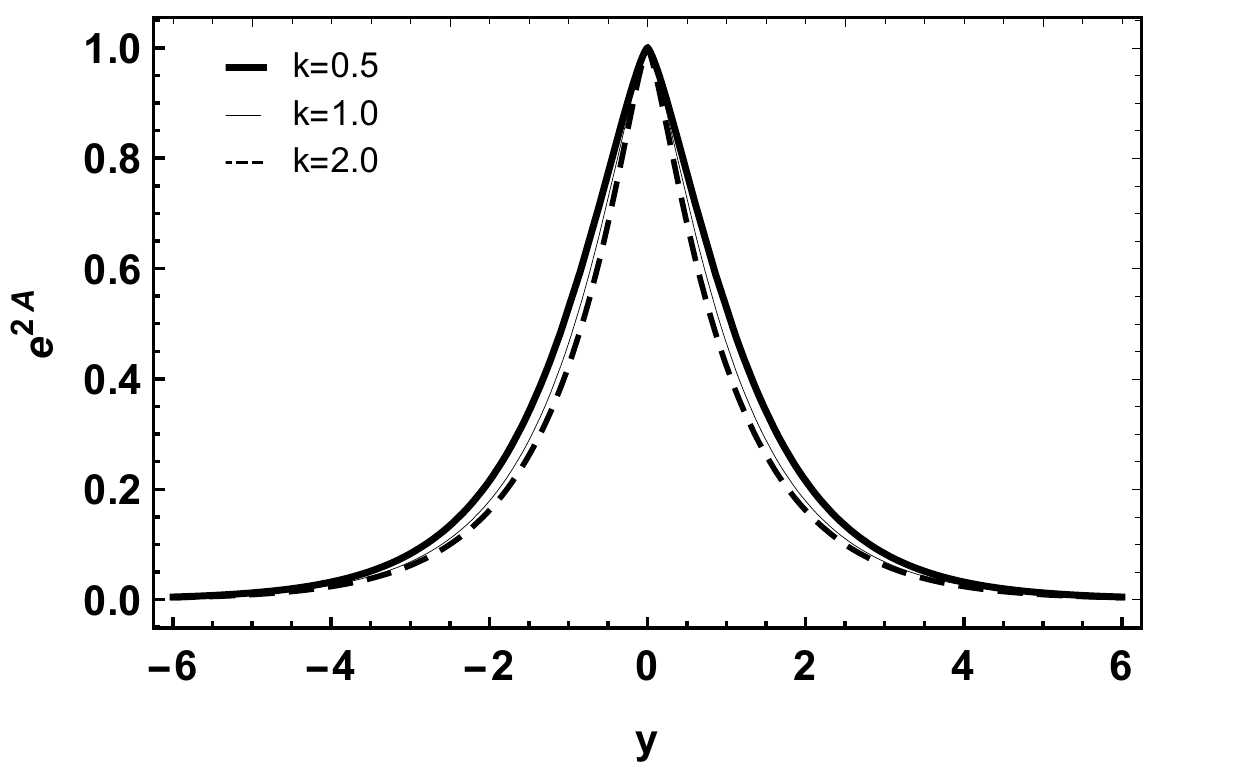}\\ 
(a) \hspace{8 cm}(b)\\
\includegraphics[height=5cm]{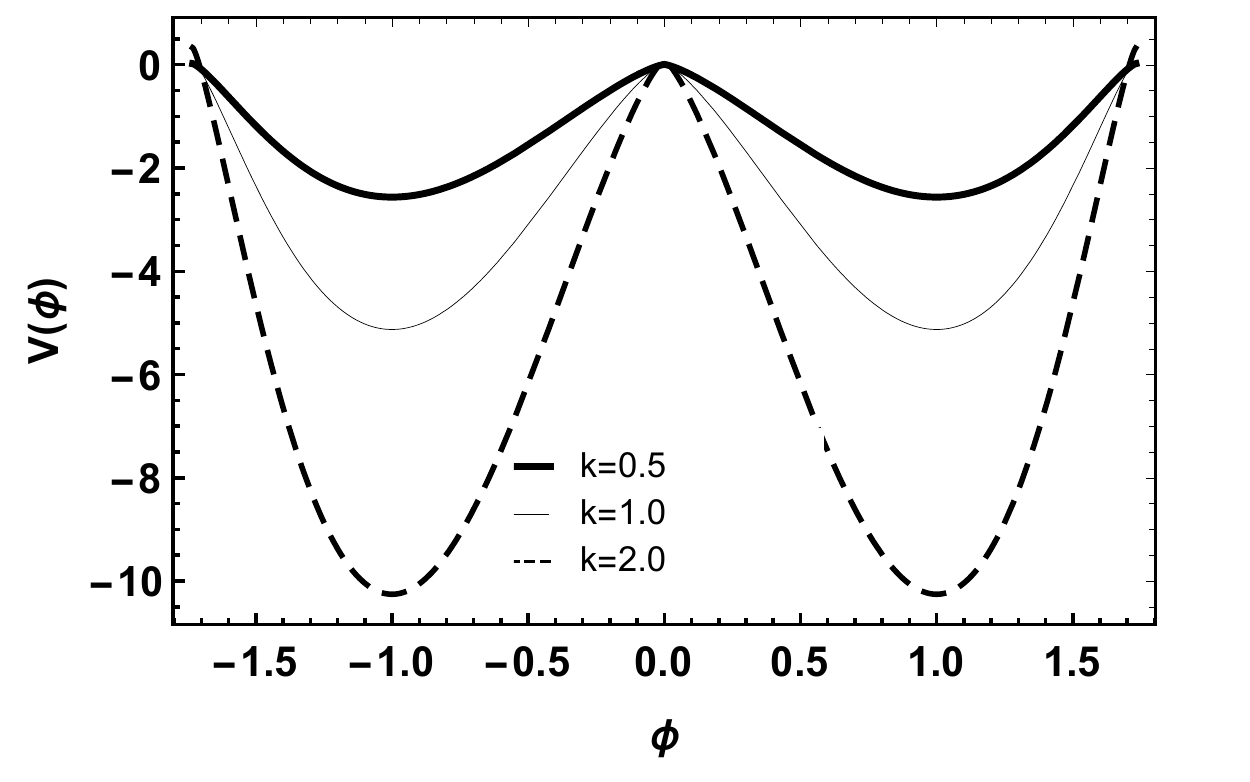}
\includegraphics[height=5cm]{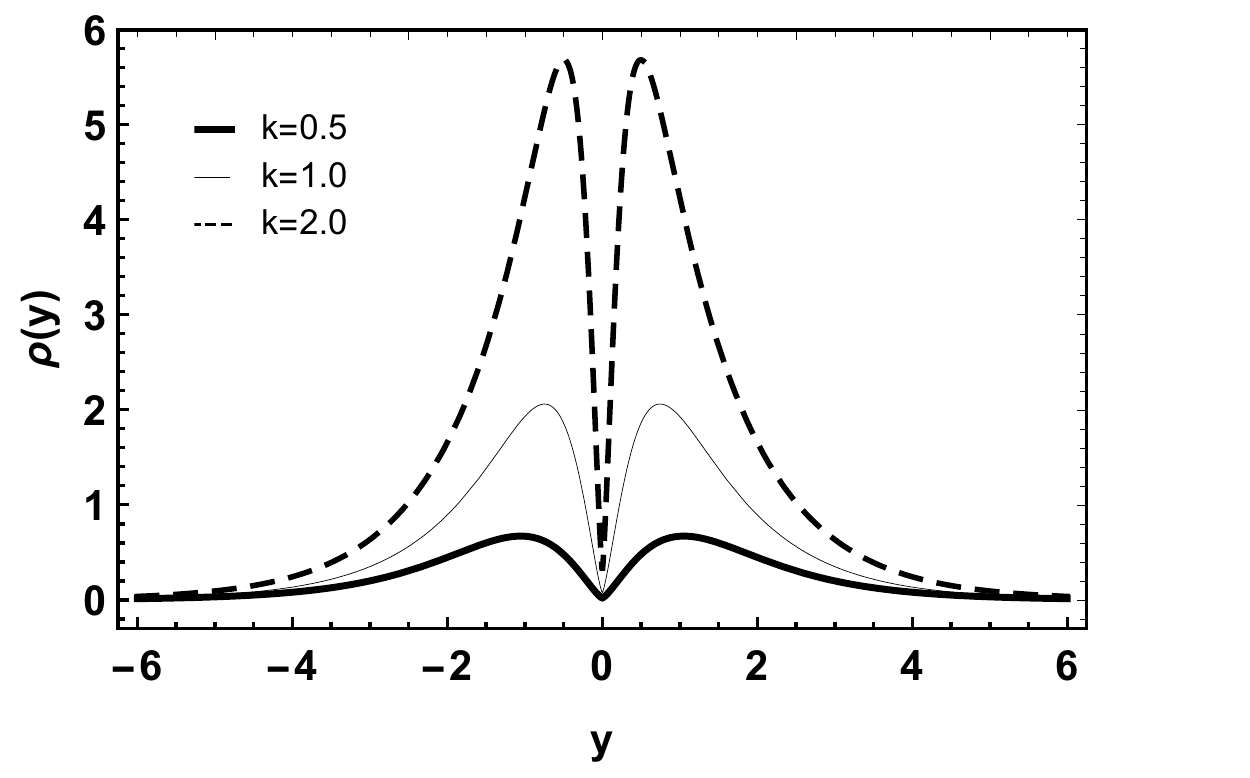}\\ 
(c) \hspace{8 cm}(d)
\end{tabular}
\end{center}
\caption{Plots of the kink solution (a), warp factor (b), potential (c), energy density (d), for $n=2$ superpotential, where $\alpha=\beta=1$. 
\label{figene4}}
\end{figure}

The behaviors of the kink $\phi(y)$ solution and the potential $V(\phi)$ are analyzed in Fig.\ref{figene4}  varying the parameter $k$. For this case we have a kink solution, where the thickness of the solution is controlled  by $k$ (Figure \ref{figene4} $a$), modifying also the potential (Figure \ref{figene4} $c$).  For positive $k$, the scalar field profile guarantees topological stability. When $k$ is negative, this topological stability is lost, giving us a physically unpleasant brane density profile.

From Eq.(\ref{14}), we can notice that when the extra dimension $y$ runs from one boundary $y \rightarrow -\infty$ to the other one $y \rightarrow \infty$, the scalar field $\phi(y)$ runs smoothly from $\phi(-\infty) \rightarrow-\beta$ to $\phi(\infty) \rightarrow\beta$ with $k\alpha>0$, where the vacuum potentials $V\left(\pm\beta\right)=-\frac{32}{3^{5/3}}k\alpha^4\beta^{8/3}$ are just located. So the scalar field is indeed a kink solution.

In this case, we do not get warp factor analytically. But we can make a numerical analysis of the behavior of warp factor through Figure \ref{figene4} ($b$), where we noticed that the warp factor narrows  as $k$ increases. Then, the energy density for the above brane system can be expressed as
\begin{eqnarray}
\rho(y)&=&\frac{e^{2A(y)}}{486}k\alpha^4\Big\{37k\alpha^2\beta^2[1-2\tanh^2(k\alpha^3y)+\tanh^4(k\alpha^3y)]\nonumber\\ &+&432(12)^{\frac{1}{3}}\Big[\beta^2[3-\tanh(k\alpha^3y)]\tanh^2(k\alpha^3y)\Big]^{\frac{4}{3}}\Big\}.
\end{eqnarray}

The energy density is also numerically analyzed, so we can see its behavior through the Figure \ref{figene4} $d$. In this case, the energy density presents two peaks, which is intensified as parameter $k$ increases, representing  a splitting of the brane. 

\subsection{$f(T,\mathcal{T})=-T-k_1T^{2}+k_2\mathcal{T}$}

For the case where $f(T,\mathcal{T})=-T-k_1T^{2}+k_2\mathcal{T}$, the equations (\ref{e.1}) and (\ref{e.2}) take the form
\begin{eqnarray}
\label{j.1}
A''(72k_1  A'^{2}-1)&=&\frac{ \phi'^2}{3}(2+k_2),\\
\label{j.2}
12A'^{2}(36k_1  A'^{2}-1)&=&-\frac{\phi'^2}{2}(4+k_2)+V(4+5k_2).
\end{eqnarray}
We introduced the first-order formalism in the same way as in the previous section, where we choose the derivative of the warp factor with respect to the extra dimension to be a function of the scalar field, i.e. $A'=-\alpha W(\phi)$. So, Eq.(\ref{j.1}) takes the form

\begin{eqnarray}\label{j.3}
\phi'=\Big(\frac{3\alpha}{2+k_2}\Big)\Big[1-72k_1 (\alpha W)^{2}\Big] W_\phi.
\end{eqnarray}
Substituting Eq.(\ref{j.3}) in Eq.(\ref{j.2}), we obtain the potential in the form
\begin{eqnarray}\label{j.4}
V(\phi)&=&\Bigg(\frac{1}{4+5k_2}\Bigg)\Bigg\{ \Big[36k_1(\alpha W)^{2}-1\Big](\alpha W)^{2}\nonumber\\&+&\Big(\frac{4+k_2}{2}\Big)\Big[\Big(\frac{3\alpha}{2+k_2}\Big)\Big(1-72k_1 (\alpha W)^{2}\Big) \Big]^2W_\phi^2\Bigg\}.
\end{eqnarray}
Note that by setting the parameters $k_{1,2}=0$ and $\alpha=\frac{1}{3}$, the default setting is restored, where equations (\ref{j.3}) and (\ref{j.4}) takes the form (\ref{oo.3}) and (\ref{oo.4}) respectively.

The energy density (\ref{05}) turns to be
\begin{eqnarray}\label{j.5}
\rho(y)&=&\frac{e^{2A}}{2}\Bigg\{\Big(\frac{1}{4+5k_2}\Big)\Big[288k_1(\alpha W)^2-8\Big](\alpha W)^2\nonumber\\&+&\Big(1+\frac{4+k_2}{4+5k_2}\Big)\Big[\frac{1-72k_1(\alpha W)^{2}}{2+k_2}\Big]^2(3\alpha W_\phi)^2\Bigg\}.
\end{eqnarray}
Now, the solutions of the thick brane system are completely determined by superpotential function $W(\phi)$ required to be specified. In Refs.\cite{Menezes,ftborninfeld}, the authors obtained analytic brane solutions of the superpotential  from an ansatz as a linear function. So, we consider a suggestion of $W(\phi)$ as a linear function, namely,
\begin{eqnarray}
W(\phi)=\beta\phi.
\end{eqnarray}

In this way the equations (\ref{j.3}) and (\ref{j.4}) give us the following thick brane solution:
\begin{eqnarray}
\phi(y)&=&\Big(\frac{1}{6\sqrt{2k_1}\alpha\beta}\Big)\tanh\Big[\Big(\frac{18\sqrt{2k_1}\alpha^2\beta^2}{2+k_2}\Big) y\Big],\\
V(\phi)&=&\Big[\frac{3\alpha^2\beta^2}{2(4+5k_2)}\Big]\Big\{8[36k_1(\alpha\beta\phi)^2-1]\phi^2+3(4+k_2)\Big[\frac{1-72k(\alpha\beta\phi)^2}{2+k_2}\Big]^2\Big\}.
\end{eqnarray}
The different form of the warp factor is
\begin{eqnarray}\label{j.6}
A(y)=-\Big(\frac{2+k_2}{216k_1\alpha^2\beta^2}\Big)\ln\Big\{\cosh\Big[\Big(\frac{18\sqrt{2k_1}\alpha^2\beta^2}{2+k_2}\Big)y\Big]\Big\}.
\end{eqnarray}
Finally, the energy density for the above brane system can be expressed as
\begin{eqnarray}
\rho(y)&=&\Big[\frac{1}{96k_1(2+k_2)^2(4+k_2)}\Big]\cosh\Big[\Big(\frac{18\sqrt{2k_1}\alpha^2\beta^2}{2+k_2}\Big)y\Big]^{-\left(4+\frac{2+k_2}{108k_1\alpha^2\beta^2}\right)}\Bigg\{5(2+k_2)^2 \nonumber\\ &+&864k_1(4+3k_2)\alpha^2\beta^2-4(2+k_2)^2\cosh\Big[2\Big(\frac{18\sqrt{2k_1}\alpha^2\beta^2}{2+k_2}\Big)y\Big]\nonumber\\&-&(2+k_2)^2\cosh\Big[4\Big(\frac{18\sqrt{2k_1}\alpha^2\beta^2}{2+k_2}\Big)y\Big]\Bigg\}.
\end{eqnarray}

\begin{figure}
\begin{center}
\begin{tabular}{ccc}
\includegraphics[height=5cm]{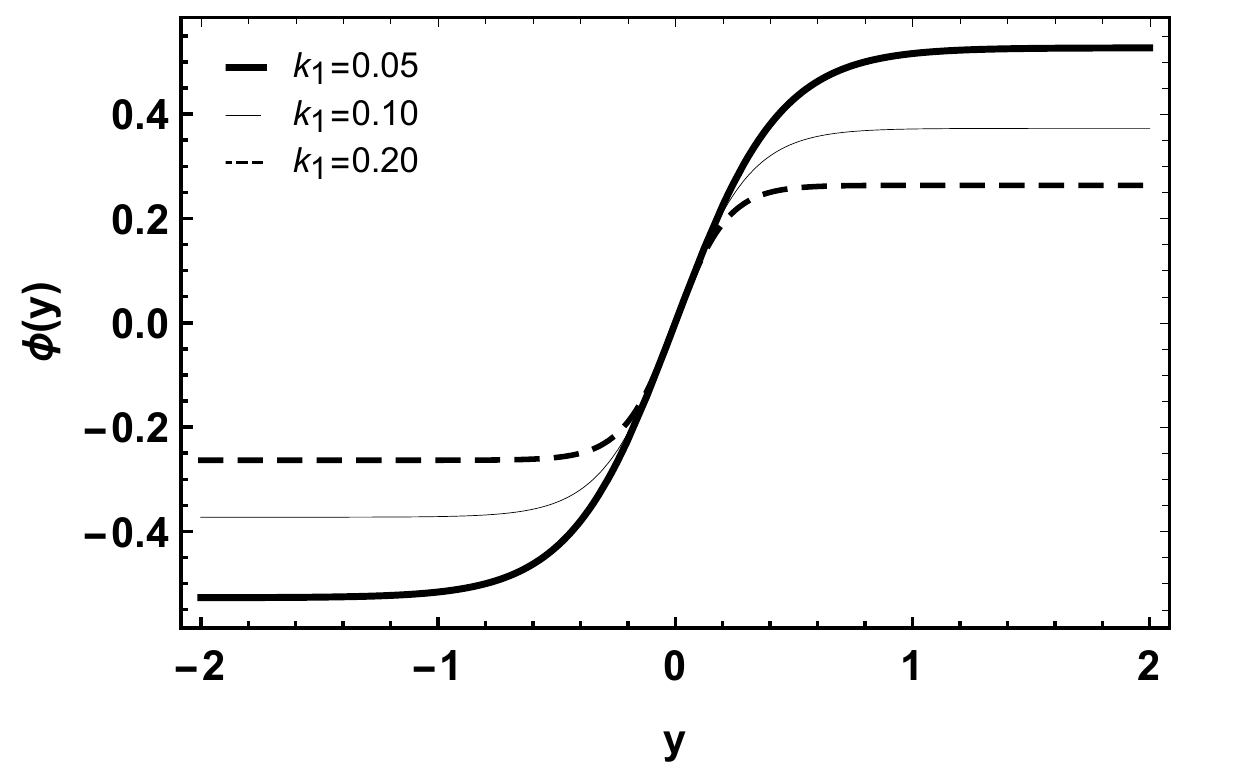}
\includegraphics[height=5cm]{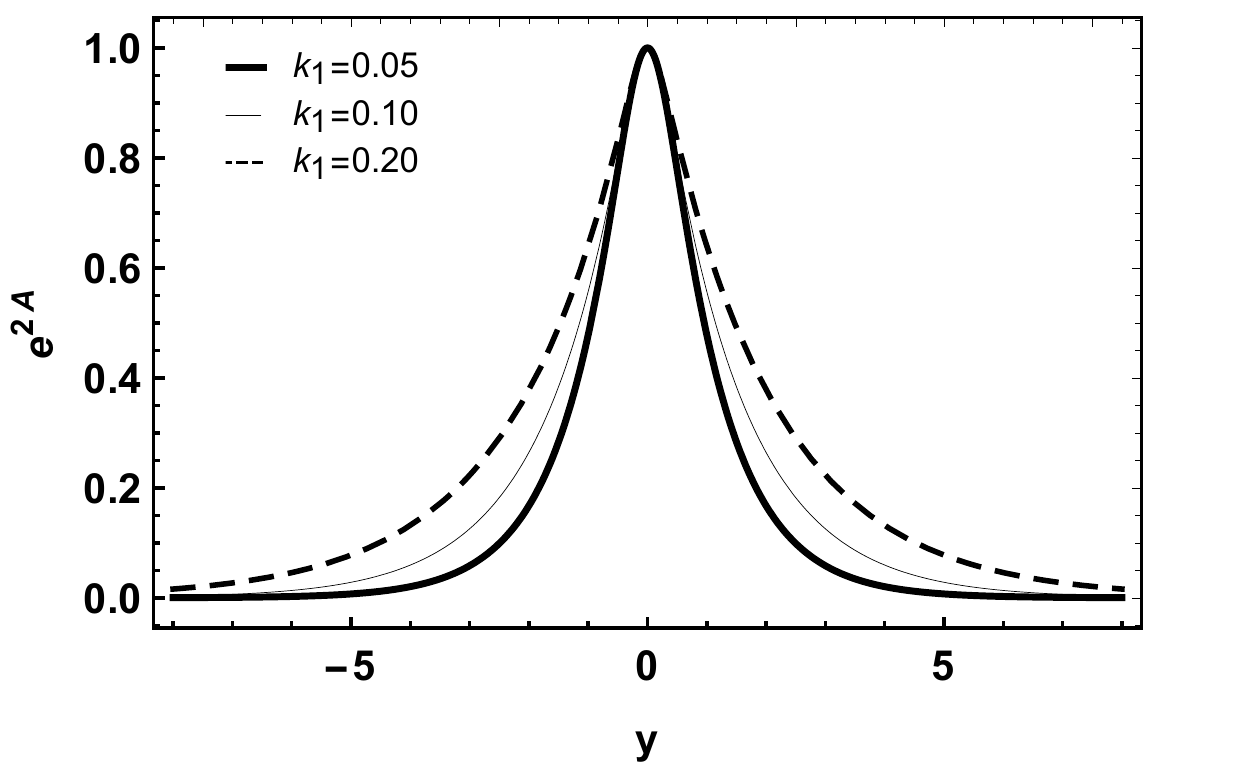}\\ 
(a) \hspace{8 cm}(b)\\
\includegraphics[height=5cm]{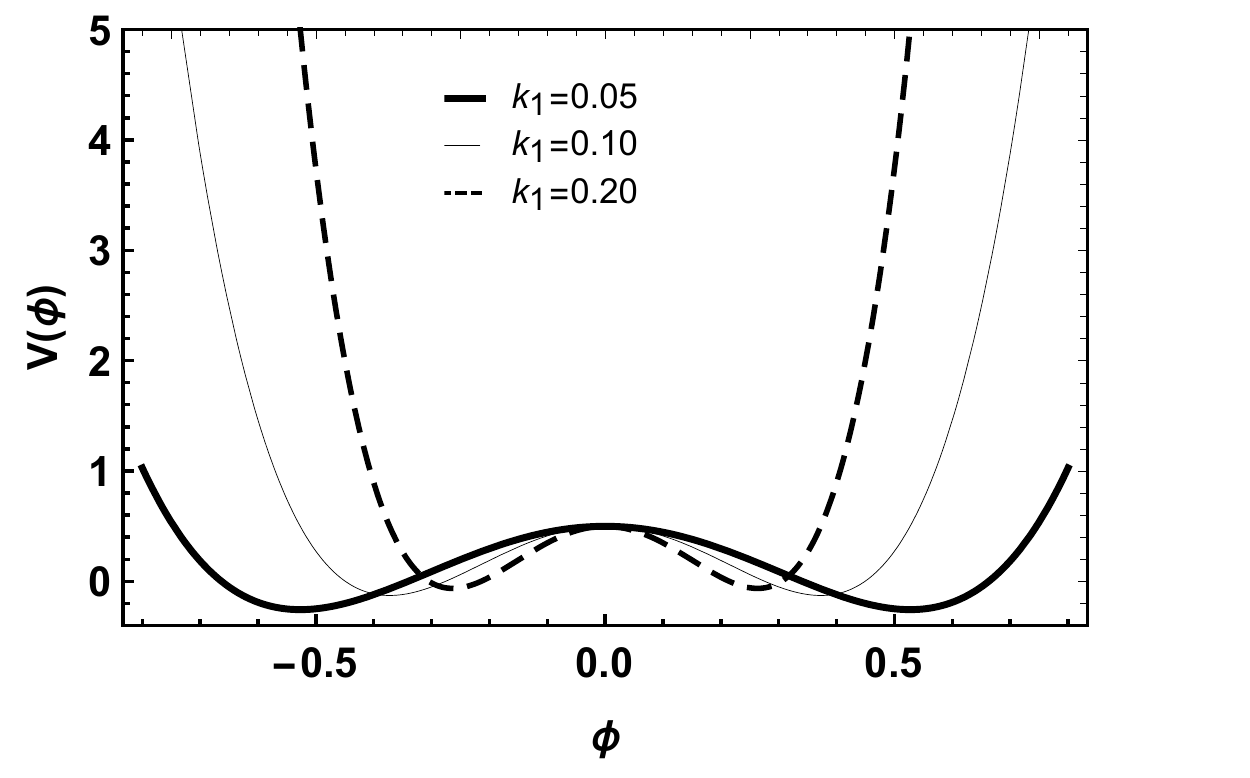}
\includegraphics[height=5cm]{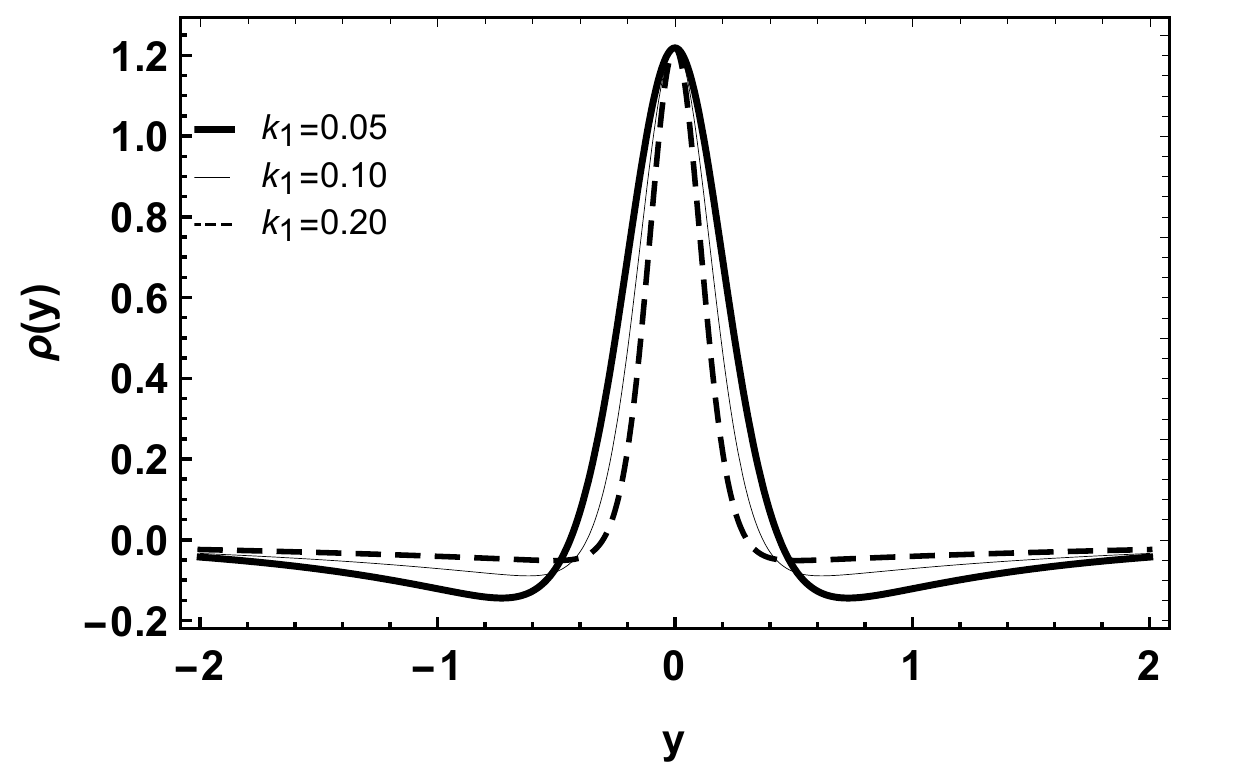}\\ 
(c) \hspace{8 cm}(d)
\end{tabular}
\end{center}
\caption{Plots of the kink solution (a), warp factor (b), potential (c), energy density (d), where $k_2=0.5 $ and $\alpha=\beta=1$. 
\label{figene5}}
\end{figure}

\begin{figure}
\begin{center}
\begin{tabular}{ccc}
\includegraphics[height=5cm]{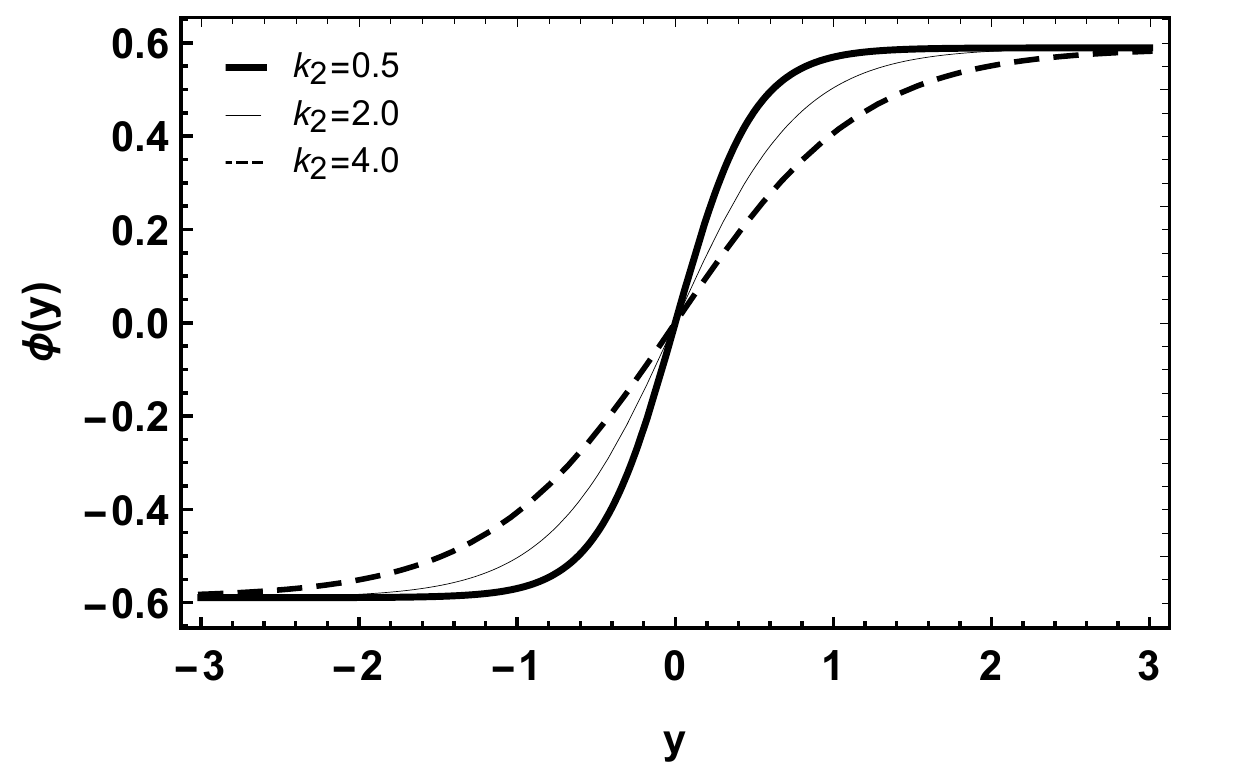}
\includegraphics[height=5cm]{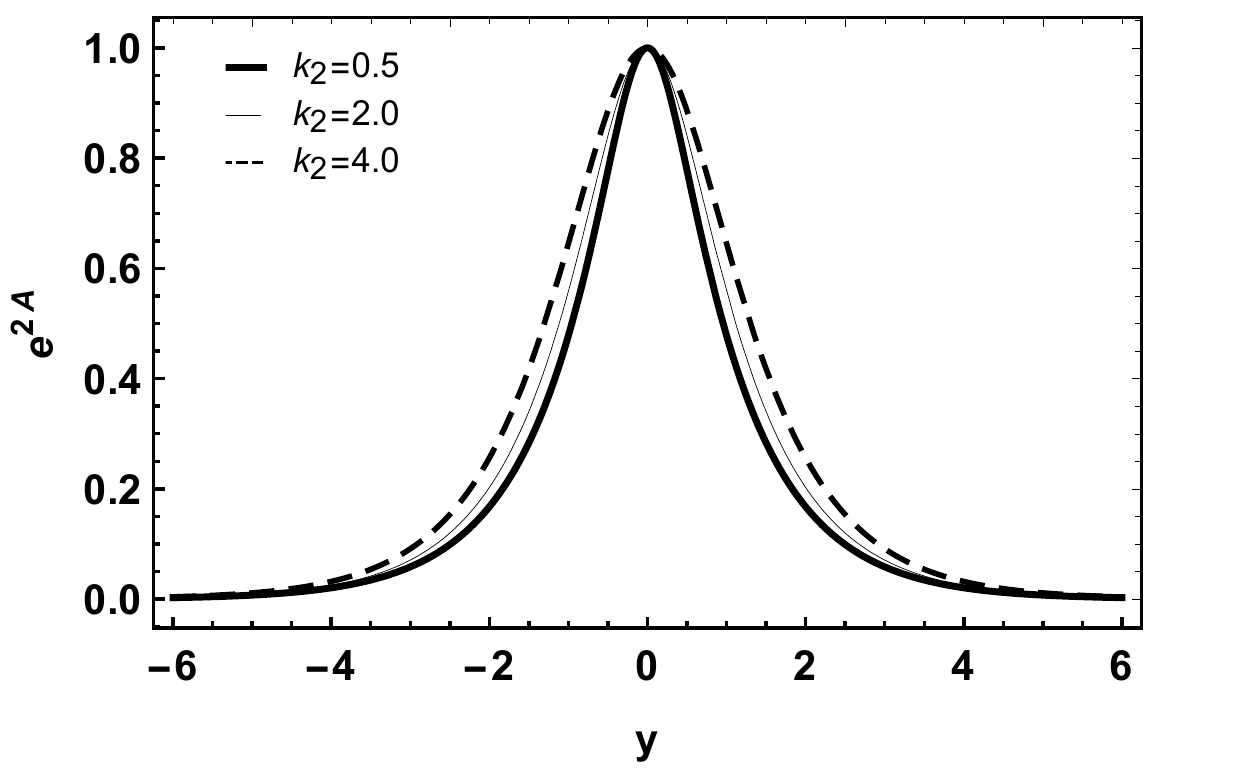}\\ 
(a) \hspace{8 cm}(b)\\
\includegraphics[height=5cm]{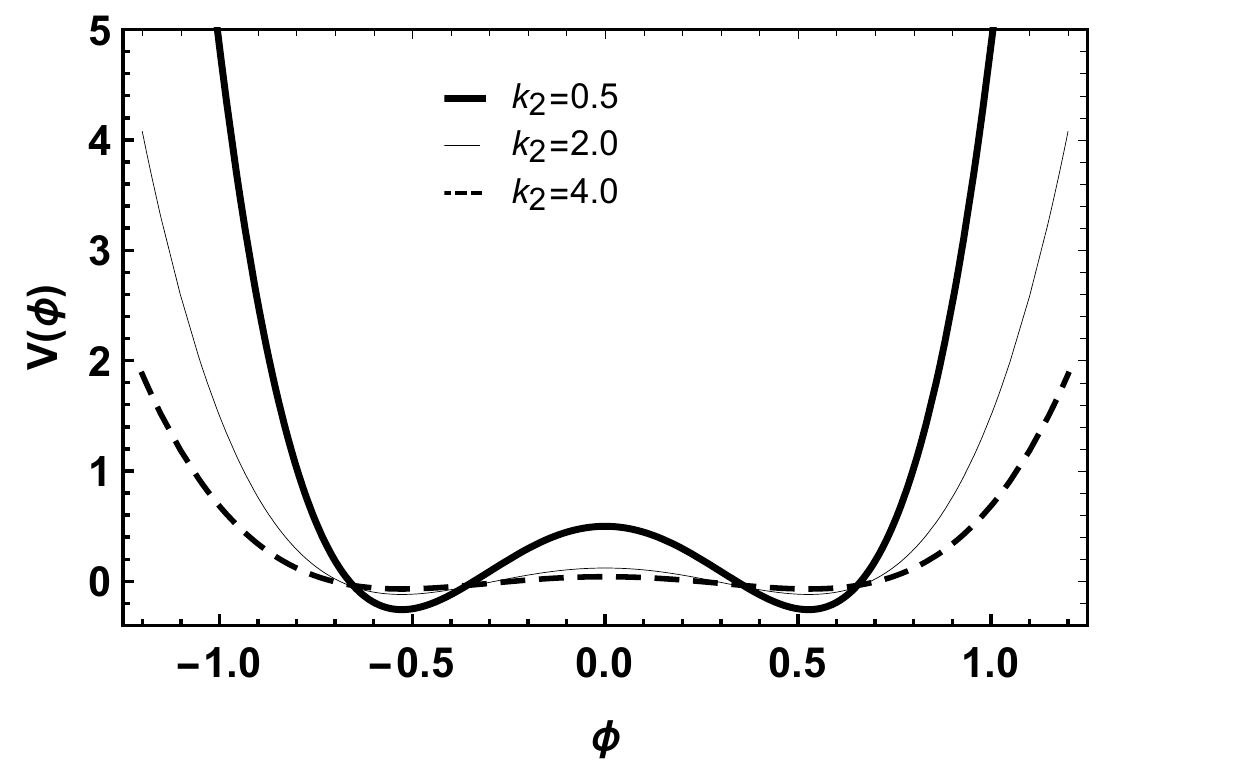}
\includegraphics[height=5cm]{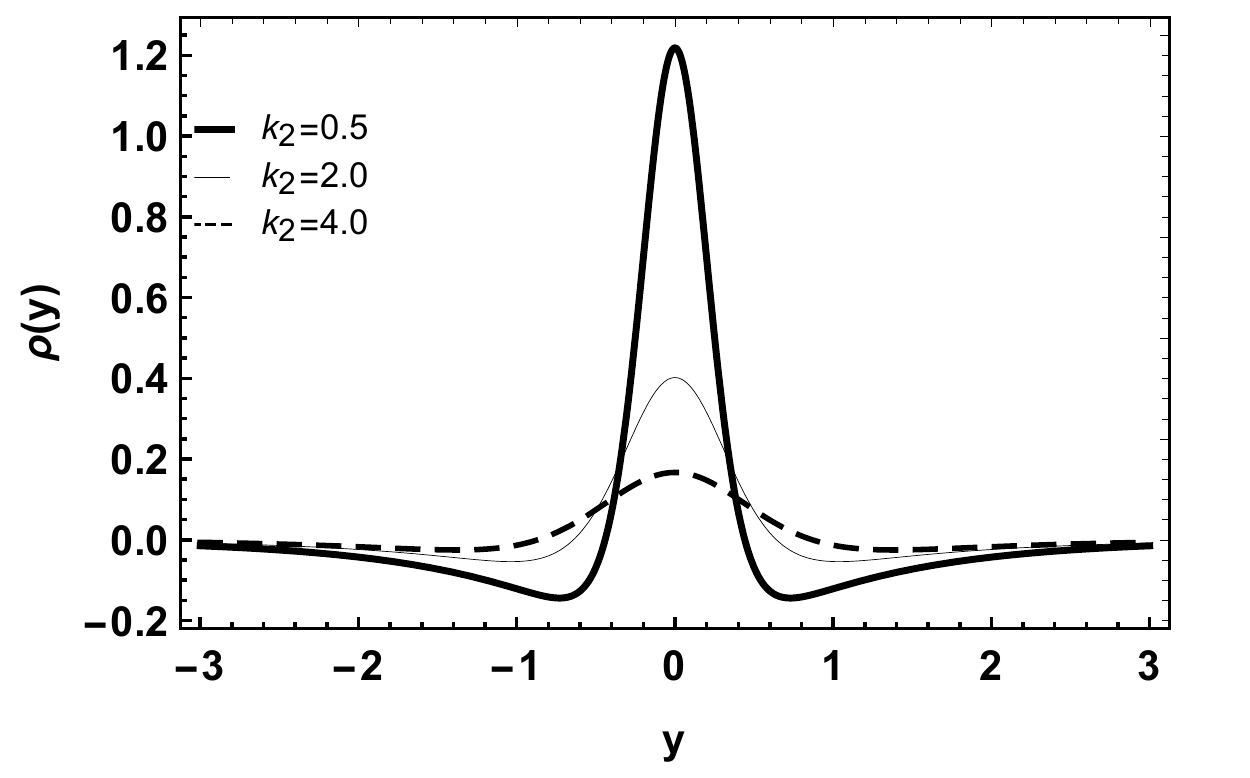}\\ 
(c) \hspace{8 cm}(d)
\end{tabular}
\end{center}
\caption{Plots of the kink solution (a), warp factor (b), potential (c), energy density (d), where $k_1=0.05 $ and $\alpha=\beta=1$. 
\label{figene6}}
\end{figure}

In Fig.\ref{figene5}, we plotted the  profile of the kink solution $\phi(y)$, the warp factor $e^{2A}$, the potential $V(\phi)$, and the energy density $\rho(y)$,  varying the parameter $k_1$ that controls the influence of the torsion. We note that by increasing the value of $k_1$, we decrease the value of convergence of $\phi(\pm\infty)$ (Figure \ref{figene5} $a$), we also increased the thickness of the warp factor (Figure \ref{figene5} $b$), we modify the potential (Figure \ref{figene5} $c$) and we make the energy density more localized around the origin (Figure \ref{figene5} $d$). When $k_1$ is negative, the topological stability of the scalar field profile is lost, giving us a physically unpleasant brane density profile.

On the other hand, in Fig.\ref{figene6}, we plotted the profile of the kink solution $\phi(y)$, the warp factor $e^{2A}$, the potential $V(\phi)$, and the energy density $\rho(y)$. In this case, we vary the parameter $k_2$ that controls the influence of the trace of the energy-momentum tensor.  
Increasing the $k_2$ value, we increase the thickness of the solution (Figure \ref{figene6} $a$) and of the warp factor (Figure \ref{figene6} $b$). Accordingly, we also modify the potential (Figure \ref{figene6} $c$) and the energy density (Figure \ref{figene6} $d$).

For that case, when the extra dimension $y$ runs from one boundary $y \rightarrow -\infty$ to the other one $y \rightarrow \infty$, the scalar field $\phi(y)$ runs smoothly from $\phi(-\infty) \rightarrow-1/6\sqrt{2k_1}\alpha\beta$ to $\phi(\infty) \rightarrow1/6\sqrt{2k_1}\alpha\beta$ with $(2+k_2)^2k_1>0$, where the vacuum potentials $V\left(\pm1/6\sqrt{2k_1}\alpha\beta\right)=-1/12k_1(4+5k_2)$ are just located. So the scalar field is indeed a kink solution. We can also observe that from the asymptotic behaviors of the warp factor
$A(y\rightarrow\pm\infty)\rightarrow-(1/6\sqrt{2k_1})|y|$. So, one can conclude that the spacetimes for the brane system are asymptotically anti-de Sitter along the fifth dimension.

\section{Tensor perturbations and localization}
\label{sec3}

In this section we investigate the effects of torsion and trace of the energy-momentum tensor on the propagation of linear perturbations on the brane system. We need to introduce a small disturbance in the metric (\ref{45.a}), which is equivalent to doing a small disturbance in the vielbein (\ref{04}). We follow closely the analysis performed in Ref.\cite{Moreira2021}.  For this, we chose the \textit{f\"{u}nfbein} perturbation \cite{Moreira2021,tensorperturbations,ftnoncanonicalscalar,ftborninfeld,ftmimetic}
\begin{eqnarray}\label{15a}
h^a\ _M=\left(\begin{array}{cccccc}
e^{A(y)}\left(\delta^a_\mu+w^a\ _\mu\right)&0\\
0&1\\
\end{array}\right),
\end{eqnarray}
where $w^a\ _\mu\equiv w^a\ _\mu(x^\mu,y)$. This choice of vielbeins perturbation (\ref{15a}) is well analyzed in Refs.\cite{Moreira2021,tensorperturbations,ftnoncanonicalscalar, ftborninfeld,ftmimetic}, proving to be a good choice. Using Eq. (\ref{15a}) we can easily get the corresponding metric
\begin{eqnarray}\label{15}
g_{MN}=\left(\begin{array}{cccccc}
e^{2A(y)}\left(\eta_{\mu\nu}+\gamma_{\mu\nu}\right)&0\\
0&1\\
\end{array}\right),
\end{eqnarray}
where the metric and the \textit{f\"{u}nfbein} perturbation are related by
\begin{eqnarray}
 \gamma_{\mu\nu}&=&(\delta^a_\mu w^b\ _\nu+\delta^b_\nu w^a\  _\mu)\eta_{ab},\nonumber\\
 \gamma^{\mu\nu}&=&(\delta_a^\mu w_b\ ^\nu+\delta_b^\nu w_a\  ^\mu)\eta^{ab}.
\end{eqnarray}
We assume transverse-traceless (TT) tensor perturbation, which is related to the gravitational wave and
four-dimensional gravitons. The TT tensor perturbation satisfies the following TT conditions: $\partial_\mu \gamma^{\mu\nu}=0=\eta^{\mu\nu}\gamma_{\mu\nu}$, which leads to the \textit{f\"{u}nfbein} 
\begin{equation}
\partial_\mu\left(\delta_a^\mu w_b\ ^\nu+\delta_b^\nu w_a\  ^\mu\right)\eta^{ab}=0,
\end{equation}
\begin{equation}
\delta_a^\mu w^a\  ^\mu=0.
\end{equation}
The perturbation of the torsion tensor, contortion tensor
and the superpotential torsion tensor $( T^P\ _{M N},\ K^P\ _{M N},\  S_P\ ^{M N})$ are given in Ref. \cite{tensorperturbations}. The energy-momentum tensor is given as Eq.(\ref{03}), where the perturbed scalar field can be expressed as $\phi=\phi_b+\phi_p$, being that $\phi_b$ is the background scalar field and $\phi_p\equiv\phi_p(x^\mu,y)$ is the perturbation. The perturbation  components of the energy-momentum tensor are \cite{tensorperturbations}
\begin{eqnarray}
\label{16}\delta\mathcal{T}_{\mu\nu}&=&-e^{2A}\Big[\Big(\frac{{\phi' _b}^2}{2}+V\Big)\gamma_{\mu\nu}+(\phi' _b{\phi' _p}+V_\phi \phi_p)\eta_{\mu\nu}\Big],\\
\delta\mathcal{T}_{y\nu}&=&\phi' _b\partial_\nu\phi' _p,\\
\delta\mathcal{T}_{y y}&=&\phi' _b{\phi' _p}-V_\phi \phi_p.
\end{eqnarray}

The perturbed modified Einstein equation (\ref{3.36}) has now the form
\begin{eqnarray}\label{27.l}
\Big(f_{TT}\partial_QT+f_{T\mathcal{T}}\partial_Q\mathcal{T}\Big)\delta S_{MN}\ ^{Q}
+\frac{1}{4}\Big(f+2p f_{\mathcal{T}}\Big)\delta g_{MN}&\nonumber\\ +\frac{1}{h}f_T\Big[\delta g_{PN}\partial_Q(h S_M\ ^{PQ})+ g_{PN}\partial_Q(h \delta S_M\ ^{PQ})&\nonumber\\
-h\Big(\delta\widetilde{\Gamma}^Q\ _{PM}S_{QN}\ ^{P}+\widetilde{\Gamma}^Q\ _{PM}\delta S_{QN}\ ^{P}\Big)\Big]&=&\delta\mathcal{T}_{MN}\Big(1-\frac{1}{2}f_{\mathcal{T}}\Big),
\end{eqnarray}
which for $\delta h=0$, $\delta T=0$ and $\delta \mathcal{T}=0$ yields 
\begin{eqnarray}\label{28.l}
\frac{e^{2A}}{4}\Big\{f' _T(6A'\gamma_{\mu\nu}-\gamma'_{\mu\nu})+
(f+2p f_{\mathcal{T}})\gamma_{\mu\nu}& &\nonumber\\ -\Big[e^{-2A}\Box \gamma_{\mu\nu}+4A'\gamma'_{\mu\nu}+\gamma''_{\mu\nu}-6(A''+4A'^2)\gamma_{\mu\nu}\Big]f_T\Big\}&=&\delta\mathcal{T}_{\mu\nu}\Big(1-\frac{1}{2}f_{\mathcal{T}}\Big),
\end{eqnarray}
where $\delta\mathcal{T}_{\mu\nu}$ is expressed in the equation (\ref{16}), $\Box=\eta^{\mu\nu}\partial_\mu \partial_\nu$.  Also, in the extra dimension, perturbations vanish, leaving only the perturbation on the brane. These features in addition to simplifying our analysis, give us the guarantee that we made a good choice for the vielbein. The gravitational field equation (\ref{3.36}) provides the condition
\begin{eqnarray}\label{30.l}
\frac{3}{2}(A''+4A'^2)f_{T}+\frac{3}{2}A'f_{T}'+\frac{1}{4}(f+2p f_{\mathcal{T}})=\Big(\frac{{\phi' _b}^2}{2}+V\Big)\Big(1-\frac{1}{2}f_{\mathcal{T}}\Big).
\end{eqnarray}
By plugging Eqs.(\ref{30.l}) and (\ref{16}) into Eq.(\ref{28.l}), we have
\begin{eqnarray}\label{17}
(e^{-2A}\Box^{(4)} \gamma_{\mu\nu}+4A'\gamma'_{\mu\nu}+\gamma''_{\mu\nu})f_T+f_{T}'\gamma'_{\mu\nu}=2\eta_{\mu\nu}(\phi' _b{\phi' _p}+V_\phi \phi_p)\Big(1-\frac{1}{2}f_{\mathcal{T}}\Big).
\end{eqnarray} 
Considering yet the vanishing trace $\phi' _b{\phi' _p}+V_\phi \phi_p=0$, it gives the perturbation equation
\begin{eqnarray}\label{32.l}
(e^{-2A}\Box^{(4)} \gamma_{\mu\nu}+4A'\gamma'_{\mu\nu}+\gamma''_{\mu\nu})f_T+f_{T}'\gamma'_{\mu\nu}=0.
\end{eqnarray}

Assuming the Kaluza-Klein decomposition $\gamma_{\mu\nu}(x^\rho,y)=\epsilon_{\mu\nu}(x^\rho)\chi(y)$ and a four-dimensional plane-wave satisfying $\left(\Box-m_0^2\right)\epsilon_{\mu\nu}=0$, the perturbed Einstein equation (\ref{32.l}) yields
\begin{eqnarray}
\label{KKequation}
\chi''+\Big(4A'+\frac{f_{T}'}{f_T} \Big)\chi'+e^{-2A}m^2\chi=0,
\end{eqnarray}
remembering that $f_{T}'\equiv f_{TT}T'+f_{T\mathcal{T}}\mathcal{T}'$. Note that, if we analyze the TEGR limit, that is $f_{T}'/f_T\rightarrow 0$, Eq.(\ref{KKequation}) is the same for GR \cite{rs}.

\subsection{Massive modes}


To get massive modes, we numerically solved Eq. (\ref{KKequation}) using the interpolation method. We adopt the usual boundary conditions $\chi'(-\infty)=\chi'(\infty)=0$. As we see in Figs.(\ref{massivemodes},  \ref{massivemodes2} and \ref{massivemodes3}), there is an asymptotic divergence of massive gravitational modes, that shows that they form a tower of non-localized states.

\subsubsection{$f(T,\mathcal{T})=k_0\mathcal{T}+kT^{n}$}

\begin{figure}
\begin{center}
\begin{tabular}{ccccccccc}
\includegraphics[height=6cm]{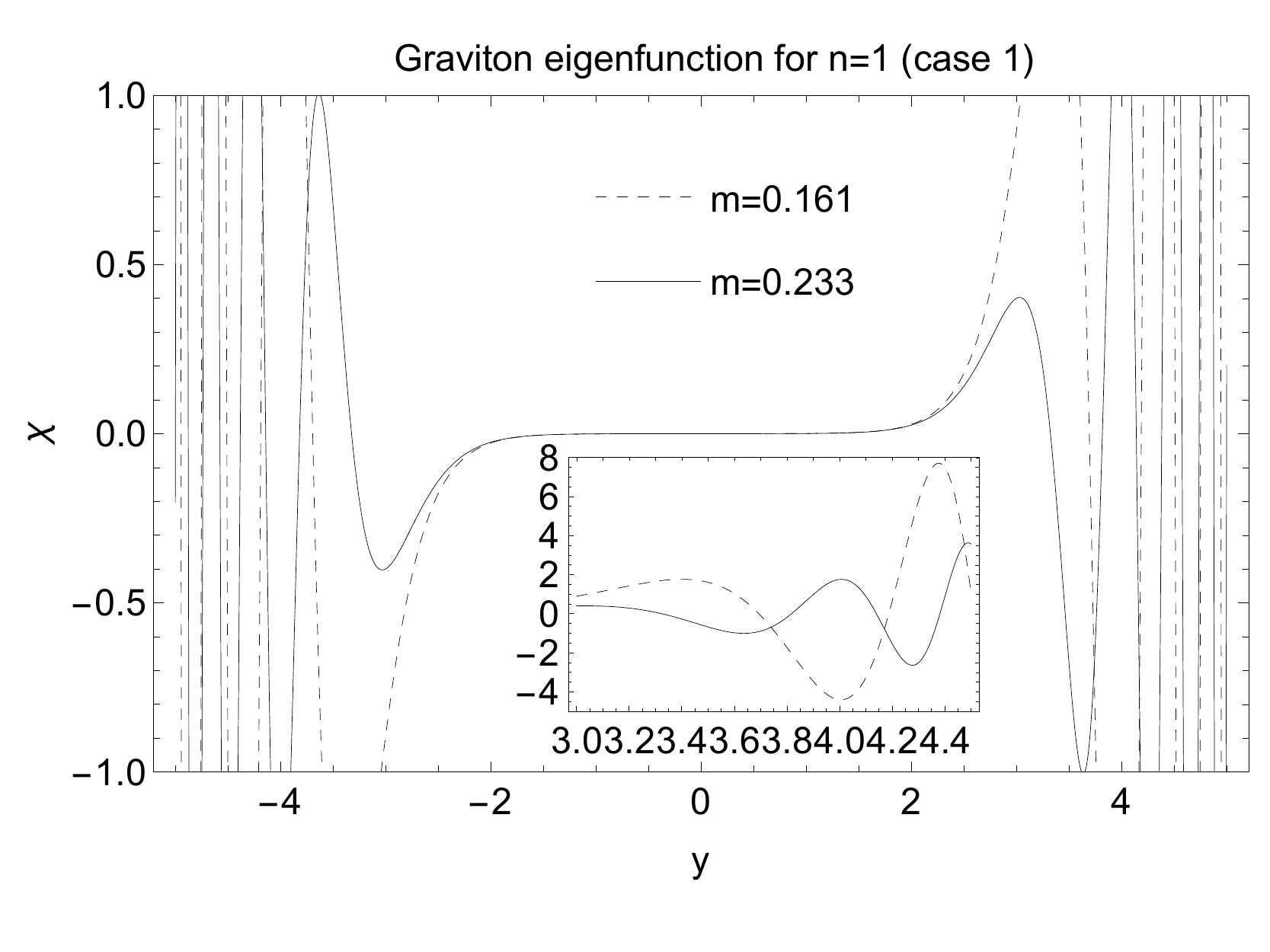}
\includegraphics[height=6cm]{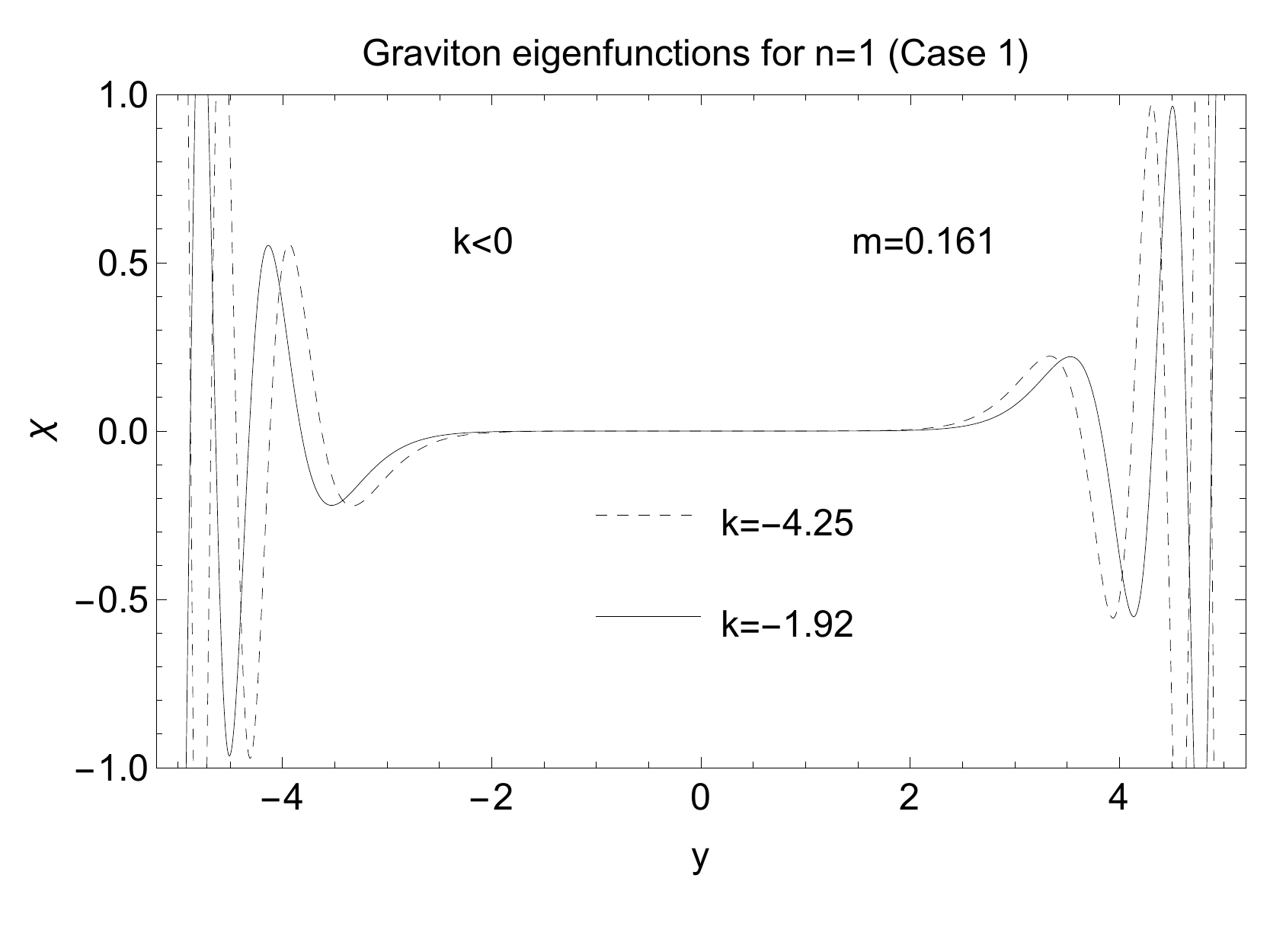}  \\
(a)\hspace{8 cm}(b)\\
\includegraphics[height=6cm]{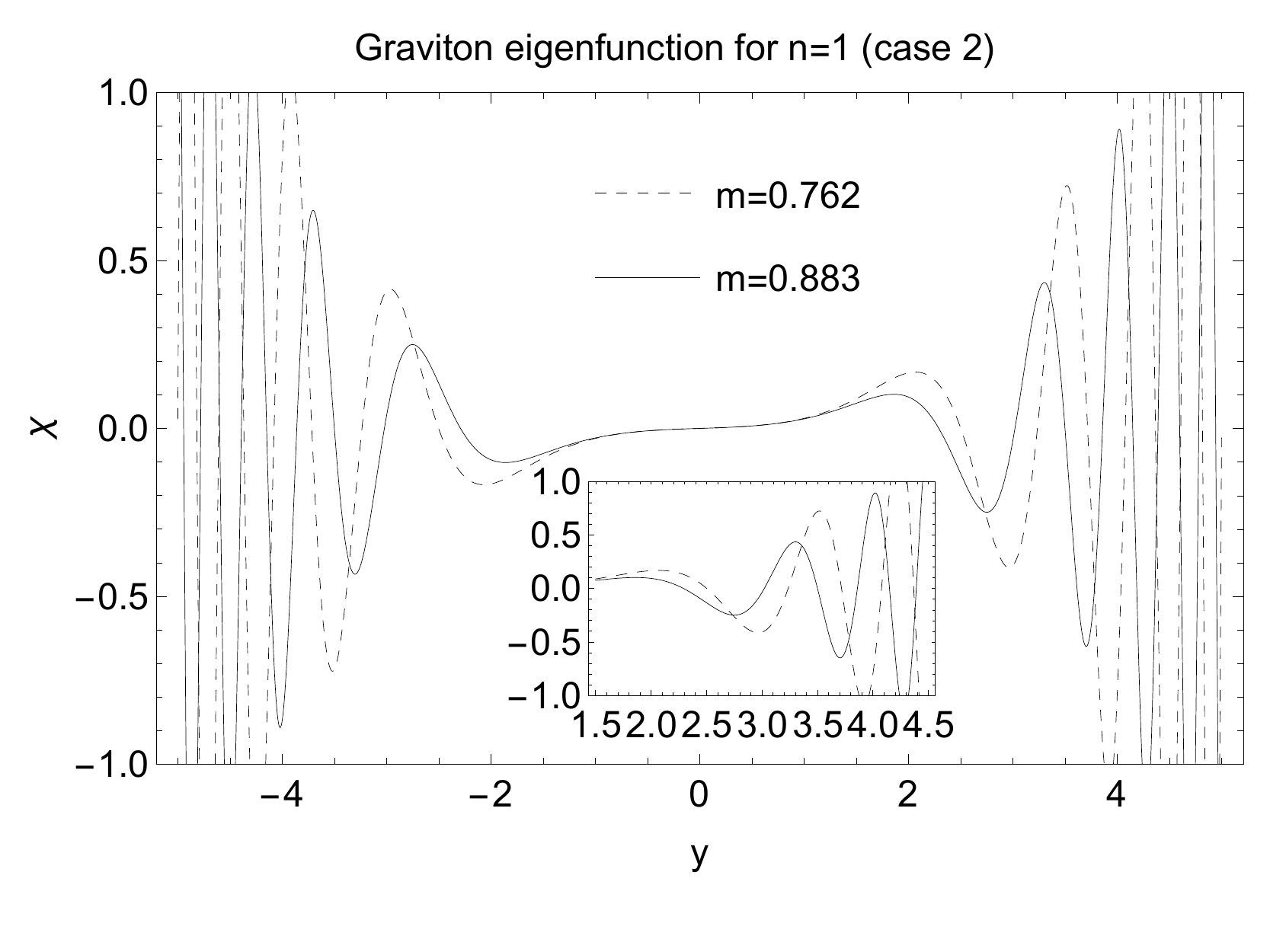} 
\includegraphics[height=6cm]{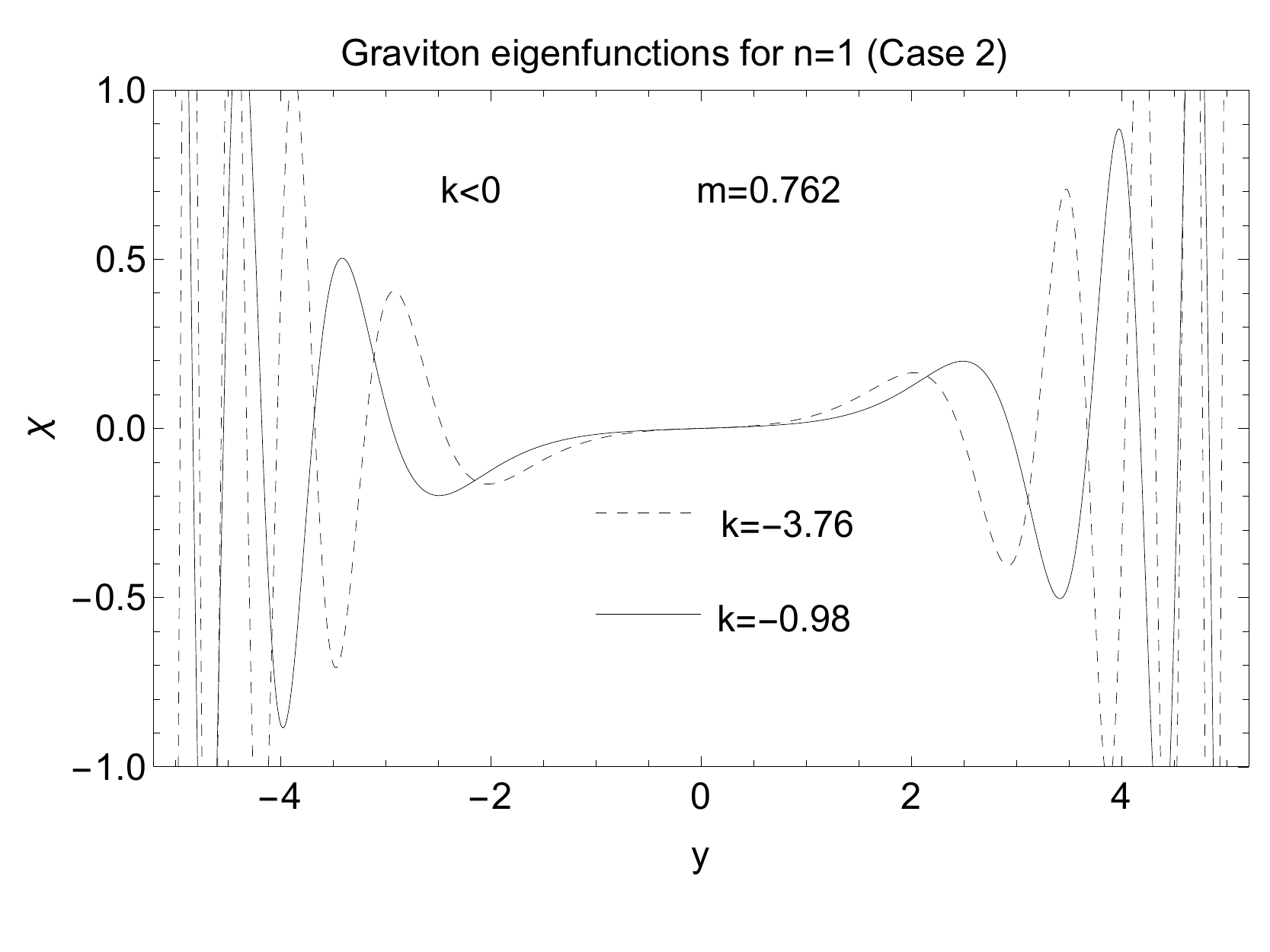}\\
(c) \hspace{8 cm}(d)\\
\includegraphics[height=6cm]{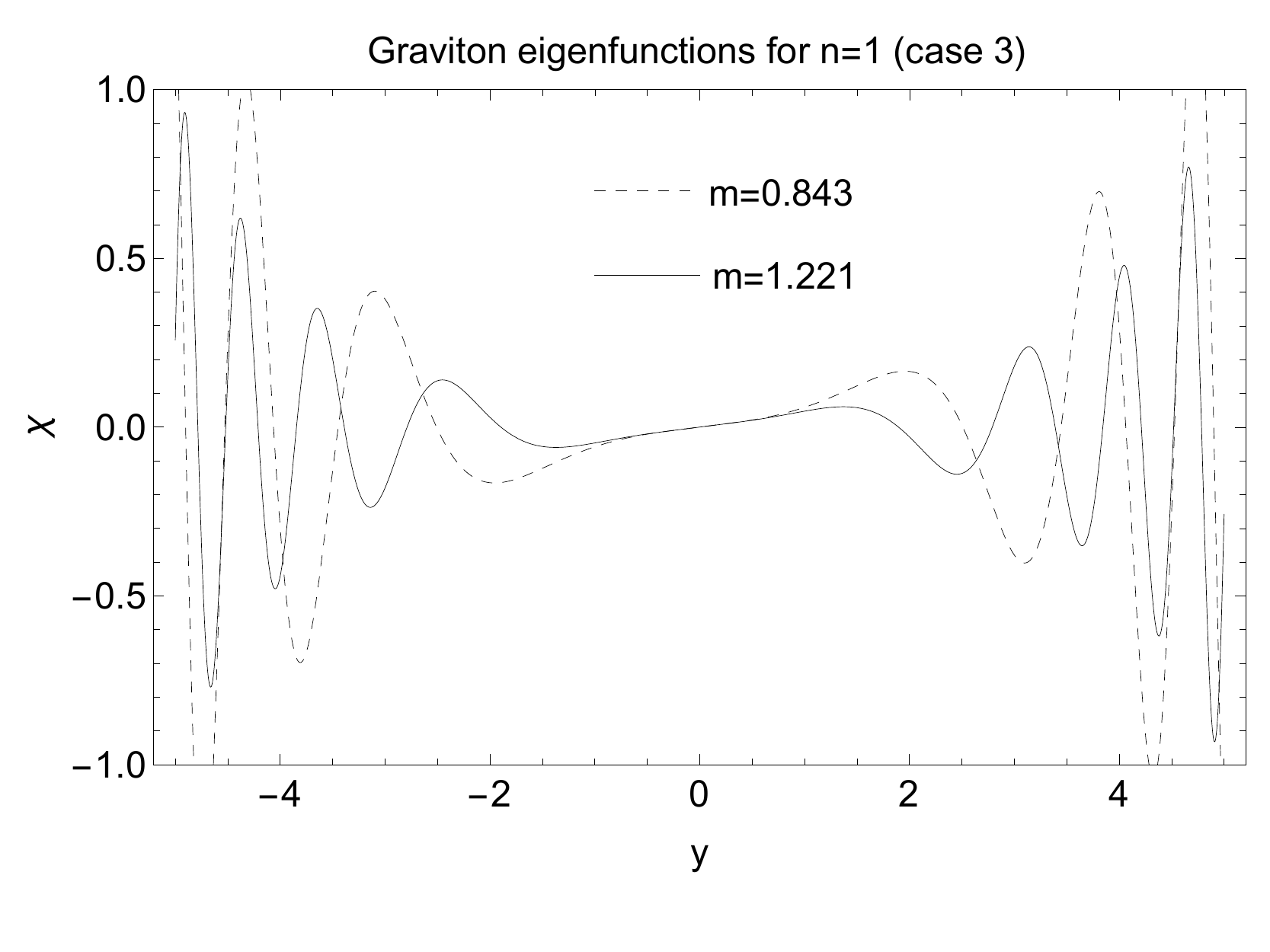} 
\includegraphics[height=6cm]{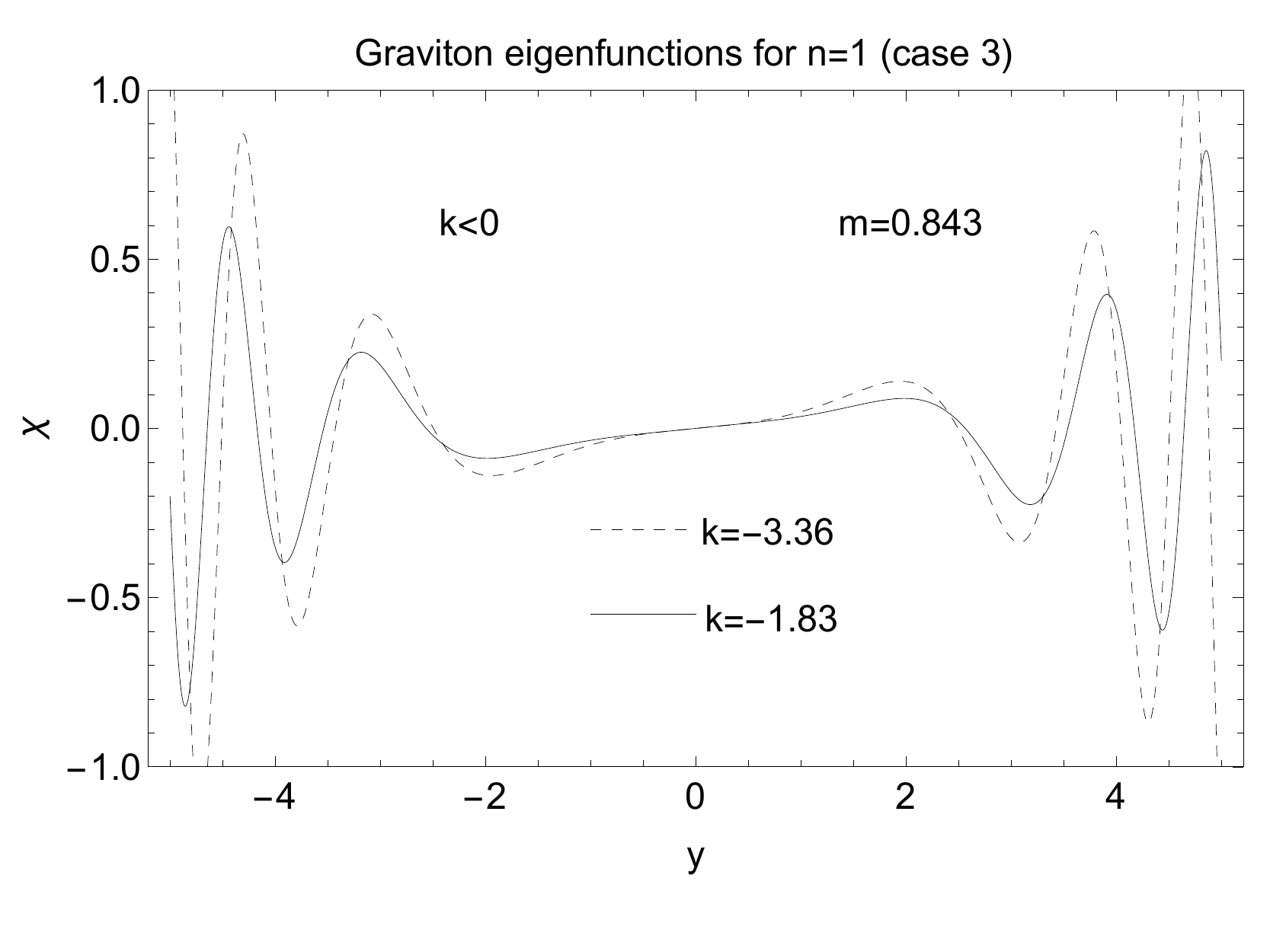}\\
(e) \hspace{8 cm}(f)
\end{tabular}
\end{center}
\caption{Massive modes for $n=1$ and $k_0=\alpha=\beta=1$. (a) and (b) periodical superpotential. (c) and (d) polynomial superpotential. (e) and (f) fractional superpotential.} 
\label{massivemodes}
\end{figure}

For all examples of the $W(\phi)$ superpotentials proposed for the case $n=1$ with $k_0=1$, increasing the eigenvalues of mass increases the oscillations and decreases the amplitudes of the oscillations, as we can see in Figures \ref{massivemodes} $(a)$ for periodical superpotential, \ref{massivemodes} $(c)$ for polynomial superpotential, and \ref{massivemodes} $(e)$ for the fractional superpotential. Taking into account the dependency on $k$ for the periodical superpotential, where there is a decreasing the value of $k$, the oscillations maintain their amplitude and move to the core of the brane (Figure \ref{massivemodes} $b$). For the polynomial superpotential, decreasing the value of $k$, decreases the amplitude of the oscillations and moves to the core of the brane (Figure \ref{massivemodes} $d$). For the fractional superpotential, decreasing the value of $k$, increases the amplitude of oscillations and moves to the core of the brane (Figure \ref{massivemodes} $f$).

\begin{figure}
\begin{center}
\begin{tabular}{ccccccccc}
\includegraphics[height=6cm]{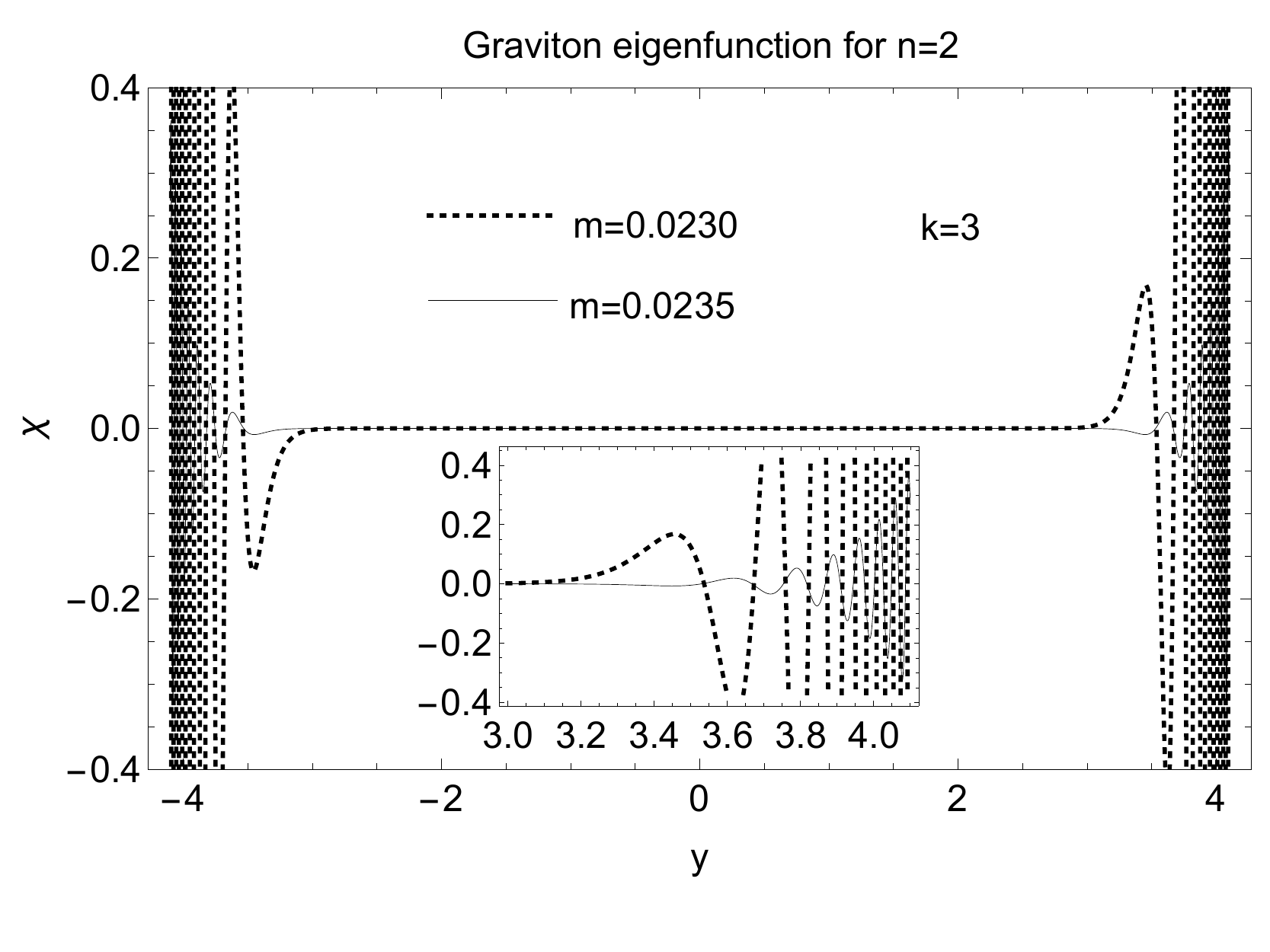}
\includegraphics[height=6cm]{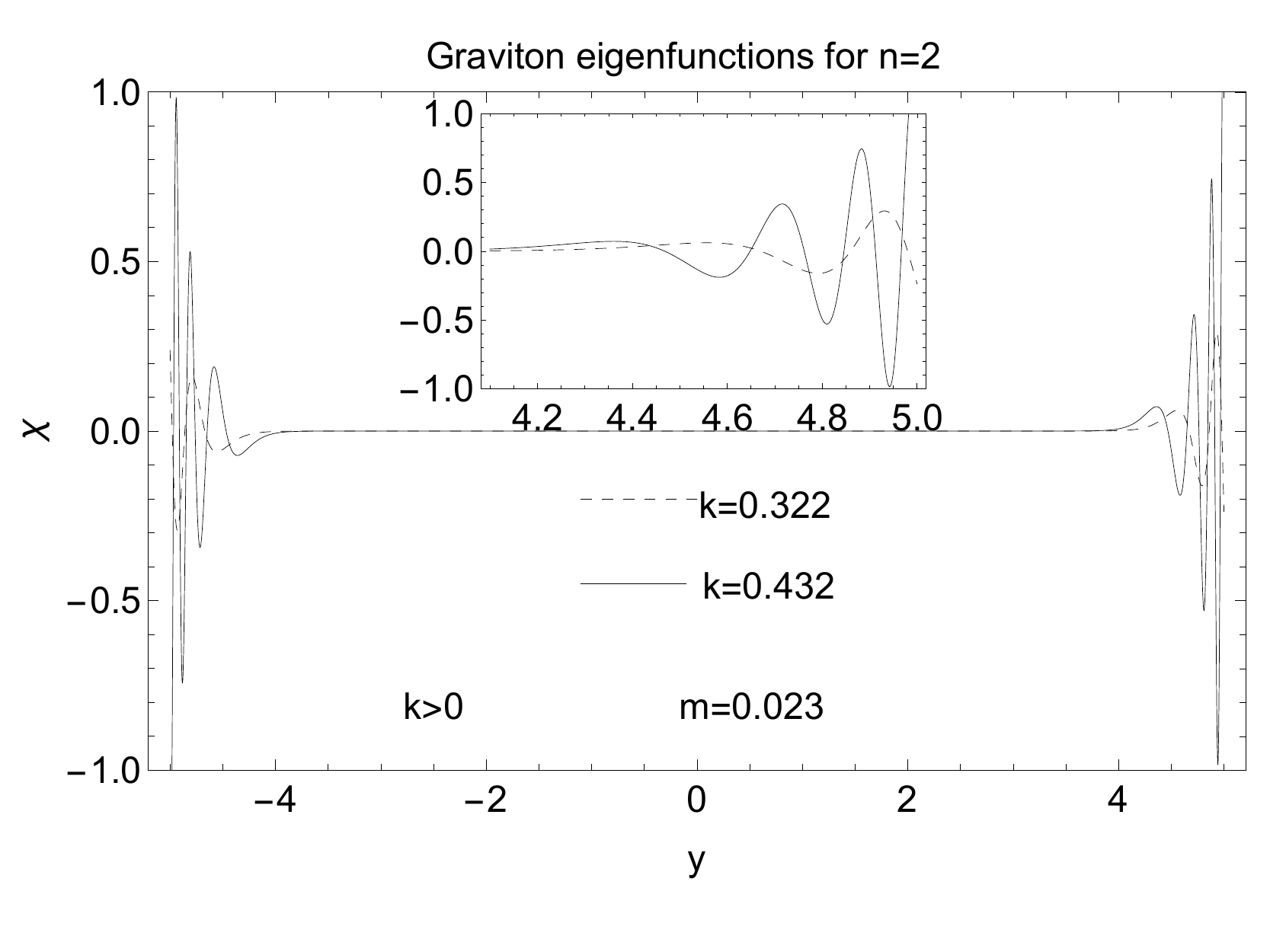}  \\
(a)\hspace{8 cm}(b)
\end{tabular}
\end{center}
\caption{Massive modes for $n=2$ and $k_0=\alpha=\beta=1$, varying the mass eigenvalue (a), and varying $k$ (b).} 
\label{massivemodes2}
\end{figure}

For the example proposed for the case $n=2$ with $k_0=1$, as well as in the examples of $n=1$, increasing the eigenvalues of mass  decreases the amplitudes of the oscillations, as we can see in Fig. \ref{massivemodes2} $(a)$. 
The dependency of $k$ is shown in Fig. \ref{massivemodes2} $(b)$, where we set the first mass eigenvalue and vary the parameter $k$.
When we increase the value of $k$,  increases the intensity and amplitude of the oscillations that moves to the core of the brane.

\subsubsection{$f(T,\mathcal{T})=-T-k_1T^{2}+k_2\mathcal{T}$}

\begin{figure}
\begin{center}
\begin{tabular}{ccccccccc}
\includegraphics[height=6cm]{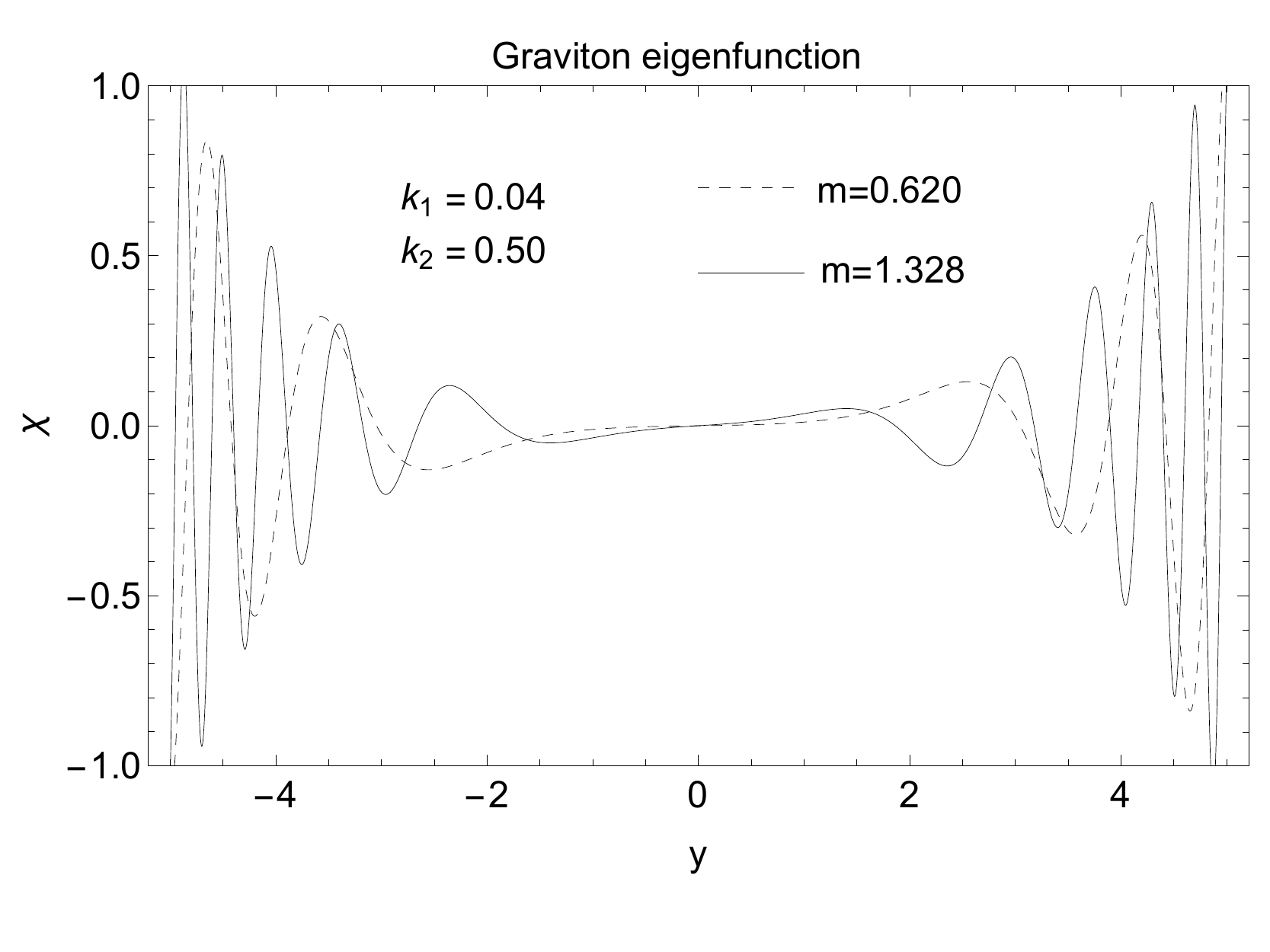}  \\
(a)\\
\includegraphics[height=6cm]{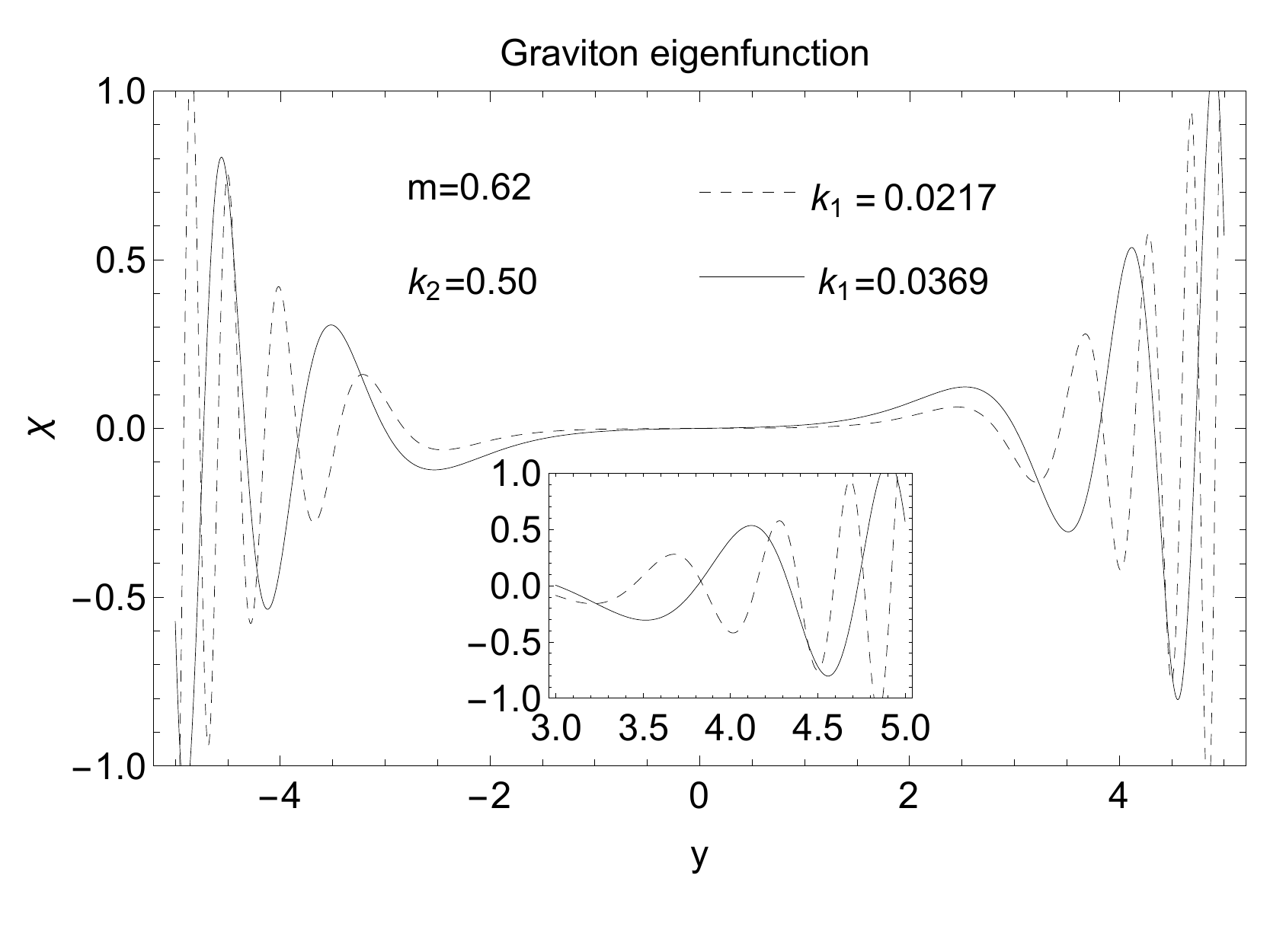}
\includegraphics[height=6cm]{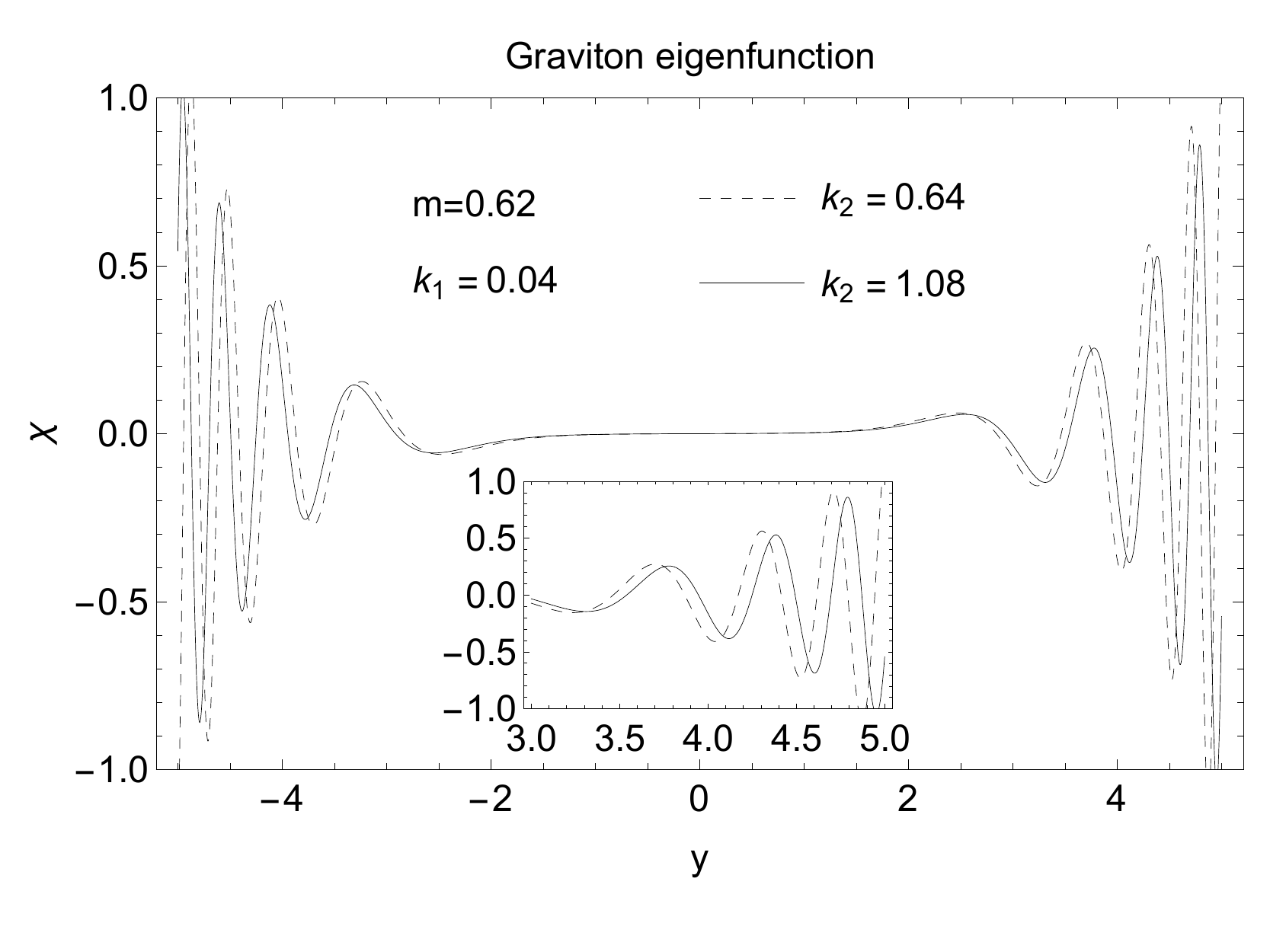}  \\
(b)\hspace{8 cm}(c)
\end{tabular}
\end{center}
\caption{Massive modes with $\alpha=\beta=1$, varying the mass eigenvalue (a), varying $k_1$ (b), and varying $k_2$ (c).} 
\label{massivemodes3}
\end{figure}

In that case, increasing the eigenvalues of mass increases the ocilations and their amplitudes, as we can see in Fig.\ref{massivemodes3} $(a)$ . Whereas is there dependency on $k_1$ and $k_2$, where decreasing the value of $k_1$ increases the ocilations and decreases their applicability (Figure \ref{massivemodes3} $b$) and decreasing the value of $k_2$, increases the amplitude of the oscillations and moves to the core of the brane (Figure \ref{massivemodes3} $c$).

\subsection{Massless modes}
Employing the change to a conformal coordinate 
\begin{equation}
dz=e^{-A}dy,
\end{equation}
 the tensor perturbation Eq.(\ref{KKequation}) is transformed as
\begin{equation}\label{w.17}
\left(\partial_z^2+2H\partial_z+m^2\right)\chi(z)=0,
\end{equation}
where
\begin{eqnarray}\label{34.l}
H=\frac{1}{2}\Big\{3\dot{A}+\frac{1}{ f_T}\Big[24e^{-2A}(\dot{A}^3-\dot{A}\ddot{A}){f}_{TT}+\dot{\mathcal{T}}{f}_{T\mathcal{T}}\Big]\Big\},
\end{eqnarray}
and the dot $(\ \dot{}\ )$  denotes differentiation with respect to $z$. With the change on the wave function $\chi(z)=F(z)\Psi(z)$ in equation Eq.(\ref{w.17}) make it possible the recast into a Sch\"{o}dinger-like equation 
\begin{eqnarray}\label{36.l}
[-\partial_z^2+U(z)]\Psi(z)=m^2\Psi(z),
\end{eqnarray}
where the potential is defined by
\begin{eqnarray}\label{potential}
U(z)=\dot{H}+H^2,
\end{eqnarray}
and 
\begin{eqnarray}
F(z)=e^{-\frac{1}{2}\left( 3A+\int K(z)dz\right)},
\end{eqnarray}
with
\begin{eqnarray}
K(z)=\frac{1}{ f_T}\Big[24e^{-2A}(\dot{A}^3-\dot{A}\ddot{A}){f}_{TT}+\dot{\mathcal{T}}{f}_{T\mathcal{T}}\Big].
\end{eqnarray}

The Schrödinger-like Eq.(\ref{36.l}) can be factorized as
\begin{eqnarray}\label{t5t5}
\left(-\partial_z+H\right)\left(\partial_z+H\right)\Psi(z)=m^2\Psi(z),
\end{eqnarray}
which means that there is no four-dimensional graviton with $m^2<0$, and so any brane solution of $f(T,\mathcal{T})$ gravity theory is stable under the TT tensor perturbation. This feature is not present in the usual formulation of the braneworld, in the framework of the curvature of gravity, proving to be a great advantage of the $f(T,\mathcal{T})$ theory.
Eq.(\ref{t5t5}) represents an equation of supersymmetric quantum mechanics. The superpotential $H$ and the supersymmetric quantum mechanical form of the potential $U$ ensures the absence of tachyonic KK gravitational modes. This behavior is also obeyed in the TEGR limit, where $H\rightarrow \frac{3}{2}\dot{A}$.

Besides the spectrum stability, the potential in Eq. (\ref{potential}) allows a massless KK mode of form 
\begin{eqnarray}
\Psi_0=N_0\ e^{\frac{1}{2}\left( 3A+\int K(z)dz\right)},
\end{eqnarray}
where $N_0$ is a normalization constant. In order to recover the four-dimensional gravity, the zero mode should be localized on the brane. Note that, in the limit tending to the TEGR, $\Psi_0\rightarrow N_0\ e^{\frac{3}{2}A}$. Let us go now investigate the massless gravity localization problem.
\subsubsection{$f(T,\mathcal{T})=k_0\mathcal{T}+kT^{n}$}

\begin{figure}
\begin{center}
\begin{tabular}{ccccccccc}
\includegraphics[height=5cm]{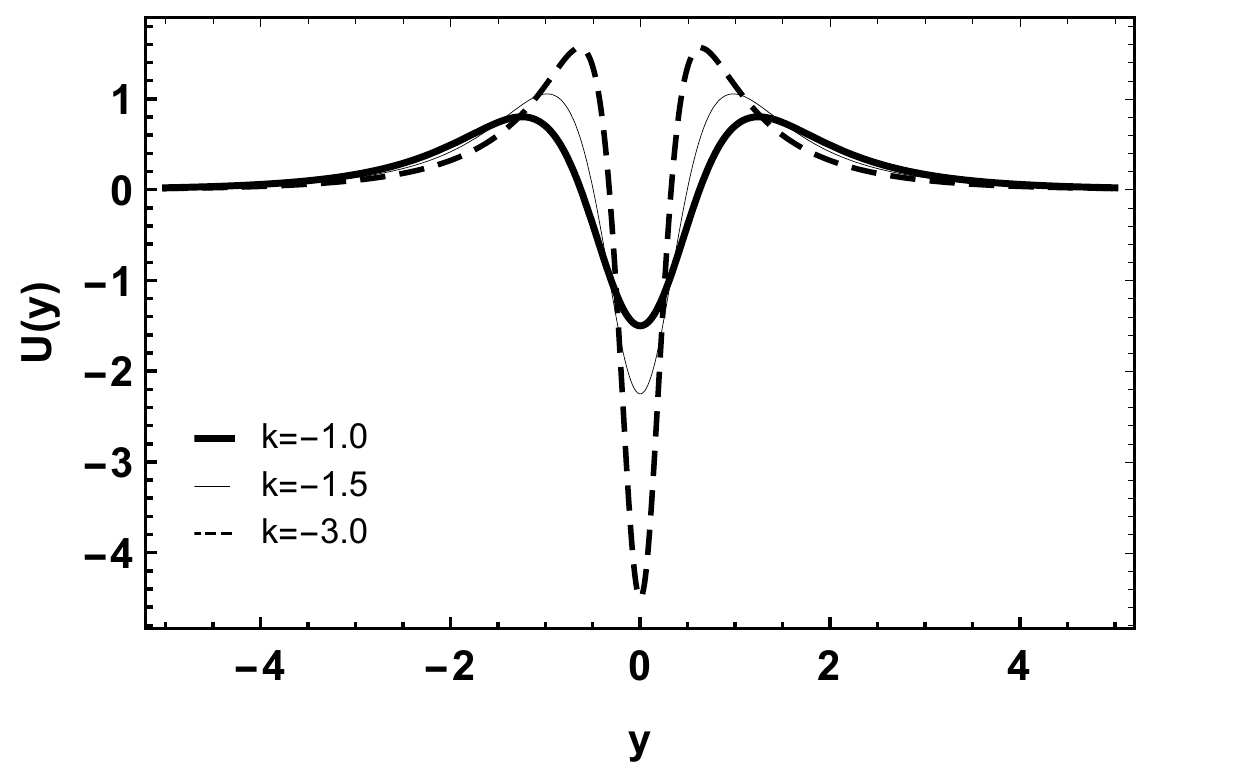}
\includegraphics[height=5cm]{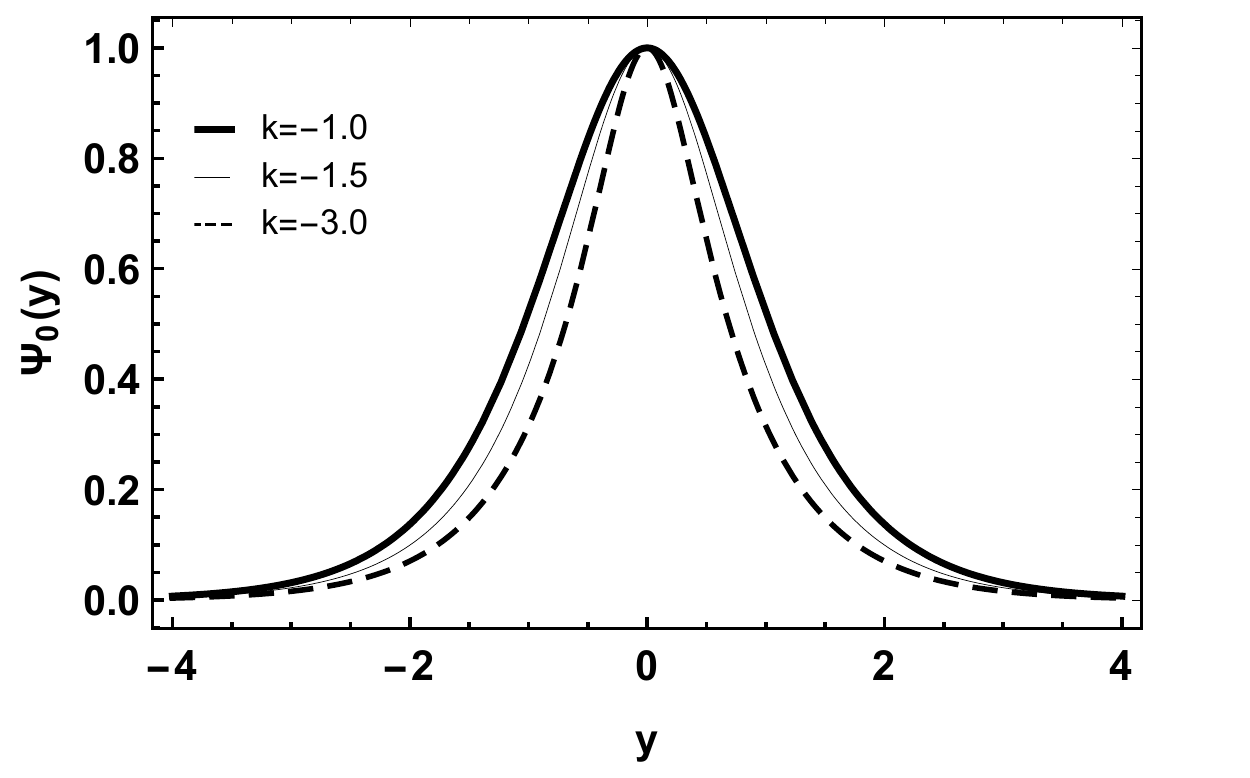}\\
(a) \hspace{8 cm}(b)\\
\includegraphics[height=5cm]{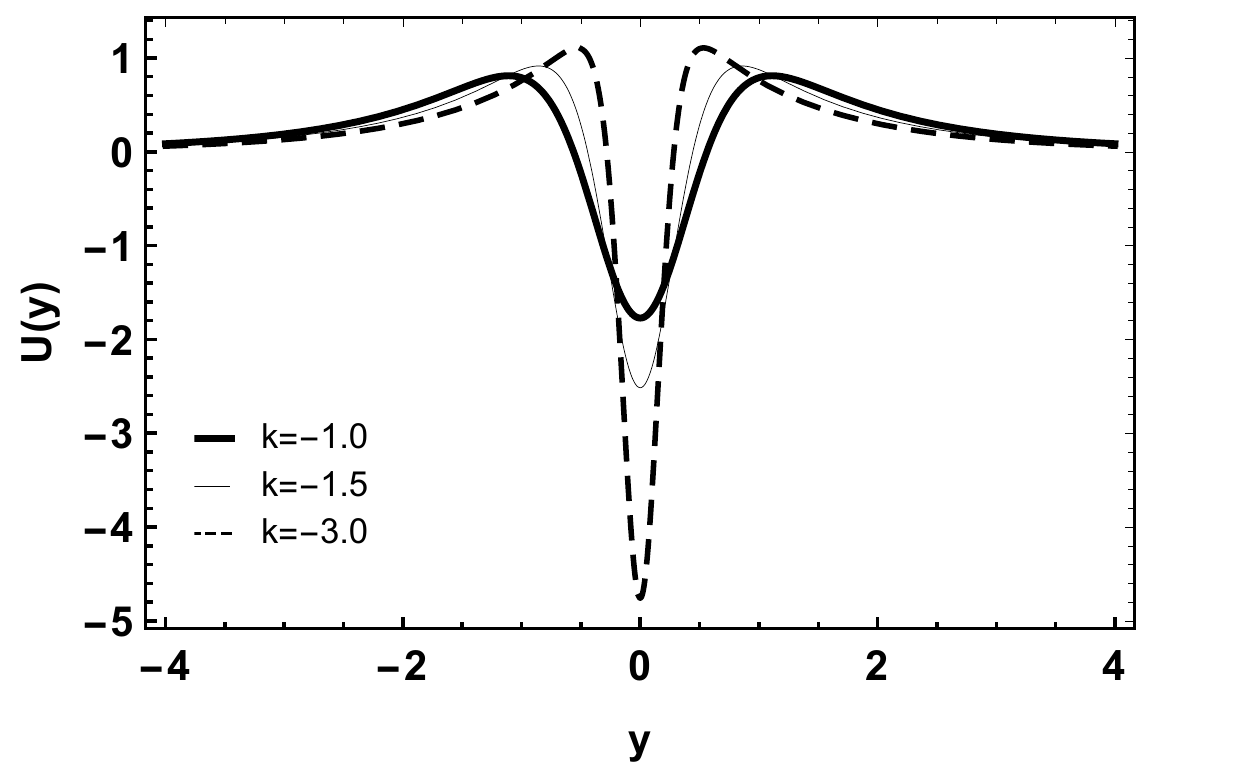} 
\includegraphics[height=5cm]{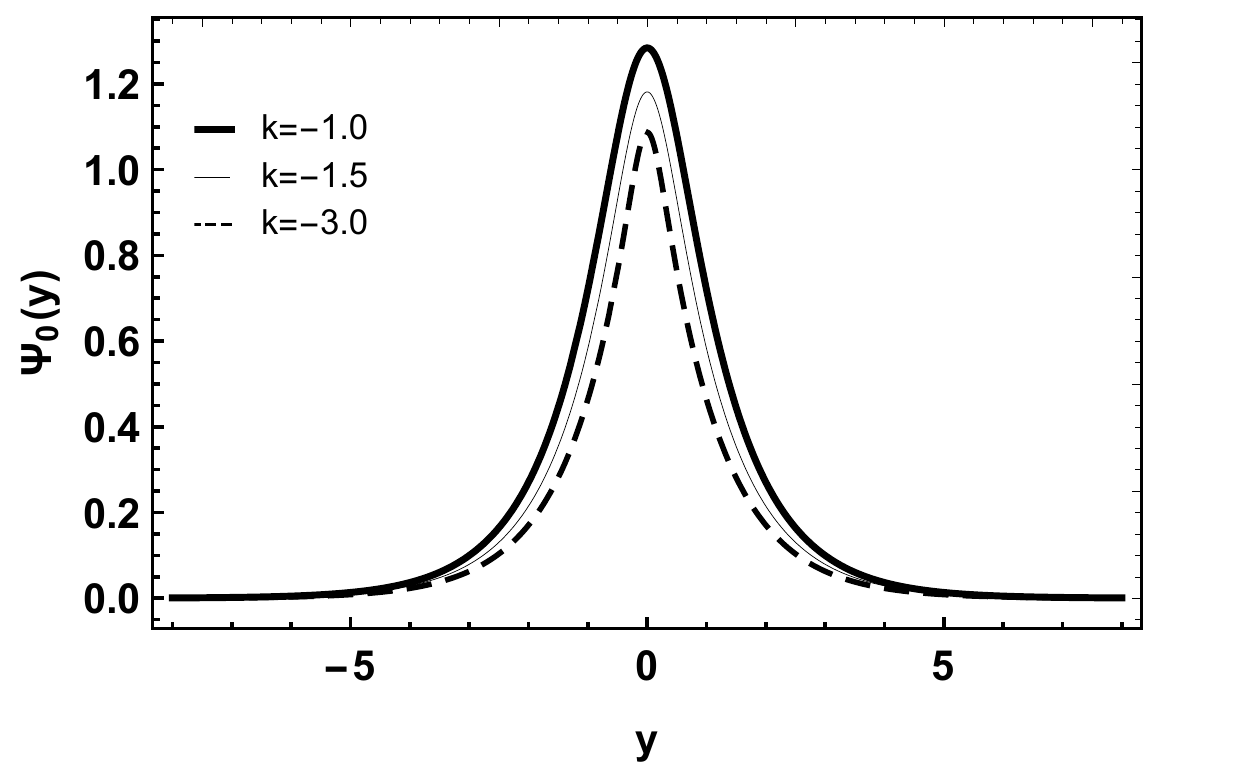}\\
(c) \hspace{8 cm}(d)\\
\includegraphics[height=5cm]{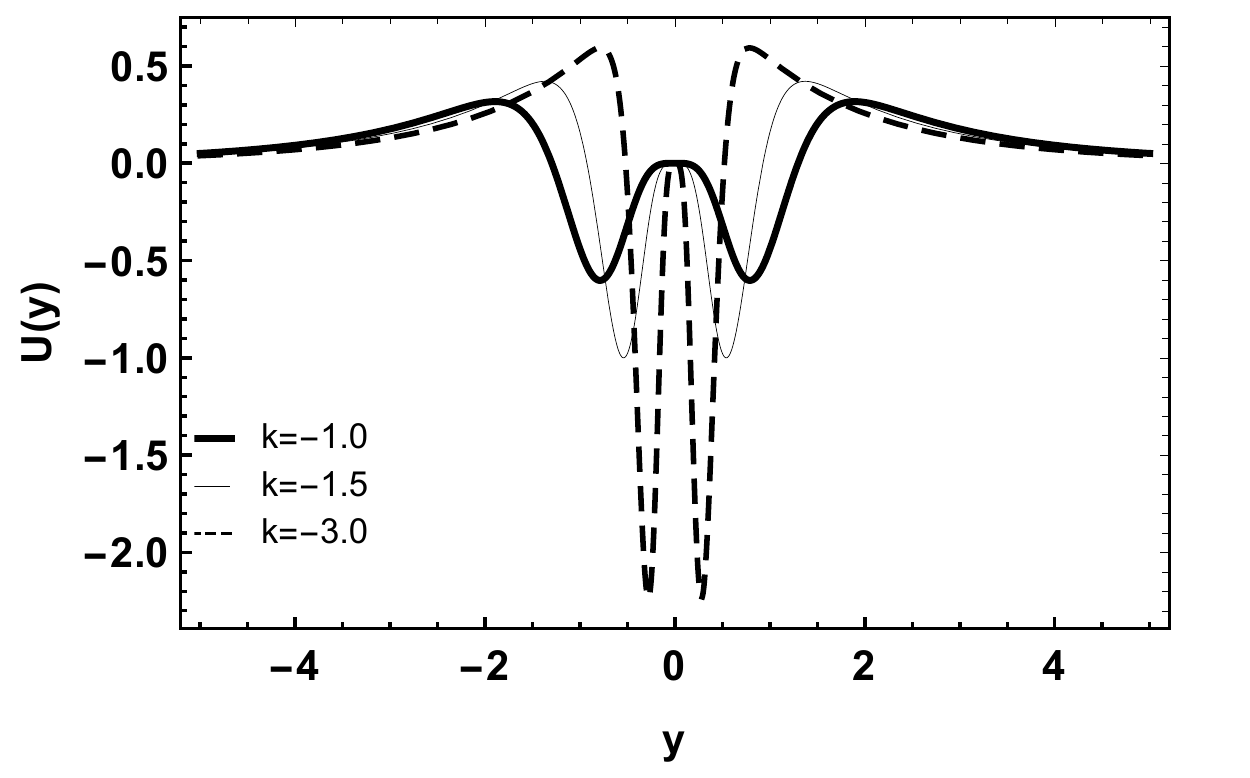} 
\includegraphics[height=5cm]{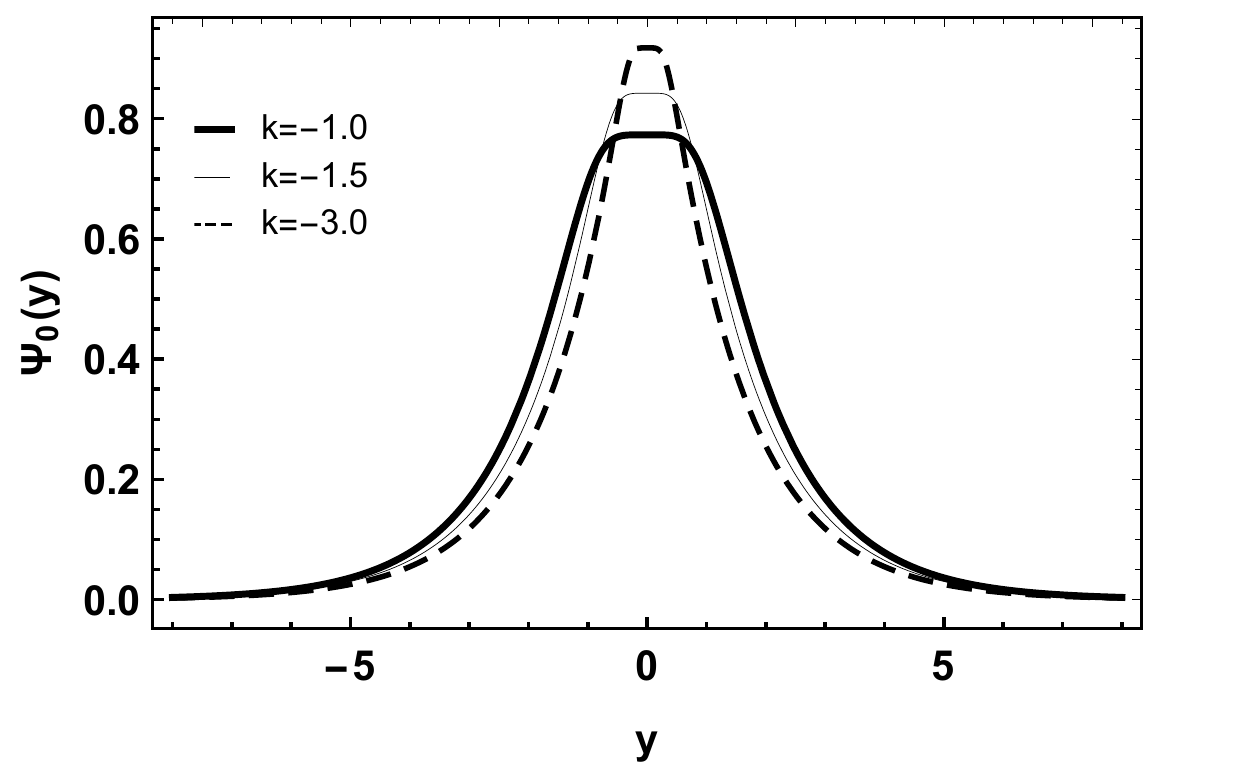}\\
(e) \hspace{8 cm}(f)
\end{tabular}
\end{center}
\caption{Plots of the effective potential, and zero mode for $n=1$ and $k_0=\alpha=\beta=1$. (a) and (b) periodical superpotential. (c) and (d)  polynomial superpotential. (e) and (f) fractional superpotential.
\label{figPE1}}
\end{figure}

For the periodical superpotential proposed for the case $n=1$ with $k_0=1$, as we can see from Fig.\ref{figPE1}($a$), when $k$ decreases, it intensifies the potential barriers away from the origin and the potential well around the origin increases. The zero mode wave function has only one peak getting more localized as seen in Fig.\ref{figPE1}($b$). Similar behavior happens for the polynomial superpotential, when $k$ decreases, it intensifies the potential barriers away from the origin and the potential well around the origin increases (figure \ref{figPE1} $c$). This behavior is reflected in the zero mode wave functions, making them more localized and decreasing their amplitude (figure \ref{figPE1} $d$).

For the fractional superpotential, as we can see from Fig.\ref{figPE1}($e$), when $k$ decreasing, it intensifies the potential barriers away from the origin and the potential well around the origin increases and splits in two. The zero mode wave function has a flattened peak, which  decreases  as $k$ decreases, as seen in Fig.\ref{figPE1}($f$).  This feature reflects the brane internal structure,
which tends to split the brane.

\begin{figure}
\begin{center}
\begin{tabular}{ccccccccc}
\includegraphics[height=5cm]{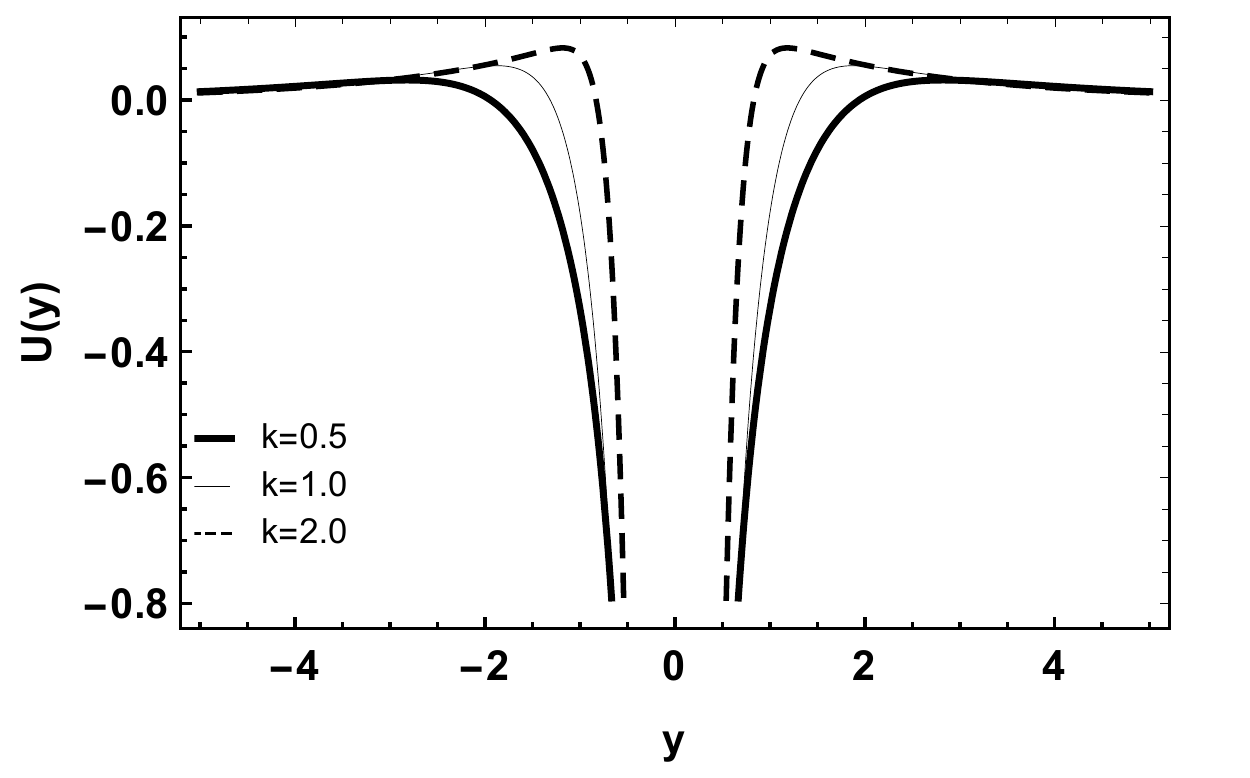}
\includegraphics[height=5cm]{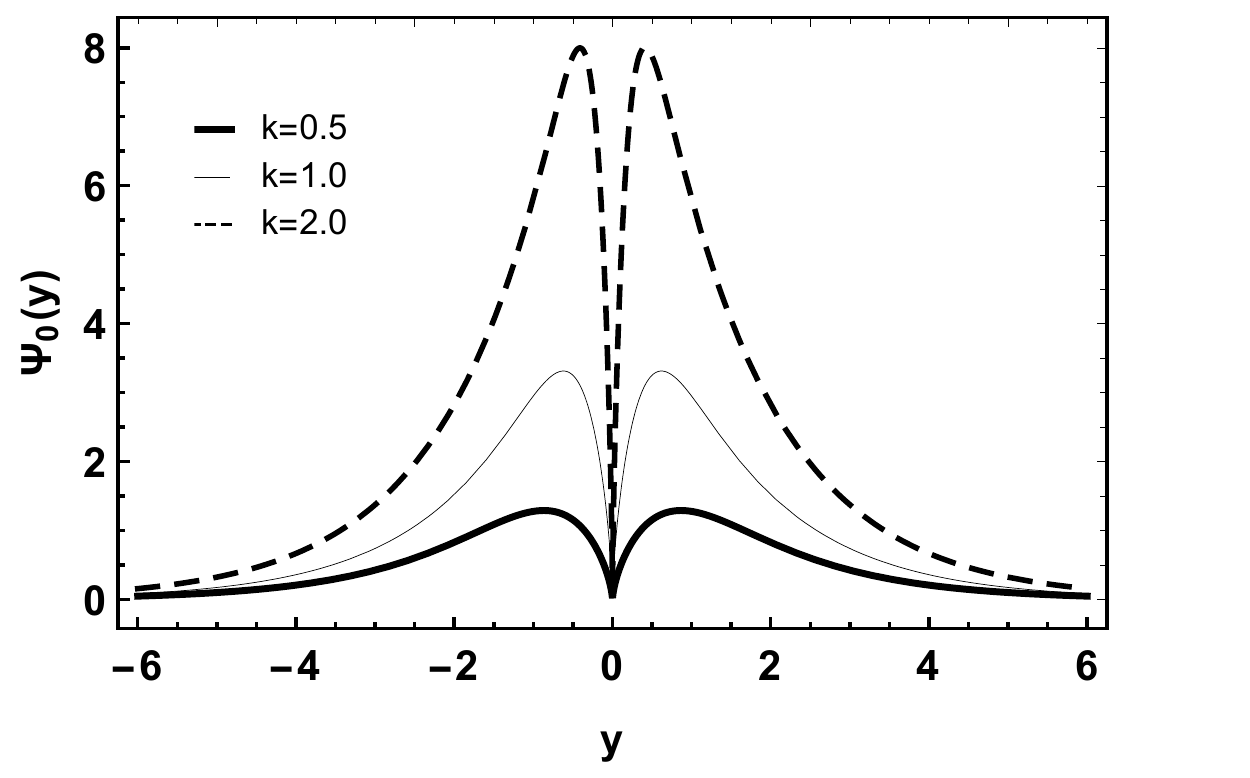}\\
(a) \hspace{8 cm}(b)
\end{tabular}
\end{center}
\caption{Plots of the effective potential (a), and zero mode (b), for $n=2$, where $k_0=\alpha=\beta=1$.  
\label{figPE2}}
\end{figure}

For the example proposed for the case $n=2$ with $k_0=1$, when $k$ increases, as we can see from Fig.\ref{figPE2}($a$), two new potential barrier appear away from the origin and the potential well around of the origin 
it has the shape of an infinite well that tends to a delta well when we increase the parameter $k$. As a result, the zero mode wave function splits into two peaks as shown in Fig.\ref{figPE2}($b$). Again, this feature reflects the brane internal structure,
which tends to split the brane.
\subsubsection{$f(T,\mathcal{T})=-T-k_1T^{2}+k_2\mathcal{T}$}

\begin{figure}
\begin{center}
\begin{tabular}{ccccccccc}
\includegraphics[height=5cm]{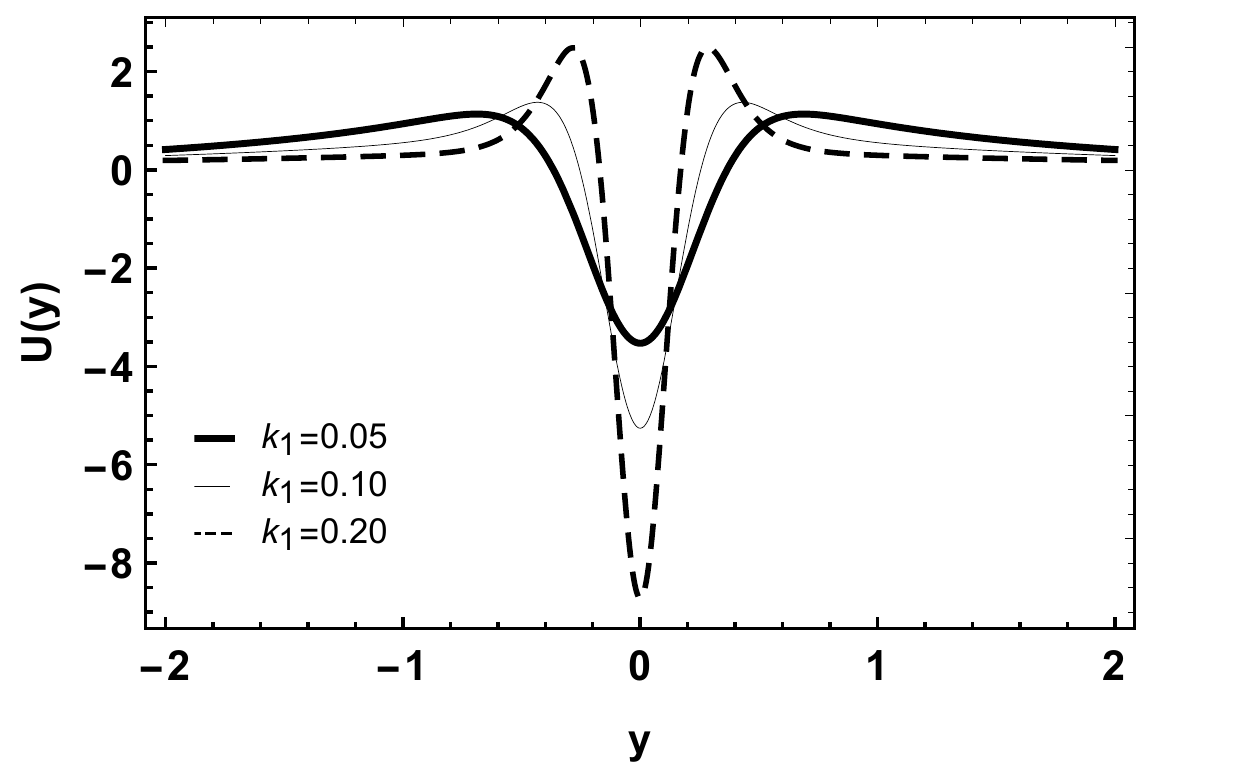}
\includegraphics[height=5cm]{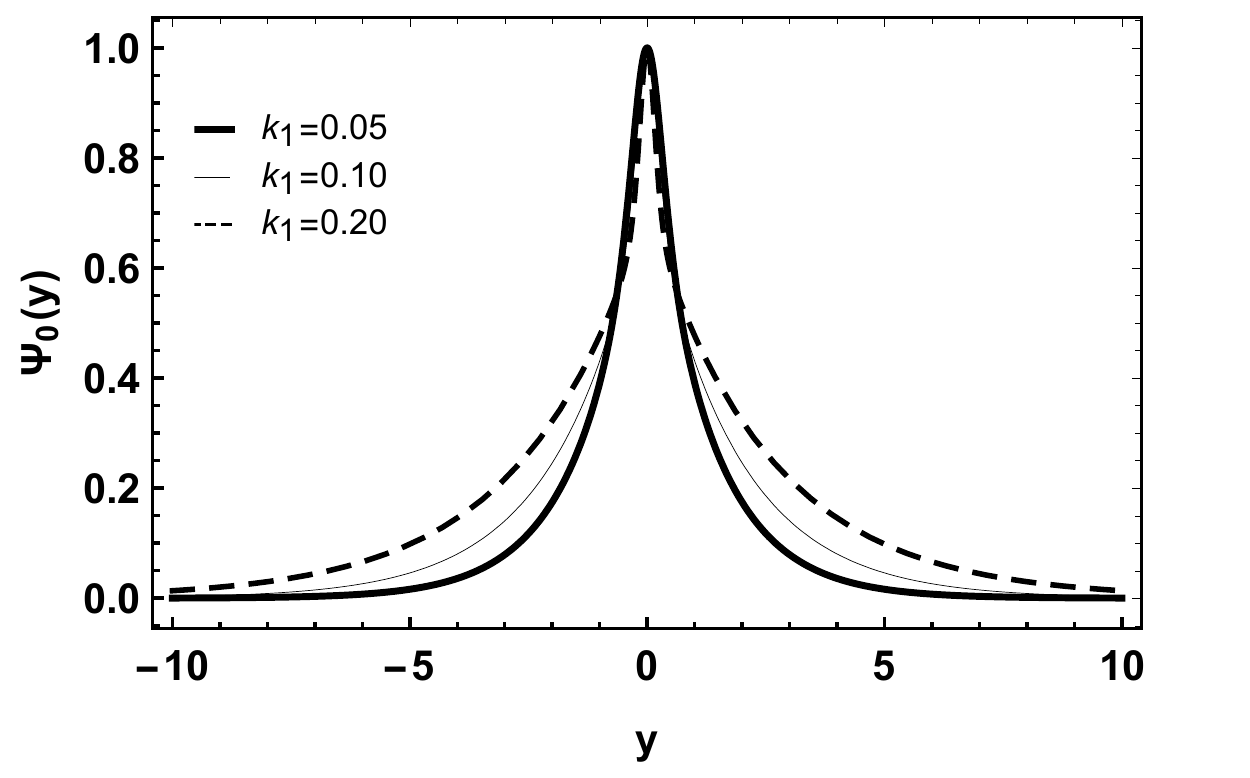}\\
(a) \hspace{8 cm}(b)
\end{tabular}
\end{center}
\caption{Plots of the effective potential (a), and zero mode (b), where $k_2=0.5 $ and $\alpha=\beta=1$.  
\label{figPE3}}
\end{figure}

\begin{figure}
\begin{center}
\begin{tabular}{ccccccccc}
\includegraphics[height=5cm]{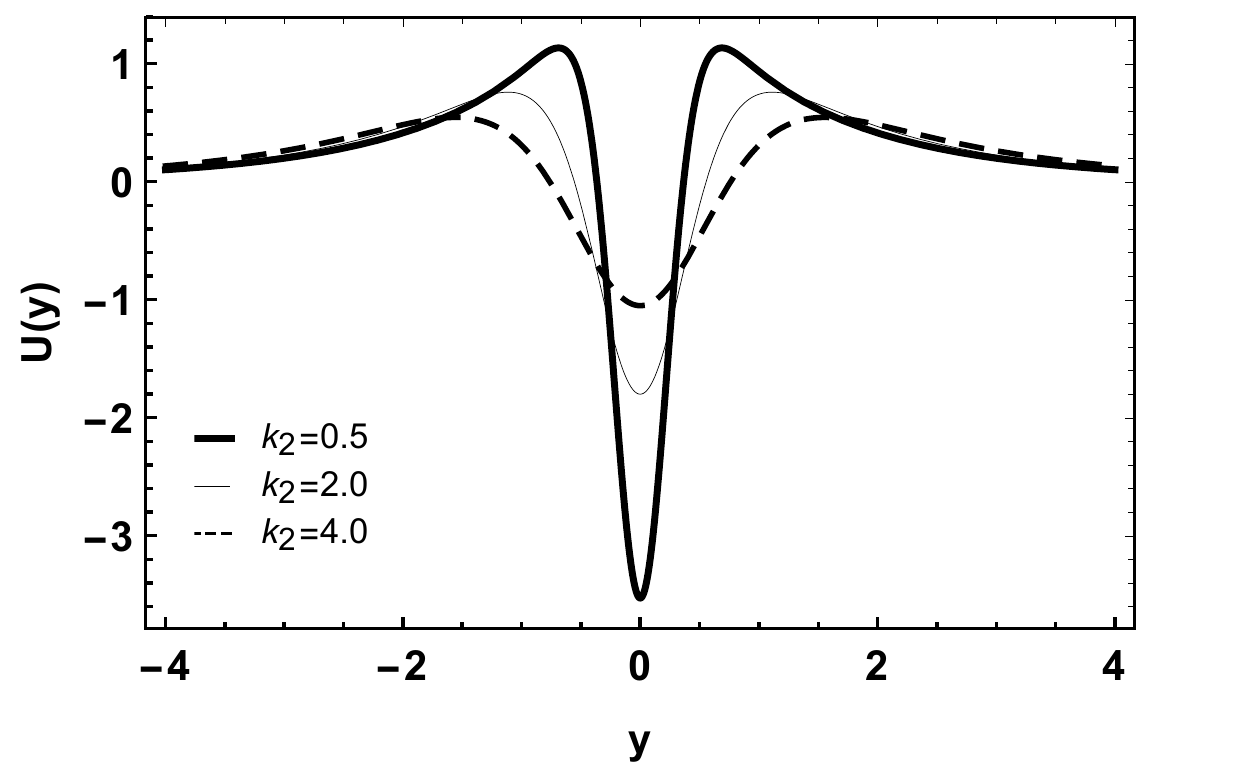}
\includegraphics[height=5cm]{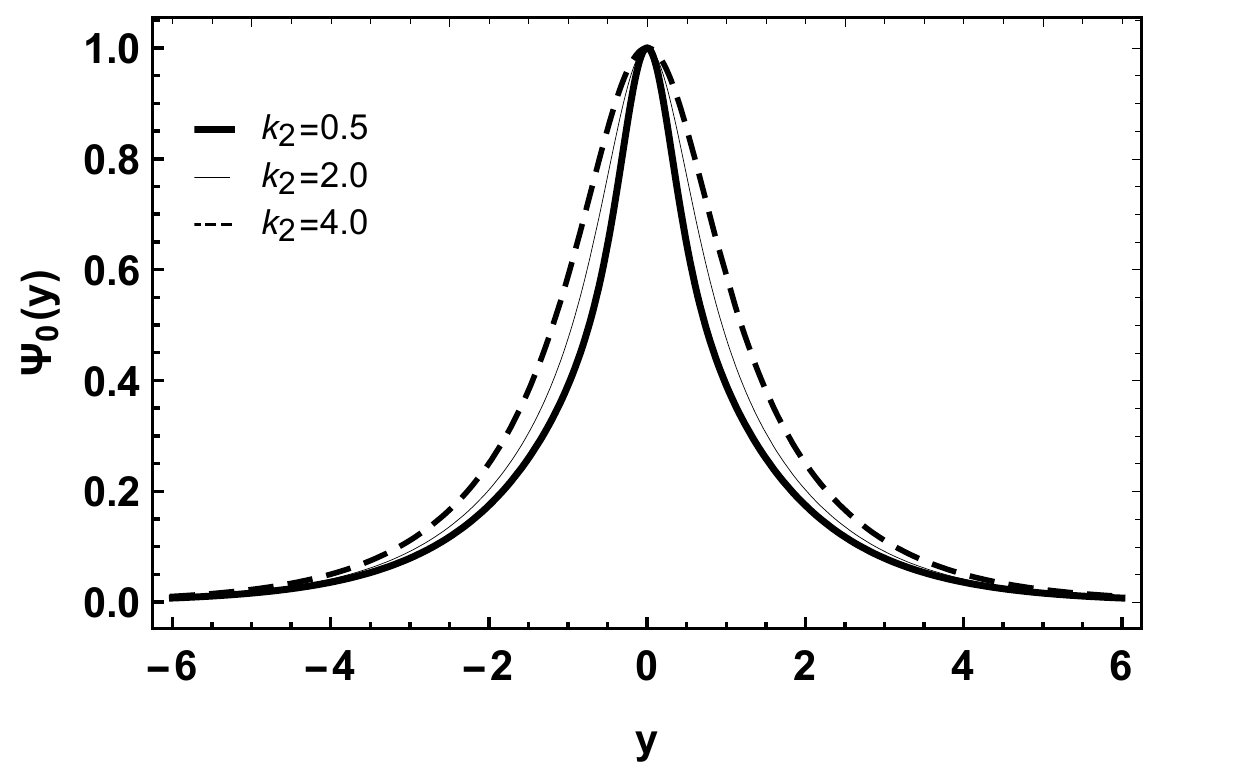}\\
(a) \hspace{8 cm}(b)
\end{tabular}
\end{center}
\caption{Plots of the effective potential (a), and zero mode (b), where $k_1=0.05 $ and $\alpha=\beta=1$.
\label{figPE4}}
\end{figure}

Parameters $k_{1,2}$ also change the massless modes. When $k_1$ increases, it intensifies the potential barriers away from the origin and the potential well around the origin increases (Fig. \ref{figPE3} $a$). This behavior is reflected in the zero mode wave functions (Fig. \ref{figPE3} $b$). In turn, when $k_2$ decreases,  it intensifies the potential barriers away from the origin and the potential well around the origin increases (Fig. \ref{figPE4} $a$). The zero mode wave function has only one peak getting more localized as seen in Fig.\ref{figPE4}($b$).

\section{Final remarks}
\label{sec4}

Through first-order formalism, we study effects of torsion and trace of the energy-momentum tensor on a braneworld in the context of the $f(T,\mathcal{T})$ modified teleparallel gravity, where we propose two particular cases $f(T,\mathcal{T})=k_0\mathcal{T}+kT^{n}$ and $f(T,\mathcal{T})=-T-k_1T^{2}+k_2\mathcal{T}$.  The solutions of the thick brane system are completely determined by the so-called superpotential function $W(\phi)$. Then, we propose some particular cases of polynomial and periodic superpotentials. The torsion  and trace of the energy-momentum tensor produces an inner brane structure modifying the behavior of the brane. In Refs.\cite{Menezes,ftborninfeld} the authors obtained something similar by studying only the influence of torsion.

The profile of the scalar field, and warp factor, and potential, and energy density are controlled by the parameters that control the torsion and trace of the energy-momentum tensor. The profile of the scalar field suggests a topological stability. For the fractional superpotential in case $n=1$ with $k_0$ from $f(T,\mathcal{T})=k_0\mathcal{T}+kT^{n}$, we obtain a double-kink solution, which generates a split in the brane, intensified by the $k$ parameter, evinced by the energy density components. Something similar was achieved to the superpotential example for which $n=2$. Although we found a kink solution, it generates a splitting in the brane, intensified by the $k$ parameter. We can clearly see the influence of the trace of the energy-momentum tensor, in the case of $f(T,\mathcal{T})=-T-k_1T^{2}+k_2\mathcal{T}$ varying the parameter $k_2$. We notice that $k_2$ controls the  thickness of the solution, where the warp factor narrows as $k_2$ decreases, modifying also the potential  and the energy density.

The torsion and the trace of the energy-momentum tensor lead to modifications of the massive gravitons. For the two chosen models $f(T,\mathcal{T})$,  the amplitude and the proximity of the oscillations to the core of the brane depend on the parameters that control the torsion and trace of the energy-momentum tensor, showing that the interaction of the massive modes with the torsion and trace of the energy-momentum tensor is more intense inside the brane core. A similar behavior was obtained in Ref.\cite{Moreira2021} where the authors study the KK massive modes in a modified teleparallel gravity $f(T,B)$, where $B$ is the so-called boundary term.

The parameters that control the torsion and the trace of the energy-momentum tensor, intensify the behavior of the Schr\"{o}dinger-like potential, modifying the KK modes. We found two interesting configurations for the $f(T,\mathcal{T})=k_0\mathcal{T}+kT^{n}$ case. The first setting is for $n=1$ with $k_0$, where  decreasing $k$,  the potential well around the origin  splits in two ones, and the zero mode wave function features a flattened peak, which  decreases  as $k$ decreases. The second setting is for  $n=2$ with $k_0$, where two new potential barrier appear away from the origin and the potential well around of the origin has the shape of an infinite well that tends to a delta well when we increase the parameter $k$. As a result, the zero mode wave function splits into two peaks. These features reflects the brane internal structure, which tends to splitting the brane.

\section*{Acknowledgments}
\hspace{0.5cm}The authors thank the Conselho Nacional de Desenvolvimento Cient\'{\i}fico e Tecnol\'{o}gico (CNPq), grants n$\textsuperscript{\underline{\scriptsize o}}$ 312356/2017-0 (JEGS) and n$\textsuperscript{\underline{\scriptsize o}}$ 308638/2015-8 (CASA), and Coordenaçao de Aperfeiçoamento do Pessoal de Nível Superior (CAPES), for financial support. The authors also thank the anonymous referees for their valuable comments and suggestions.


\end{document}